\documentclass[twocolumn,showpacs,reprint,amsmath,amssymb,aps,prb]{revtex4-1}
\usepackage{graphicx}
\usepackage{dcolumn}
\usepackage{bm}
\usepackage{subfigure}
\usepackage{natbib}
\usepackage{multirow}
\usepackage{soul}
\usepackage[usenames,dvipsnames]{xcolor}
\usepackage{soul}
\usepackage{floatrow}
\usepackage{amsmath}
\usepackage{amsfonts}
\usepackage[hyperfootnotes=false]{hyperref}
\hypersetup{
 colorlinks=true,
 citecolor=blue,
 linkcolor=blue,
 urlcolor=Blue}
\begin{document}
\title{First principle design of  new thermoelectrics from TiNiSn based pentanary alloys based on 18 valence electron rule}
\author{Mukesh K. Choudhary\textit{$^{1,2}$}  and P. Ravindran\textit{$^{1,2}$} }
\email{raviphy@cutn.ac.in}

\affiliation{$^{1}$Department of Physics, School of Basics and Applied Sciences, Central University of Tamil Nadu, Thiruvarur, India}
\affiliation{$^{2}$Simulation Center for Atomic and Nanoscale MATerials (SCANMAT) Central University of Tamil Nadu, Thiruvarur, India.}
\begin{abstract}
In this study we have reported electronic structure, lattice dynamics, and thermoelectric (TE) transport properties of a new family of pentanary substituted TiNiSn systems using the 18 valence electron count (VEC) rule. We have modeled the pentanary substituted TiNiSn by supercell approach with the aliovalent substitution, inspite of the traditional isoelectronic substitution. Structural optimization and electronic structure calculations for all the pentanary substituted TiNiSn systems were performed using the projected augmented wave method and the TE transport properties were investigated using the full potential linearized augmented plane wave method, with the semi$-$classical Boltzmann transport theory, under the constant relaxation time approximation. We have performed the detailed analysis of electronic structure, lattice dynamics, and TE transport properties selected systems from this family. From our calculated band structures and density of states we show that by preserving the 18 VEC through aliovalent substitutions at Ti site of  TiNiSn semiconducting behavior can be achieved and hence one can tune the band structure and band gap to maximize the thermoelectric figure of merit (\textit{ZT}) value. Two approaches have been used for calculating the lattice thermal conductivity ($\kappa_{L}$),  one by fully solving the  linearized phonon Boltzmann transport  (LBTE) equation from first$-$principles anharmonic lattice dynamics calculations implemented in Phono3py code and other using Slack's equation with calculated Debye temperature and Gr\"{u}neisen parameter  using the calculated elastic constant values. At high temperatures the calculated $\kappa_{L}$ and \textit{ZT} values from both these methods show very good agreement. The calculated $\kappa_{L}$ values decreases from parent TiNiSn to pentanary substituted TiNiSn systems as expected due to fluctuation in atomic mass. The substitution of atoms with different mass creates more phonon scattering centers and hence lower the $\kappa_{L}$ value. 
The calculated $\kappa_{L}$ for Hf containing systems La$_{0.25}$Hf$_{0.5}$V$_{0.25}$NiSn and non Hf containing system La$_{0.25}$Zr$_{0.5}$V$_{0.25}$NiSn calculated from Phono3py (Slack's equation) are found to be 0.37 (1.04) and 0.16 (0.95) W/mK, at 550\,K, respectively and the corresponding  \textit{ZT} value are found to be 0.54 (0.4) and 0.77 (0.53). Among the considered systems, the calculated phonon spectra and heat capacity show that La$_{0.25}$Hf$_{0.5}$V$_{0.25}$NiSn  has more optical$-$acoustic band mixing which creates more phonon$-$phonon scattering and hence lower the $\kappa_{L}$ value and maximizing the \textit{ZT}. Based on the calculated results we conclude that one can design  high efficiency thermoelectric materials by considering 18 VEC rule with aliovalent substitution.   
\end{abstract}
\maketitle

\section{Introduction}
In  21$^{st}$ century, the global demand for energy consumption has increased tremendously. Increase in trend of  global warming due to the growth of  industries and electronic devices has widespread concern for developing new technologies and strategies to efficiently use natural resources and convert waste heat energy into useful  clean forms of energy without wasting the future availability of natural resources. With the increase of population and modern lifestyle with economic growth, the requirement of energy for an individual has increased tremendously. Providing the energy needs of every individual has become a challenging task. Thermoelectric (TE) materials serve a great potential to convert the waste form of heat into a clean form of energy. The TE materials are also very usful for efficient  cooling, one of the another emerging areas to minimize climate change~\cite{bell2008cooling}. In TE devices, the thermoelectric effect can be used to convert waste heat into electricity, measure temperature, and change the temperature of the object by changing the polarity of the applied voltage. There are three main types of  TE effects~\cite{jaumot1958thermoelectric}, namely the Seebeck effect which converts temperature differences directly into electricity, the Peltier effect produces heat or cold at an electrical junction where two different conductors are connected and the Thomson effect produces heat or cold in a current$-$ carrying conductor with a temperature gradient.  Most of the research so far focused on the Seebeck effect for TE power generators and the Peltier effect for cooling~\cite{zhao2014review}. 
The efficiency of TE materials can be calculated by the dimensionless TE figure of merit  
\begin{equation} \label{eqn1} 
\textit{ZT} = {S^{2}\sigma T}/{\kappa}
\end{equation}
 where $S$ is the Seebeck coefficient, $\sigma$ is the electrical conductivity, T is the absolute temperature, and $\kappa$ is the total thermal conductivity comprise of  electronic part $\kappa_{e}$ and the lattice part $\kappa_{L}$.  

\begin{figure*}
\centering
\includegraphics[height=5.5cm]{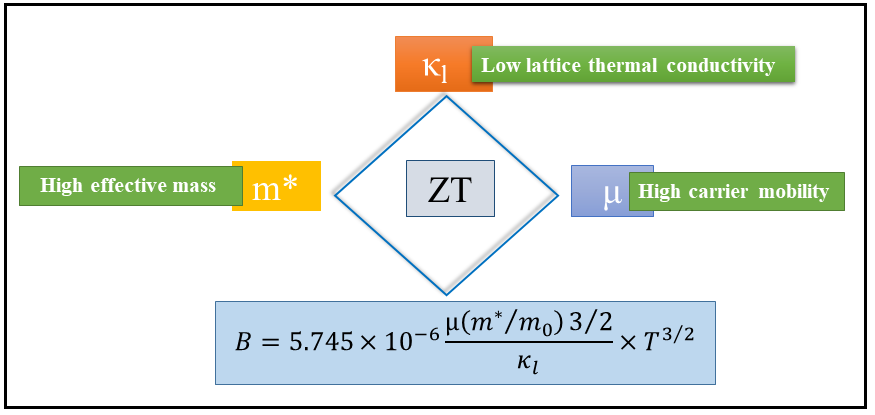}      
\caption{Schematic diagram for designing  high performance TE materials. These strategies and concepts are used to maximize the thermoelectric figure of merit \textit{ZT} value. Here B refer as a dimensionless material parameter}
\label{fgr:1}
\end{figure*}
In this study we have focused on the principle of the Seebeck effect, which is used in many areas of research, such as in the automotive industry to increase fuel efficiency, in power plants to convert waste heat to electricity, to convert solar heat to electricity and in radioisotope TE generators for space probes. TE materials have low maintenance, long life cycle, and high reliability, which make possible to use them on Earth, in space, and deep in the ocean. TE materials can be used from low to high temperature  applications such as electronic circuits/microchips (portable cooling) and radioactive TE generators  which are mainly used by the National Aeronautics and Space Administration (NASA). Though there are many advantages in using TE materials, there is also some disadvantages such as high thermal conductivity, poor electrical conductivity, and low Seebeck coefficient those will reduces the \textit{ZT} value, i.e heat to energy conversion efficiency. The main focus in the TE research society is to optimize the \textit{ZT} value by reducing the thermal conductivity. The three main strategies used by scientists to reduce the thermal conductivity of the lattice to increase the  \textit{ZT} values are alloying, introducing complex crystal structures, and  nanoengineering~\cite{joshi2011enhancement,kanatzidis2009nanostructured,pei2011high,xie2010identifying,
poudeu2009high,joshi2008enhanced,liu2012recent,alam2013review}. 
\par
Alloying method works on the strategy of phonon glass and electron crystal (PGEC) behavior~\cite{slack1995new}. The PGEC suggests an ideal state in which the phonon should see the material as amorphous with a large number of phonon scattering center and the electron should see the material as a crystalline structure with minimal electron scattering. In this method, one can introduce  defects by introducing two or more atoms on the same sites of the system, thus changing the crystal structure and  achieving high \textit{ZT} value by reducing the lattice thermal conductivity. 
The idea of the second strategy, i.e complex crystal structure, is to make the optical phonon mode flat, which increases the Umklapp scattering~\cite{toberer2011phonon} and thus decreases the thermal conductivity. This intensifies the search for new bulk TE with complex unit cell to discover new phases such as clathrates~\cite{cohn1999glasslike},  skutterudites~\cite{nolas1998effect,caillat1996properties} and complex Zintl phases~\cite{toberer2009zintl,kauzlarich2007zintl,toberer2010electronic}. 
In recent years, new developments in nanostructure materials have attracted much interest in the TE community to develop new thermoelectric nanomaterials with high power factor (PF) and high \textit{ZT} value due to the enhanced density of states (DOS) near Fermi level via quantum confinement.  The thermal conductivity due to phonons can be suppressed in nanostructures in order to get better electrical conductivity and \textit{ZT}~\cite{zhang2010nanostructures,hicks1993effect,venkatasubramanian2001thin,xie2009high}.
 Band structure calculations  show that the  half$-$Heusler (HH) alloys with 18 VEC are narrow band semiconductors~\cite{kuentzler1992gap,tobola1998crossover}, resulting in high Seebeck coefficient and large effective mass. 
 Isoelectronic substitution, i.e the substitution of elements from the same group of The Periodic Table in TE materials, is well known and has attracted  much attention for reducing thermal conductivity. The substitution of two or more atoms of different size and mass introduce more phonon scattering centers into the system via mass fluctuation~\cite{ren2020establishing}. This substitution approach has been adopted by many research groups to develop new multinary TE compounds. Figure. \ref{fgr:1} shows some of the strategies and approaches proposed by many researchers to maximize the \textit{ZT} value. A dimensionless material parameter B, which is directly related to the ratio between the effective mass of charge carriers and the mass of  free electron charge, carrier mobility, temperature and inversely related to the thermal conductivity of the lattice at a given temperature, is also considered for achieving the maximum \textit{ZT} value. Considering these material parameters one can design  high efficiency TE materials with high \textit{ZT} value. One can achieve the heavier effective mass, high carrier mobility and increase in phonon scattering by well$-$known strategies such as band convergence, band alignment, alloying and nanostructuring.
 \par 
In the 1950s, Ioffe~\cite{ioffe1957semiconductor} suggested that alloying (forming solid solutions) was an effective way to optimize the TE performance. Alloying technique is widely used in chalcogenides, silicides, selenides, zintl phases, and HH compounds ~\cite{jaldurgam2021low} But, selecting effective alloying elements to reduce the lattice thermal conductivity is a great challenge for the researcher due to the fact that alloying reduces the value of lattice thermal conductivity while it also reduces the charge carrier mobility. Wang \textit{et al}.~\cite{wang2013material} proposed that elements with heavy mass and small alloying scattering potential reduce the lattice thermal conductivity without greatly affecting the charge carrier mobility. Furthermore, it was found that the alloying atoms with heavier mass and smaller radius difference from the host atoms is the most reasonable choice to reduce the lattice thermal conductivity. 
To achieve high TE conversion efficiency, one must achieve high \textit{ZT} value~\cite{zhu2015high,bulman2014high}. One of the most common strategies to achieve this is to reduce the total thermal conductivity, which consists of both electronic and lattice  parts (i.e. $\kappa$ $=$ $\kappa_e+\kappa_L$). Most work focuses on reducing the lattice portion of the thermal conductivity to increase the TE performance~\cite{xie2014intrinsic,kutorasinski2014application,chaput2006electronic}. 
One of the widely used approaches to reduce the lattice thermal conductivity in HH alloys is to  introduce mass/stress fluctuation effect by isoelectronic or aliovalent doping/substitution. Experimental studies also show that the multinary compounds can be formed with more than 5 or more elements while maintaining the semiconducting behaviour in these systems~\cite{xie2013beneficial,lee2010high,simonson2011introduction,xie2012recent,xie2008preparation,sekimoto2007high}. 

The mass and size differences between the host atoms and the substituted atoms create point defects that scatter more phonons and thus reduce the lattice thermal conductivity~\cite{yang2004strain}. Numerous studies show that doping at Ti site of  TiNiSn or TiCoSb can effectively reduce the lattice thermal conductivity~\cite{sakurada2005effect,yan2013thermoelectric}. However, the maximum \textit{ZT} value achieved for doped/substituted TiNiSn is still in the range of 0.3$-$0.6 only~\cite{katayama2003effects,gelbstein2011thermoelectric,birkel2012rapid,douglas2012enhanced} so far due to the high lattice thermal conductivity. The recent study by  G\"{u}rth\textit{et al}.~\cite{gurth2016thermoelectric} showed that they have achieved a \textit{ZT} value of 0.98 for TiNiSn by using a modified preparation route. In addition, they also achieved \textit{ZT} value of 1.2 for multi$-$component systems. It was found that the Hf atom is an effective dopant at the Ti site that lowers the thermal conductivity of the lattice~\cite{fu2015realizing}. 
In this work, despite the traditional isoelectronic substitution approach, we report our recent progress in aliovalent substitution  by substituting III$^{rd}$ (Sc, Y, La) IV$^{th}$ (Ti, Zr, Hf) and V$^{th}$ (V, Nb, Ta) group elements at the Ti site of  TiNiSn. To model the pentanary systems, we have substituted 25\% of the III$^{rd}$, 50\% of the IV$^{th}$ and 25\% of the V$^{th}$ group elements at the Ti sites of  TiNiSn. These aliovalent substitutions are chosen in such a way that the total number of  VEC is always 18 to preserve the semiconducting behavior (see Table \ref{tbl:1}). For simplicity, we divide these systems into Sc, Y and La series. This article is divided into the following parts. The first part contains the methodology to compute the structural and the transport properties of these pentanary systems. The second part involve results and analysis and that is presented in two parts where the first part is the analysis of the electronic structure and the second part is a brief description of the TE transport properties including lattice dynamics of some of the selected systems, which could be considered as potential candidates for high$-$efficiency TE materials. The last section is the conclusions.  
\section{Computational Details}
Density functional theory (DFT) calculations for structural optimization and the electronic structure calculations were performed using projector$-$augmented plane$-$wave (PAW)~\cite{kresse1999ultrasoft} method, as implemented in the Vienna \textit{ab initio} simulation package (VASP)~\cite{kresse1996efficiency}. For the exchange$-$correlation potential in all our calculations we have used the generalized gradient approximation (GGA)~\cite{perdew1996generalized} proposed by Perdew$-$Burke$-$Ernzerh. The Brillouin zone (BZ) was sampled using a Monkhorst Pack scheme~\cite{monkhorst1976special} for structural optimization and employed a 12$\times$12$\times$8 \textbf{k}$-$mesh. A plane$-$wave energy cutoff of 600~eV is used for geometry optimization for all the pentanary substituted TiNiSn. The convergence criterion for energy was taken to be 10$^{-6}$ eV/cell for total energy minimization and that for the Hellmann$-$Feynman force acting on each atom was taken less than 1 meV/\AA~ for ionic relaxation. We have used the tetrahedron method with Bl\"{o}chl correction~\cite{blochl1994improved} for BZ integrations for calcuating the density of states. Our previous study shows that the computational parameters used for the present study are sufficient enough to accurately predict the equilibrium structural parameter for HH alloys~\cite{choudhary2020thermal}.
\par
The full potential linearized augmented plane wave method as implemented in WIEN2k code~\cite{blaha2001wien2k,schwarz2003solid} was used for calculating the accurate band structure for TE transport properties calculations. We have used a very high density of \textbf{k}$-$point 31$\times$31$\times$31 with R$_{MT}$K$_{max}$ $=$ 7, where R$_{MT}$ is the smallest atomic sphere radii of all the atomic spheres and K$_{max}$ represent the maximum reciprocal lattice vector in the plane wave expansion and the convergence criteria is set to be 1 mRy/cell for all our calculations in order to obtain accurate eigenvalue. 
We have then used the calculated eigen energy in the BoltZTraP code~\cite{madsen2006boltZTrap} for calculating the TE transport properties such as Seeback coefficient, electrical conductivity, and the electronic part of the thermal conductivity. For calculating the lattice dynamic properties a finite displacement method implemented in the VASP$-$Phonopy~\cite{togo2015first} interface was used with supercell approach. In all the phonon calculations we have used relaxed primitive cells to create supercell of dimension 2$\times$2$\times$2 with the displacement distance of 0.01~\AA. The third order anharmonic force constants were calculated using VASP$-$Phono3py~\cite{togo2015distributions} with a 2$\times$2$\times$2 supercell incorporating interactions out to 5th nearest neighbors. Finally, we calculated lattice thermal conductivity by explicitly solving the phonon Boltzmann transport equation.

\begin{figure}[b]
\centering
\includegraphics[width=1.0\linewidth]{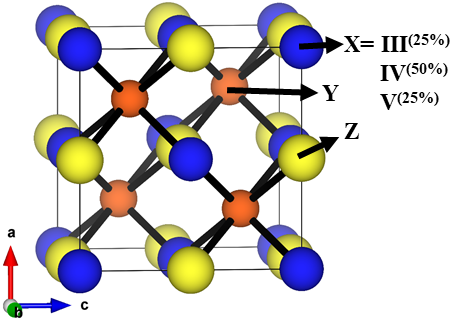}
\caption{The conventional unit cell of half$-$Heusler alloy TiNiSn (\textit{XYZ}). The pentanary substituted TiNiSn is modeled by substituting the III$^{rd}$, IV$^{th}$ and V$^{th}$ group atoms with the composition of 25\%, 50\% and 25\% at Ti site of TiNiSn, respectively.}
\label{fgr:2}
\end{figure}
\section{Results and Discussion}
\subsection{Structural Description}
The electronic structure of full Heusler~\cite{kubler1983formation} (FH) and HH~\cite{pierre1997properties,tobola2000electronic,offernes2007electronic} alloys are vary with the VEC. Their properties can easily be predicted by counting their valence electrons~\cite{graf2011simple}. The HH alloys can be described by the general formula  \textit{XYZ} with a composition 1:1:1 and they crystallize in a non$-$centrosymmetric cubic structure with space group  F$\bar{4}$3m (no.216), while the FH alloys are generally described by the formula \textit{X$_{2}$YZ} with a composition of 2:1:1 and those crystallize in the cubic space group  Fm$\bar{3}$m (no. 225). In these compounds the \textit{X} and \textit{Y} are transition metals or rare earth elements and \textit{Z} is usually a main group element. The FH alloys show all kinds of multi$-$functional magnetic properties, such as magnetocaloric~\cite{krenke2005inverse,levin2017tuning,liu2012giant}, magneto$-$optical~\cite{buschow1981magnetic,picozzi2006magneto,sanvito2017accelerated} and half metallic ferromagnetic~\cite{kundu2017new,blum2009highly,wurmehl2006investigation} behaviors. The half metallic ferromagnet shows semiconducting behavior for electrons of one spin orientation while metallic for electrons with the opposite spin. The emerging new physical properties such as thermoelectricity~\cite{huang2018dramatically,krishnaveni2016band,yang2008evaluation,gofryk2011magnetic}, superconductivity~\cite{radmanesh2018evidence,pavlosiuk2016antiferromagnetism,xu2014weak,pan2013superconductivity,nakajima2015topological}, magnetic ordering~\cite{pavlosiuk2018magnetic,suzuki2016large} and topological transitions~\cite{manna2018heusler,shi2018tunable,liu2016observation,xiao2010half} of HH alloys have attracted more attention.
\par
Figure. \ref{fgr:2} shows the crystal structure of TiNiSn with four formula units, where the atom \textit{X} is located at 4a (0 0 0), the \textit{Y} atom is located at 4b ($\dfrac{1}{2}$, $\dfrac{1}{2}$, $\dfrac{1}{2}$) and the \textit{Z} atom is located at 4c ($\dfrac{1}{4}$, $\dfrac{1}{4}$, $\dfrac{1}{4}$) position, respectively. The most electropositive element \textit{X} in \textit{XYZ} transfers its electron to the more electronegative elements \textit{Y} and \textit{Z} forming a closed shell configuration, i.e., \textit{d}$^{10}$ for \textit{Y} and \textit{s}$^{2}$\textit{p}$^{6}$ for \textit{Z}. Therefore, the HH alloys with 18 VEC are considered as  non$-$magnetic and semiconducting alloys~\cite{jung2000study}. 
The structural analysis of the  pentanary substitution show that the symmetry of the HH alloys reduces  from cubic to tetragonal by pentanary alloying. The primitive unit cell of all  pentanary substituted TiNiSn contains 12 atoms and has a tetragonal structure with space group P$\bar{4}$2m (No.111). The optimized equilibrium lattice parameters, heat of formation ($\Delta$H$ _{f} $) and PBE$-$GGA band gap values  for the pentanary substituted TiNiSn systems are summarized in Table \ref{tbl:2}. Also, our lattice dynamics calculations (discussed latter)  for all these pentanary systems show no negative phonon mode indicating the thermodynamic stability of these systems.

\begin{table}[t]
\centering
\begin{tabular}{cccccccc}
\hline\hline
 Group &&&&Composition&Ti&Ni&Sn\\
\hline\\ [-0.5 ex]
III$^{rd}$&IV$^{th}$&V$^{th}$&&III$_{0.25}$IV$_{0.5}$ V$_{0.25}$&&&\\
\hline
Sc&Ti&V&ScTiV&ScZrV&ScHfV&\checkmark &\checkmark\\
&Zr&Nb&ScTiNb & ScZrNb& ScHfNb& \checkmark&\checkmark \\
&Hf&Ta& ScTiTa & ScZrTa& ScHfTa&\checkmark &\checkmark\\
\hline
Y&Ti&V&YTiV&YZrV&YHfV&\checkmark &\checkmark\\
&Zr&Nb&YTiNb & YZrNb& YHfNb& \checkmark &\checkmark \\
&Hf&Ta& YTiTa & YZrTa& YHfTa&\checkmark &\checkmark\\
\hline  
La&Ti&V&LaTiV&LaZrV&LaHfV&\checkmark&\checkmark\\
&Zr&Nb&LaTiNb & LaZrNb& LaHfNb& \checkmark &\checkmark \\
&Hf&Ta& LaTiTa & LaZrTa& LaHfTa&\checkmark &\checkmark \\
  \hline
  \hline
\end{tabular}
 \caption{The list of predicted pentanary HH alloys formed from the combination of the III$^{rd}$, IV$^{th}$ and V$^{th}$ group atoms with the composition of 25\%, 50\% and 25\% at the Ti site of TiNiSn, respectively.}
\label{tbl:1}
\end{table}

\begin{table*}
\small
\caption{\ The structural parameters for transition metals substituted at Ti site in TiNiSn where Z=1;  A,  B, C, D, E, F and G are in the Wyckoff  position 1a (0, 0, 0), 2f (0, 1/2, 1/2), 1d (1/2, 1/2 0), 4n (x$^{'}$, y$^{'}$, z$^{'}$) 2e(1/2, 0, 0), 1b (1/2, 1/2, 1/2) and 1c (0, 0, 1/2). The lattice parameters a and c (in \AA), the internal structure parameter (z) obtained from our structural optimization, PBE$-$GGA band gap value E$_{g}$ (in eV), heat of formation ($\Delta$H$ _{f} $) (in kJ mol$^{-1}$ are listed. All the pentanary substituted systems have space group P$\bar{42}$m (No.111). }
\label{tbl:2}
\begin{tabular*}{\textwidth}{@{\extracolsep{\fill}}cccccccccc}            
\hline
Compound&\multicolumn{2}{c}{Unit$-$cell dimension (\AA)} &&\multicolumn{3}{c}{Positional parameters}&$\Delta$H$ _{f}$ (kJ mol$^{-1}$) &E$_{g}$ (eV)\\ [1.0 ex]
\cline{2-3} \cline{5-7} \\
&a&c&&x$^{'}$&y$^{'}$&z$^{'}$&&\\ [0.5 ex]
 \hline \\ [1.0 ex]
 TiNiSn&5.94&5.94&&&&&$-$56.02&0.45\\ [1.0 ex]
{La$_{0.25}$Ti$_{0.5}$V$_{0.25}$NiSn}&6.108&6.118&&0.735&0.735&0.256&$-$53.91&0.21\\ [1.0 ex]
{La$_{0.25}$Zr$_{0.5}$V$_{0.25}$NiSn}&6.204&6.217&&0.732&0.732&0.247&$-$60.66&0.17\\ [1.0 ex]
{La$_{0.25}$Hf$_{0.5}$V$_{0.25}$NiSn}&6.178&6.189&&0.735&0.735&0.251&$-$58.04&0.27\\ [1.0 ex]
{La$_{0.25}$Ti$_{0.5}$Nb$_{0.25}$NiSn}&6.141&6.154&&0.738&0.738&0.26& $-$54.30&0.20\\ [1.0 ex]
{La$_{0.25}$Zr$_{0.5}$Nb$_{0.25}$NiSn}&6.230&6.244&&0.735&0.735&0.252&$-$60.86&0.22\\ [1.0 ex]
{La$_{0.25}$Hf$_{0.5}$Nb$_{0.25}$NiSn}&6.206&6.221&&0.736&0.736&0.255&$-$58.25&0.23\\ [1.0 ex]
{La$_{0.25}$Ti$_{0.5}$Ta$_{0.25}$NiSn}&6.134&6.145&&0.736&0.736&0.257&$-$55.27&0.24\\ [1.0 ex]
{La$_{0.25}$Zr$_{0.5}$Ta$_{0.25}$NiSn}&6.224&6.236&&0.734&0.734&0.249&$-$61.92&0.28\\ [1.0 ex]
{La$_{0.25}$Hf$_{0.5}$Ta$_{0.25}$NiSn}&6.201&6.213&&0.734&0.734&0.252&$-$59.31&0.32\\ [1.0 ex]
 \hline
 \end{tabular*} 
 \end{table*}
\begin{figure*}
\centering
\includegraphics[width=1.0\linewidth]{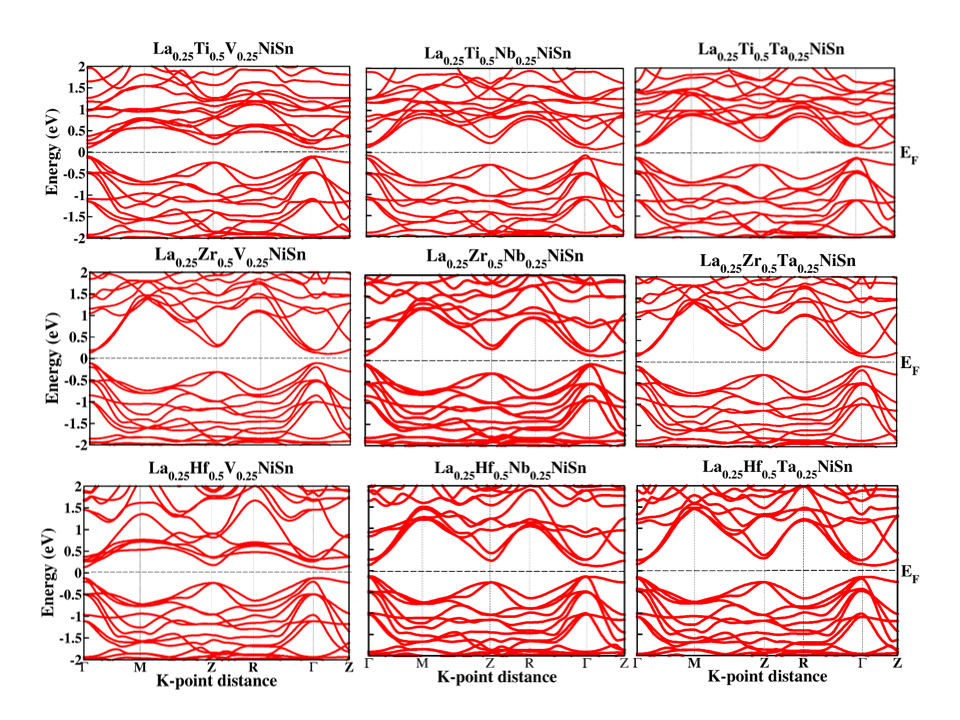}
\caption{The calculated band structures for pentanary systems La$_{0.25}$X$^{IV}_{0.5}$X$^{V}_{0.25}$NiSn where (X$^{IV}$ $=$ Ti, Zr and Hf, and X$^{V}$ $=$ V, Nb and Ta) substituted at Ti site in TiNiSn obtained from PBE$-$GGA calculation.}
\label{fgr:3}
\end{figure*}
\subsection{Analysis of the electronic structure of pentanary substituted TiNiSn.}
 The good TE materials are considered to have a small band gap value. However, there is no set of rules to determine the exact band gap value to optimize the \textit{ZT}. This, of course gives us the freedom to design efficient TE by modifying their band structures including band gap values, band edge positions, and carrier mobilities. Band offset and band convergence~\cite{hao2019computational,lee2020band,xiao2018charge,zhu2018discovery} play an important role in TE materials for the design of n$-$type or p$-$type materials and their effect on the TE properties. These aspects were investigated by comparing the contributions of the light and heavy conduction bands to the electrical resistance and  Seebeck coefficient. The pentanary  substitution is made under the assumption of a rigid band approximation, where the shape of the DOS curve is not going to change by the substitution and there is only a shift in the Fermi level due to the electron/hole doping if we vary the VEC to analyse the role of electron/hole doping/substitution on the transport properties. However, a large atomic size mismatch between Zr, Hf and V enhances the TE transport properties, which will be discussed in the next section of this paper.
\par
The concept of multiband model to describe the TE properties was first introduced by Simon~\cite{ simon1964maximum} in late 1964. He  showed very clearly the relationship between the maximization of the dimensionless material parameter \textit{ZT} value by focusing on the two$-$band model. The two$-$band model, i.e., a conduction band (CB) and a valence band (VB), was considered a very effective approach to tune the Fermi energy, which is a function of the DOS, the effective mass of the charge carriers, the temperature and the optimal doping conditions (donors and acceptors). Later, Slack \textit{et al}~\cite{slack1991maximum}, liu \textit{et al}~\cite{liu2008enhanced} and pei \textit{et al}~\cite{pei2011convergence} predicted the maximum energy conversion efficiency by utilizing three band model.
\par
 Figure. \ref{fgr:3} shows the calculated electronic band structures obtained with the PBE$-$GGA function for La$_{0.25}$X$^{IV}_{0.5}$X$^{V}_{0.25}$NiSn where (X$^{IV}$ $=$ Ti, Zr and Hf, and X$^{V}$ $=$ V, Nb and Ta) series close to their band edge, i.e., from $-$2 eV to 2 eV. The band structure is plotted along the high symmetry directions of the first BZ of the simple tetragonal lattice for these pentanary substituted systems. Our PBE$-$GGA calculations show that all these systems possess semiconducting behavior with the band gap values varying from 0.17 to 0.32~eV. Also, all these materials show indirect band gap behavior with the valence band maximum (VBM) at the $\Gamma$ point and the conduction band minimum (CBM) between the $\Gamma$ to Z points. 
\par
\begin{figure}
 \centering
\includegraphics[height=6cm]{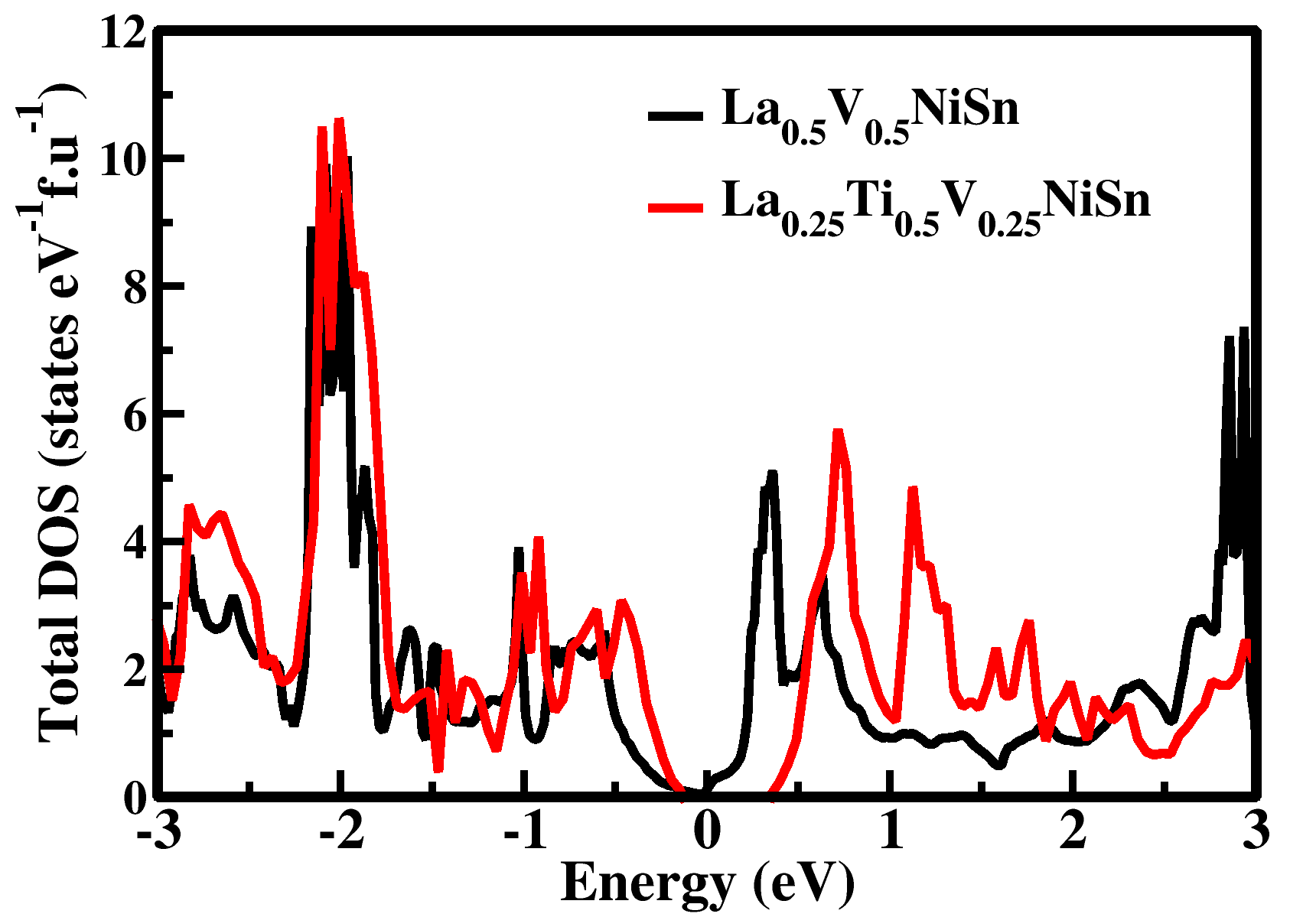}
\caption{The calculated total density of states for  the quaternary  (La$_{0.25}$V$_{0.25}$NiSn) and pentanary (La$_{0.25}$Ti$_{0.5}$V$_{0.25}$NiSn) obtained from PBE$-$GGA calculation.}
\label{fgr:4}
\end{figure}
In our previous study~\cite{choudhary2020thermal}, it was found that the quaternary systems with (La, V)$_{0.5}$, (La, Nb)$_{0.5}$, and (La, Ta)$_{0.5}$ substituted at the Ti site in TiNiSn exhibit semi$-$metallic behavior. In this work, we show that a semiconducting state can be achieved by inserting additional atoms from the IV$^{th}$ group element of The Periodic Table in the Ti sites of TiNiSn i.e La$_{0.25}$X$^{IV}_{0.5}$X$^{V}_{0.25}$NiSn. The introduction of this additional transition metal atoms increases the hybridization between the \textit{d}$-$ and \textit{p}$-$orbitals resulting a band gap opening. Further, the electrons donated by these additional electropositive elements stabilizes the system by filling up the bonding states and opening the band gap between the VB and CB states. A broad comparison is needed here to show the differences in the electronic structure between these systems. In the case of  La$_{0.25}$X$^{IV}_{0.5}$V$_{0.25}$NiSn where (X$^{IV}$ $=$ Ti, Zr, Hf) series, the band gap increases upon substitution of Ti with Zr or Hf. The calculated band gap values for La$_{0.25}$Ti$_{0.5}$V$_{0.25}$NiSn, La$_{0.25}$Zr$_{0.5}$V$_{0.25}$NiSn, and La$_{0.25}$Hf$_{0.5}$V$_{0.25}$NiSn, are 0.17, 0.2, and 0.21 eV, respectively. Similarly, the band gap values increase in the case of La$_{0.25}$X$^{IV}_{0.5}$Nb$_{0.25}$NiSn with the order of 0.2, 0.22 and 0.23 eV, and for La$_{0.25}$X$^{IV}_{0.5}$Ta$_{0.25}$NiSn 0.24, 0.28 and 0.32 eV, when X$^{IV}$ $=$ Ti, Zr, Hf), respectively.

\begin{figure*}
 \centering
\includegraphics[width=0.85\linewidth]{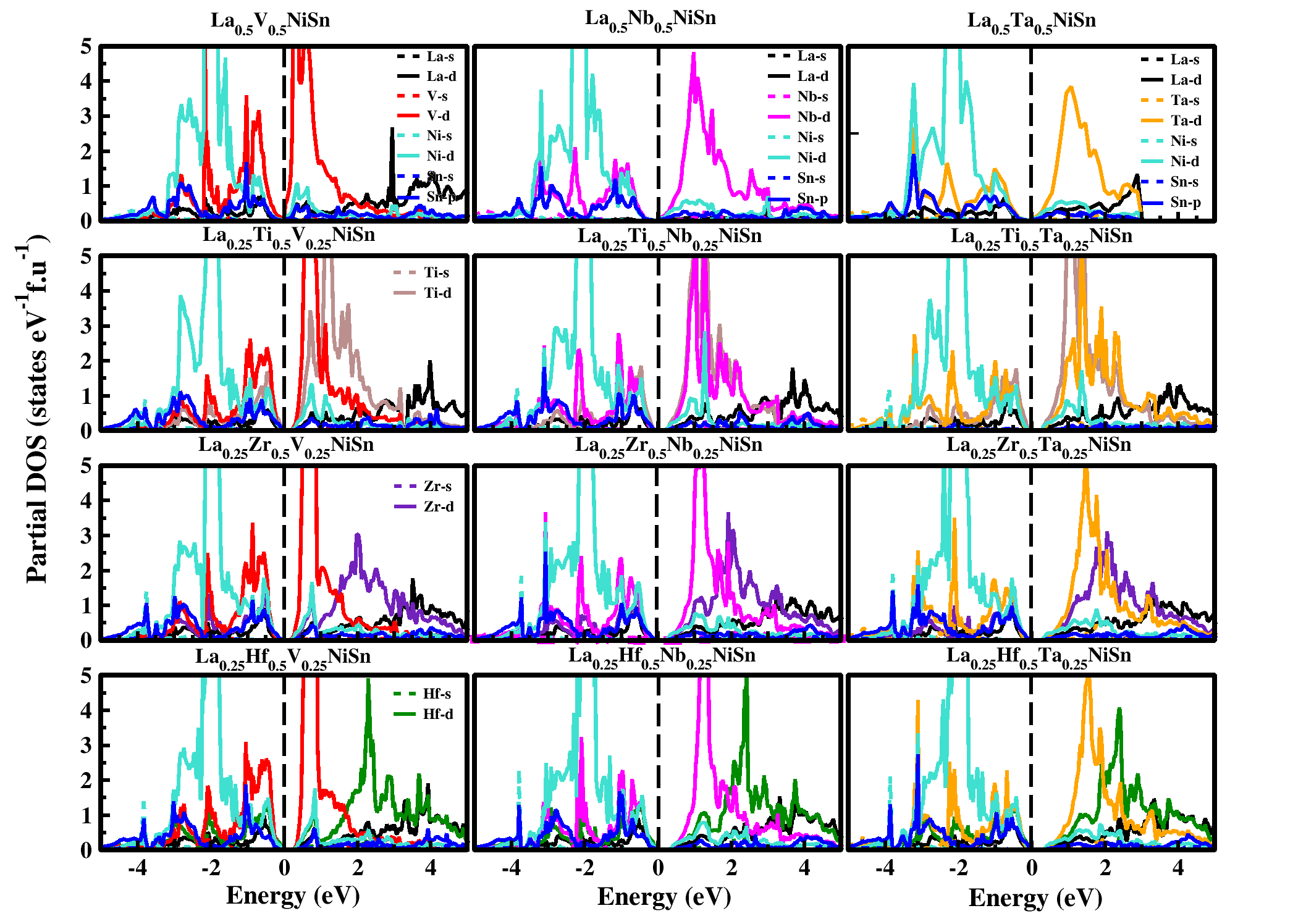}
\caption{The calculated partial density of states for  the pentanary  La$_{0.25}$X$^{IV}_{0.5}$X$^{V}_{0.25}$NiSn where (X$^{IV}$ $=$ Ti, Zr and Hf, and X$^{V}$ $=$ V, Nb and Ta) substituted at the Ti site in TiNiSn obtained from PBE$-$GGA calculation.}
\label{fgr:5}
\end{figure*}
\begin{figure*}
\centering
\includegraphics[width=0.85\linewidth]{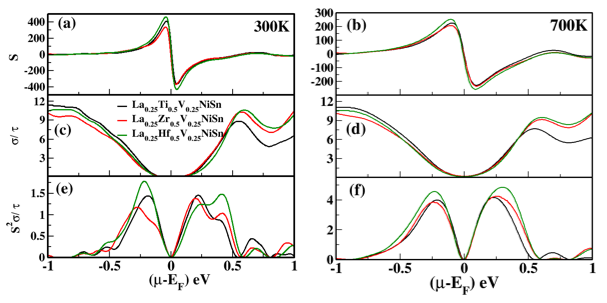}
  \caption{The calculated transport properties (a) and (b) Seebeck coefficient (in $\mu$V/K) , (c) (d)  electrical conductivity (in 10$^{19}$/$\Omega$ms)  and (e) and (f) power factor (in 10$^{11}$W/mK$^{2}$s ) as a function of chemical potential $\mu$ in the range of $-$1 to 1 eV at 300 and 700\,K obtained from PBE$-$GGA calculation.}
  \label{fgr:6}
\end{figure*} 

There is a cluster of narrow bands around $-$2\,eV, in the band diagram in Fig. \ref{fgr:3} and are originating from the Ni$-$3\textit{d }electrons and they show sharp peak in the VB in the DOS curve. Reasonably dispersed bands present in the VBM are equally contributed by all the five atoms in all the pentanary compounds considered in the present study. The electronic band structure of all these systems have several adjacent low$-$lying bands (LLB) around the CBM and are get degenerate at the $\Gamma$ point  in all these compounds as shown in Fig. \ref{fgr:3}. Though these materials are having indirect band feature, the energy difference between the direct bandgap and the indirect band gap is very small. Hence, though these materials are having indirect band behavior, the energy loss due to phonon assisted optical excitation will be very small.  The lowest conduction band disperses rapidly from the $\Gamma$ point and flattens out as it approaches the \textit{Z} point, giving rise to the high effective electron mass. The electronic band structure of all the pentanary substituted TiNiSn has small band gap values compared to TiNiSn, which is also responsible for enhancing the \textit{ZT}. 
\par 
In order to understand the role of pentanary addition in introducing the semiconducting behaviour we have plotted the total DOS for the quaternary La$_{0.25}$V$_{0.25}$NiSn and pentanary La$_{0.25}$Ti$_{0.5}$V$_{0.25}$NiSn systems and are shown in Fig.~\ref{fgr:4}. From the total DOS analysis we can see that the addition of extra atom shift the CBM in to high energy region  and open up the band gap. On the other hand, the quaternary system show the semi$-$metallic behaviour. If one go from quaternary to pentanary substituted systems i.e. La$_{0.5}$V$_{0.5}$NiSn to (La$_{0.25}$Zr$_{0.5}$V$_{0.25}$NiSn), our electronic structure calculation predict that a band gap is open up. For example, the substitution of Ti, Zr and Hf as a donors lead to an expected upward shift of the CBM. 
\par 
To understand the contribution of electronic states responsible for the electronic transport, we have calculated the partial DOS for La$_{0.25}$X$^{IV}_{0.5}$X$^{V}_{0.25}$NiSn and are displayed in Figure. \ref{fgr:5}. The detail analysis of the partial DOS curve of  La$_{0.25}$X$^{IV}_{0.5}$V$_{0.25}$NiSn where (X$^{IV}$ $=$ Ti, Zr, Hf) show that the \textit{d}$-$states of all the four transition metals and Sn$-$\textit{p} states are equally contributing to the VBM. However, the CBM is mainly contributed equally by V, Ni, and Ti$-$\textit{d} states with small contribution from both La$-$\textit{d} and Sn$-$\textit{p} states equally. If we replace Ti with Zr or Hf then the \textit{d}$-$states from these atoms systematically shifted to higher energy in the CB and this could explain why the band gap increases when the Ti is replaced with Zr or Hf in these pentanary systems. However, in all these systems the contribution from the La$-$\textit{d} and the Sn$-$\textit{p} states to the VBM is very small compared with the electronic \textit{d}$-$states contributed by the other three transition metals. Though the Ni$-$\textit{d} electrons are mainly localized around $-$2\,eV in the VB its contribution at the band edges are almost same as the contribution from the\textit{d}$-$states of  V/Nb/Ta. While the V$^{th}$ group elements (V, Nb and Ta)$-$\textit{d} states show the main contribution in CM between the energy range of 0.5$-$2.5 eV, and are also responsible for the semi$-$metallic nature in La$_{0.5}$X$_{0.5}$NiSn (X$=$V, Nb, Ta) systems. From these analysis one can conclude that the hole transport in these systems are equally contributed by all the five atoms. However, the electron transport is dominated by the \textit{d}$-$states of  Ti/Zr/Hf and V/Nb/Ta with moderate contribution from Ni$-$\textit{d} states and small contribution equally by the La$-$\textit{d} and Sn$-$\textit{p} states.  It may be noted that the contribution from the\textit{ s}$-$ states of the constituents are negligibly small in both VBM and CBM and hence these electrons will not participate significantly in the transport properties of these systems. 

\begin{figure*}
\centering
\includegraphics[width=0.79\linewidth]{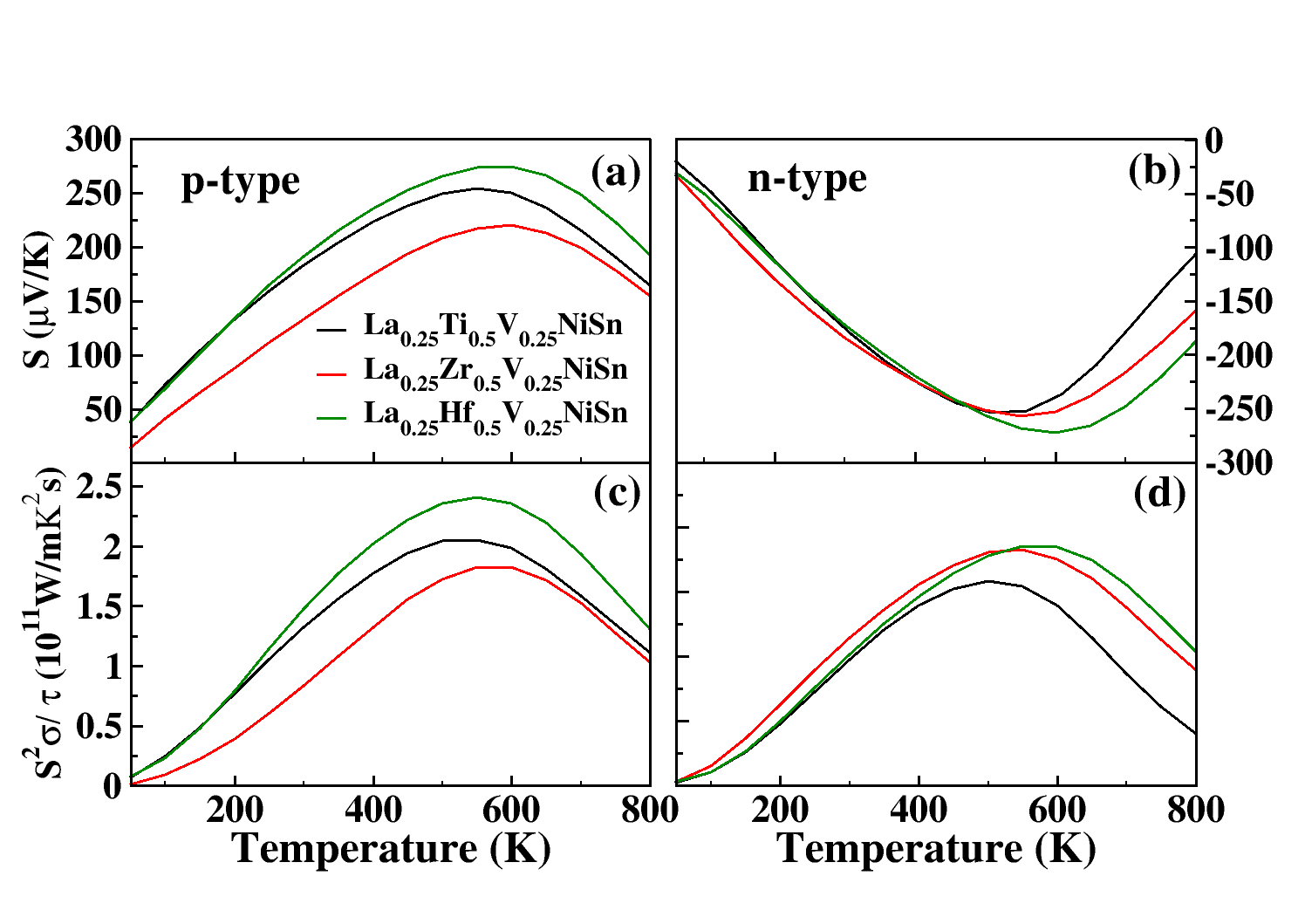}
\caption{The temperature dependent transport properties at the optimal doping carrier concentrations 
under p$-$type and  n$-$type conditions (a) and (b) Seebeck coefficient (S) and (c) and (d) power factor (${S^{2}\sigma}$) obtained from PBE$-$GGA functional calculation.}
\label{fgr:7}
\end{figure*}
\subsection{Thermoelectric transport properties }
For metals or degenerate semiconductors Seebeck coefficient and electrical conductivity are given by the equation 
\begin{equation}
  S = \frac{8\pi^2k^{2}_{B}}{3eh^2}\Big(\frac{\pi}{3n}\Big)^{2/3} m^{*}T
\end{equation}
\begin{equation}
 \sigma = \frac{ne^{2}\tau}{m^{*}}
\end{equation}
where  k$_{B}$, h, e, T, n,  and m$^{*}$, $\sigma$ and $\tau$ are the Boltzmann constant, Planck constant, electrical charge, absolute temperature, carrier concentration, carrier effective mass, electrical conductivity and averaged relaxation time of electron, respectively. 
Seebeck coefficient and electrical conductivity are inversely proportional to each other, also these quantities are strongly dependent on temperature and chemical potential. With the increase of doping concentration and temperature, the electrical conductivity increases and Seebeck coefficient decreases. To maximize the PF, one needs to increase both $S$ and $\sigma$.
\par 
Fig. \ref{fgr:6}. shows the $S$, $\sigma$/$\tau$ and PF (${S^{2}\sigma}$/$\tau$) for La$_{0.25}$X$^{IV}_{0.5}$V$_{0.25}$NiSn where (X$^{IV}$ $=$ Ti, Zr, Hf) as a function of chemical potential in the interval range of $\mu-E_{F}=\pm$1 eV at two constant temperatures 300\,K, and 700\,K. From the Fig. \ref{fgr:6}. (a) and (b), one can see that the Seebeck coefficient decrease as the Fermi level shifts down towards the VB or shifts up towards the CB with hole/electron doping. It can be seen that at 300\,K and 700\,K, the Seebeck coefficient exhibits two peaks, one corresponds to p$-$type and other corresponds to n$-$type conditions. We can also see that these peaks are much closer to the VBM and CBM. It can be seen that at 300\,K, the maximum values of Seebeck coefficients are 423.82, 352.23 and 473.92 $\mu$V/K at $\mu-E_{F}=$ $-$0.037, $-$0.044 and $-$0.046 eV in p$-$type condition  and $-$349.32, $-$377.95 and $-$435.85 $\mu$V/K at $\mu-E_{F}=$ 0.040, 0.041 and 0.044 eV in n$-$type condition for La$_{0.25}$Ti$_{0.5}$V$_{0.25}$NiSn, La$_{0.25}$Zr$_{0.5}$V$_{0.25}$NiSn, and La$_{0.25}$Hf$_{0.5}$V$_{0.25}$NiSn, respectively. However, at higher temperature ( 700\,K) the peak values decreases to 235.08, 210.26 and 255.77 $\mu$V/K  at $\mu-E_{F}=$ $-$ 0.098, $-$0.095 and $-$0.097 eV in p$-$type and $-$224.13, $-$232.27 and $-$257.22 $\mu$V/K at $\mu-E_{F}=$ 0.091, 0.086 and 0.081 eV in n$-$type condition, respectively for the above mentioned compounds. In general, the Seebeck coefficient in HH compounds starts decreasing with increase of temperature and this is due to the fact that at higher temperatures the charge carrier hole/electron conductivity increases with the increase of thermal energy.
Figure. \ref{fgr:6} (c) and (d), show  the electrical conductivity per relaxation time ($\sigma$/$\tau$)  as a function of chemical potential. There are two peaks in these curves, one corresponds to p$-$type and other corresponds to n$-$type conditions. The maximum electrical conductivity in p$-$type and n$-$type conditions for La$_{0.25}$Ti$_{0.5}$V$_{0.25}$NiSn, La$_{0.25}$Zr$_{0.5}$V$_{0.25}$NiSn, and La$_{0.25}$Hf$_{0.5}$V$_{0.25}$NiSn are 11.53, 10.87 and 9.95 10$^{19}$/$\Omega$ms at $\mu-E_{F}=$ $-$1, $-$0.92 and $-$1 eV and 8.97, 10.75 and 10.5 10$^{19}$/${\Omega}$ms at $\mu-E_{F}=$ 0.54, 0.56 and 0.55 eV at 300\,K, respectively. Whereas, the highest value of $\sigma$/$\tau$ at 700\,K in p$-$type and n$-$type conditions for  La$_{0.25}$Ti$_{0.5}$V$_{0.25}$NiSn, La$_{0.25}$Zr$_{0.5}$V$_{0.25}$NiSn, and La$_{0.25}$Hf$_{0.5}$V$_{0.25}$NiSn are 11.24, 10.69 and 10.23 10$^{19}$/$\Omega$ms at $\mu-E_{F}=$ $-$1 eV and 7.83, 9.95 and 10.14 10$^{19}$/${\Omega}$ms at $\mu-E_{F}=$ 1 eV,  respectively. However, one can see that the electrical conductivity is less affected by temperature compared to Seebeck coefficient. Moreover, we have noticed that the electrical conductivity of all the investigated compounds is relatively low at low chemical potential at 300\,K compared to 700\,K and it increases rapidly with the increase of chemical potential since the electrical conductivity is directly proportional to charge carrier density.
\par
Let us now discuss about the power factor (${S^{2}\sigma}$/$\tau$) which is an important parameter for searching the high efficiency TE materials. Figure. \ref{fgr:6}. (e) and (f), shows the calculated PF for La$_{0.25}$X$^{IV}_{0.5}$V$_{0.25}$NiSn where (X$^{IV}$ $=$ Ti, Zr, Hf). From the Fig. \ref{fgr:6}. (e) and (f) one can see that the calculated PF for all the selected compounds are lower value at 300\,K. and reach high value at higher temperature i.e at 700\,K. For example, at 300\,K one can see the shift in the chemical potential for all the investigated compounds with respect to the maximum PF. In the case of p$-$type condition, the maximum PF for La$_{0.25}$Ti$_{0.5}$V$_{0.25}$NiSn, La$_{0.25}$Zr$_{0.5}$V$_{0.25}$NiSn, and La$_{0.25}$Hf$_{0.5}$V$_{0.25}$NiSn are 1.45, 1.20 and 1.81 10$^{11}$W/mK$^{2}$s at $\mu-E_{F}=$ $-$0.19, $-$0.28 and $-$0.22 eV, respectively. The highest PF at low value of chemical potential for La$_{0.25}$Ti$_{0.5}$V$_{0.25}$NiSn indicating that one can maximize the PF by small hole doping. However, for La$_{0.25}$Zr$_{0.5}$V$_{0.25}$NiSn, and La$_{0.25}$Hf$_{0.5}$V$_{0.25}$NiS one need to do heavy hole doping to  maximize the PF. In the case of  n$-$type condition, for all these three compounds, the maximum value of  PF achieved are 1.47, 1.41  and 1.5 10$^{11}$W/mK$^{2}$s at $\mu-E_{F}=$ 0.21, 0.19 and 0.4 eV, respectively. At higher temperature i.e at 700\,K the PF  has almost doubled for all the investigated compounds compared to that at 300\,K and the  maximum PF for  La$_{0.25}$Ti$_{0.5}$V$_{0.25}$NiSn, La$_{0.25}$Zr$_{0.5}$V$_{0.25}$NiSn, and La$_{0.25}$Hf$_{0.5}$V$_{0.25}$NiSn in p$-$type condition are 4.02, 3.97 and 4.61 10$^{11}$W/mK$^{2}$s at $\mu-E_{F}=$ $-$0.21,$-$0.23 and $-$0.24) and in n$-$types condition are 4.22, 4.32 and 4.91 10$^{11}$W/mk$^{2}$s at $\mu-E_{F}=$ 0.23, 0.26 and 0.29 eV), respectively.
\par
Figure. \ref{fgr:7} shows the temperature dependence TE transport properties at the fixed charge carrier concentration n $=$ 10$^{20}$ cm$^{-3}$ for both hole$-$doped and electron$-$doped (p$-$type and n$-$type) for La$_{0.25}$X$^{IV}_{0.5}$V$_{0.25}$NiSn where (X$^{IV}$ $=$ Ti, Zr, Hf). Figure. \ref{fgr:7} (a) and (b) show the $S$ for the selected systems. The positive and  negative values of $S$ indicate that holes/electrons are the dominant charge carriers, suggesting p$-$type and n$-$type condition. It can be seen that Seebeck coefficient increases rapidly with temperature and it reaches a maximum value at 550\,k for all the investigated compounds and then decreases at higher temperature. The calculated Seebeck coefficient values for La$_{0.25}$Ti$_{0.5}$V$_{0.25}$NiSn, La$_{0.25}$Zr$_{0.5}$V$_{0.25}$NiSn, and La$_{0.25}$Hf$_{0.5}$V$_{0.25}$NiSn in p$-$type and n$-$type conditions are 253.54, 219.14 and 273.53 and -258.82, -256.84 and -268.57 $\mu$V/K  at 550\,k respectively. In both p$-$type and n$-$type condition La$_{0.25}$Hf$_{0.5}$V$_{0.25}$NiSn shows the maximum Seebeck coefficient value of 273.53 and -268.57 $\mu$V/k at 550\,K. 
Figure.\ref{fgr:7} (c) and (d), shows the calculated power factor. It follows the similar trend as Seebeck coefficient i.e the power factor increases with temperature and reaches maximum value at 550\,k and then decreases at higher temperature. The PF for La$_{0.25}$Hf$_{0.5}$V$_{0.25}$NiSn exhibits the highest pick for the p$-$type condition and the has maximum value of 2.44 10$^{11}$W/mk$^{2}$s at 550\,K. \\
\par
\begin{figure*}
\centering
\includegraphics[width=0.9\linewidth]{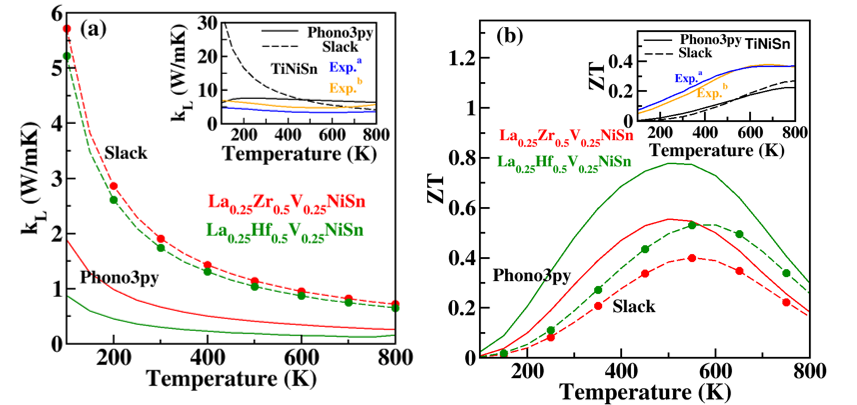}
\caption{The comparison of calculated (a) lattice thermal conductivity ($\kappa_{L}$) and (b) thermoelectric figure of merit (\textit{ZT}) between ternary (TiNiSn) and pentanary (La$_{0.25}$Zr$_{0.5}$V$_{0.25}$NiSn and La$_{0.25}$Hf$_{0.5}$V$_{0.25}$NiSn), half Heusler alloys as a function of temperature obtained from PBE$-$GGA  functional calculation. The inset figures show the calculated and the available experimental $\kappa_{L}$ and  \textit{ZT} values for TiNiSn ( Exp.$^{a}$~\cite{birkel2013improving}  Exp.$^{b}$~\cite{kim2007high}). }
\label{fgr:8}
\end{figure*}
 \begin{table}[h]
\caption{The calculated single crystal elastic constants C$_{ij}$ (in GPa), the  bulk modulus (B in GPa), shear modulus (G in GPa), Young modulus (E  in GPa), longitudinal, transverse, average elastic wave velocity ($ \nu$\textsubscript{l} ,$\nu$\textsubscript{t},$\nu$\textsubscript{m} in m/s), the Debye $\theta_{D}$ (in K) and acoustic Debye $\theta_{a}$ (in K) temperature, primitive unit cell volume (V in (\AA $^{3}$)), average mass per atom $\overline{M_{a}}$ (in amu), Poisson's ratio ($\nu$) and  Gr\"{u}neisen parameter ($\gamma$) for TiNiSn, La$_{0.25}$Zr$_{0.5}$V$_{0.25}$NiSn and La$_{0.25}$Hf$_{0.5}$V$_{0.25}$NiSn systems obtained from PBE$-$GGA calculation.}
\label{tab:3}
\begin{tabular*}{1\textwidth}{cccc}
\hline \hline
Parameters&TiNiSn & La$_{0.25}$Zr$_{0.5}$V$_{0.25}$NiSn&La$_{0.25}$Hf$_{0.5}$V$_{0.25}$NiSn\\ 
\hline
\textit{C}\textsubscript{11} & 238.09& 216.31 & 220.29   \\[0.5 ex]
\textit{C}\textsubscript{12} & 82.39 & 56.82& 60.10\\[0.5 ex]
\textit{C}\textsubscript{13}& & 59.21& 62.30\\[0.5 ex]
\textit{C}\textsubscript{33} & & 217.75 & 222.60\\[0.5 ex]
\textit{C}\textsubscript{44} &64.30  & 46.86 & 47.30\\[0.5 ex]
\textit{C}\textsubscript{66} & & 37.64  & 40.30\\[0.5 ex]
B & 134.29& 117.17 & 114.73 \\[0.5 ex]
G&69.42 & 55.39 & 56.52\\[0.5 ex]
E& 177.65& 142.50& 145.64\\[0.5 ex]
$ \nu$\textsubscript{l} & 5618.45& 7627.20 &7041.41\\[0.5 ex]
$\nu$\textsubscript{t} & 3108.03 & 4173.71& 3840.42\\[0.5 ex]
$\nu$\textsubscript{m}&3462.74 &4653.93 &4283.36\\[0.5 ex]
V &52.05&239.33 &236.30\\ [0.5 ex]
$\nu$ & 0.279 & 0.286 & 0.288\\ [0.5 ex]
$\gamma$ & 1.64& 1.68 & 1.7\\ [0.5 ex]
$\overline{M_{a}}$ &75.09&91.69 &109.14\\ [0.5 ex]
$\theta_{D}$&398.22&381.6 &352.71\\ [0.5 ex]
$\theta_{a}$&277.12&168.06 &155.34\\ \hline \hline
\end{tabular*}
 \end{table}
 
\begin{table*}
\begin{center}
\caption{The Calculated values of lattice thermal conductivity ($\kappa_{L}$ in W/mK) obtained from phonon dispersion curve (Phono3py) obtained from elastic constants (Slack's equation) and the corresponding thermoelectric figure of merit \textit{ZT}$-$Phono3py  and \textit{ZT}$-$Slack,  respectively.}
\label{tab:3}
\begin{tabular*}{\textwidth}{c @{\extracolsep{\fill}}ccccc}
\hline
Compound & $\kappa_{L}-$Phono3py & $\kappa_{L}-$Slack & ZT$-$Phono3py  & \textit{ZT}$-$Slack  \\
 & 300\,K\hspace{0.2in} 550\,K & 300\,K \hspace{0.2in}  550\,K & 300\,K \hspace{0.2in}  550\,K  & 300\,K \hspace{0.2in}  550\,K \\
 \hline
 TiNiSn & 7.54 \hspace{0.2in}  6.87 & 11.04\hspace{0.2in}  6.02 & 0.05 \hspace{0.2in}  0.15 & 0.03  \hspace{0.2in} 0.16 \\
  La$_{0.25}$Zr$_{0.5}$V$_{0.25}$NiSn & 0.66 \hspace{0.2in} 0.37 & 1.91 \hspace{0.2in}  1.04 & 0.29 \hspace{0.2in}  0.54 & 0.14 \hspace{0.2in}  0.40 \\
  La$_{0.25}$Hf$_{0.5}$V$_{0.25}$NiSn  & 0.31 \hspace{0.2in}  0.16 & 1.74 \hspace{0.2in}  0.95 & 0.48 \hspace{0.2in}  0.77 & 0.19 \hspace{0.2in}  0.53 \\
 \hline
\end{tabular*}
\end{center}
 \end{table*}
 
 \begin{figure*}
\centering
\includegraphics[width=0.99\linewidth]{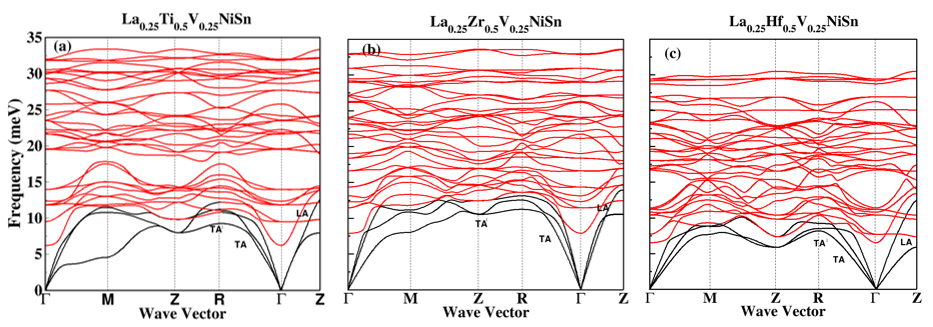}
  \caption{The calculated phonon dispersion curves for selected pentanary substituted systems such as La$_{0.25}$Ti$_{0.5}$V$_{0.25}$NiSn, La$_{0.25}$Zr$_{0.5}$V$_{0.25}$NiSn and La$_{0.25}$Hf$_{0.5}$V$_{0.25}$NiSn obtained from finite difference methods and are shown in (a), (b) and (c), respectively along the high symmetry lines. The optical and acoustic modes are highlighted with red and black colour, respectively}.
\label{fgr:9}
\end{figure*}

\subsection{Lattice thermal conductivity and Thermoelectric figure of merit}
  Thermal conductivity plays an important role to study the behavior of the atom within a crystal lattice when it is heated or cooled. The low or high thermal conductivity of the materials are very useful to use them in wide range of applications to achieve the best performance of the systems. Therefore it is necessary to investigate the thermal conductivity of the materials as an important parameter for designing  new materials. The DFT calculation of thermal conductivity is computationally expensive and time consuming. In the 19$^{th}$ century Debye~\cite{debye1912theorie} proposed the concept of phonon which is the lattice vibration of the solid crystal. In his work he used the quantized normal mode of atomic vibrations to explain the specific heat capacity of the crystalline solids. The concept of phonon provide a new way to describe the lattice vibration of the solids from which many thermodynamic properties can be calculated like lattice thermal conductivity, Debye temperature, G\"{u}neisen parameter, specific heat capacity, and lattice thermal expansion. Peierls ~\cite{peierls1955quantum,ziman1960electrons} was among the first to use the idea of Boltzmann transport theory (BTE) for calculating the lattice phonon life times and lattice thermal conductivity. Since then, solving BTE is considered as a best method for accurately predicting the $\kappa_{L}$~\cite{broido2007intrinsic, ward2010intrinsic, tang2010lattice}.  But, solving the BTE is a complicated task. Therefore to solve the complex set of equations a variety of approximation have thus been used for understanding the phonon driven thermal properties~\cite{callaway1959model, allen2013improved, deinzer2003ab}. Nowadays with more computational power and open source code such as Phono3py ~\cite{togo2015distributions}, PhonTS~\cite{chernatynskiy2015phonon}, almaBTE ~\cite{tadano2014anharmonic} and ShengBTE~\cite{li2014shengbte} one can perform the  \textit{ab initio } calculations to solve the BTE from the third order anharmonic force constant for calculating the $\kappa_{L}$. However, calculation of third order interatomic force constant is time$-$consuming and computationally expensive. The simplest method or formula was developed by Slack~\cite{slack1991maximum} for calculating the $\kappa_{L}$ by measuring the speed of sound which carry the energy of phonon and anharmonicity in term of Debye temperature and average Gr\"{u}neisen parameter, which is given as
 \begin{equation} \label{eu_eqn2}
\kappa_{L}=A\dfrac{\overline{M_{a}}\delta\theta\textsubscript{a}^{3}}{\gamma^{2}Tn^{2/3}}
\end{equation}
where $\overline{M_{a}}$ is the average mass per atom in the crystal, $\theta_{a}$ is the acoustic Debye temperature, $\delta$ is the cube root of the average volume per atom, $n$ is the number of atoms in the primitive unit cell, $\gamma$ is the Gr\"{u}neisen parameter.
and $A$ is a physical quantity which can be calculated as
A$=\dfrac{2.43\times10^{-8}}{1-0.514/\gamma+0.228/\gamma^{2}}$
The bulk modulus (B) and shear modulus (G) are obtained from the Voigt$-$Reuss$-$Hill$-$(VRH) theory as 
\begin{equation} \label{eu_eqn3}
B_v = 1/9(C_{11}+C_{22}+C_{33})+2/9(C_{12}+C_{23}+C_{13})
\end{equation}
\begin{equation} \label{eu_eqn4}
\begin{split}
     G_V&=1/15(C_{11}+C_{22}+C_{33})-1/15(C_{12}+C_{23} +C_{13})\\
     & +1/5(C_{44}+C_{55}+C_{66}) 
     \end{split}
\end{equation} 
 \begin{equation} \label{eu_eqn5}
1/B_R = (S_{11} + S_{22} + S_{33}) + 2(S_{12} + S_{23} + S_{13}) 
\end{equation}
 \begin{equation} 
 \begin{split}
     1/G_R& =4/15 (S_{11} + S_{22} + S_{33}) -4/15 (S_{12} + S_{23} +S_{13}) \\
 & +1/5 (S_{44} + S_{55} + S_{66})
 \label{eu_eqn6}
 \end{split}
\end{equation}

here the compliance constants S$_{ij}$ is the inverse matrix of the single crystal elastic constant C$_{ij}$. Finally the B and G are obtained by averaging the B$_{V}$ and  B$_{R}$, G$_{V}$ and G$_{R}$.
\begin{equation}
B=\frac{B_{V}+B_{R}}{2}
\end{equation}

\begin{equation}
G=\frac{G_{V}+G_{R}}{2}
\end{equation}

The Young's modulus (E) and the Poisson's ratio ($\nu$) were calculated as follows

\begin{equation} \label{eu_eqn7}
E = \frac{9BG}{3B+G} 
\end{equation} 
\begin{equation} \label{eu_eqn8} 
\nu = \frac{3B-2G}{6B+2G} 
\end{equation} 
Furthermore the transverse (v$_{t}$) , longitudinal  (v$_{l}$) and average (v$_{m}$) sound velocities are calculated by the following equations 
  
 \begin{equation} \label{eu_eqn9} 
 v_t = \sqrt{\dfrac{E}{2\rho(1+\nu)}} = \sqrt{\frac{G}{\rho}}
 \end{equation} 
 
 \begin{equation} \label{eu_eqn10}
 v_{l}=\dfrac{E(1-\nu)}{\rho(1+\nu)(1-2\nu)} = \sqrt{(\frac{3B+4G}{3\rho})} 
 \end{equation} 
 
\begin{equation} \label{eu_eqn11}
 v_{m}=\sqrt{[\frac{1}{3}(\frac{2}{v\textsubscript{t}\textsuperscript{3}}+\frac{1}{v\textsubscript{l}\textsuperscript{3}})]} 
 \end{equation} 
 here $\rho$ is the density of the material.
 
 From the  calculated v$_{t}$, v$_{l}$ and v$_{m}$ using the above approach one can calculate the Gr\"{u}neisen parameter ($\gamma$), Debye temperature ($\theta_{D}$) and acoustic Debye temperature ($\theta_{a}$) by the following equations 

  \begin{equation} \label{eu_eqn12}
 \gamma=\dfrac{9-12(v\textsubscript{t}/v\textsubscript{l})^{2}}{2+4(v\textsubscript{t}/v\textsubscript{l})^{2}}
 \end{equation} 

 \begin{equation} \label{eu_eqn12}
 \theta\textsubscript{D}=\frac{h}{k}[ \frac{3n}{4\pi}(\frac{N\textsubscript{A}\rho}{M})]\textsuperscript{1/3} v\textsubscript{m} 
 \end{equation} 
 
 \begin{equation} \label{eu_eqn12}
 \theta\textsubscript{a}=\theta\textsubscript{D}n^{-1/3}
 \end{equation}

These parameters are easily calculated from the elastic properties such as bulk and shear moduli. The elastic constants and moduli can reveal the important thermal and lattice dynamic properties such as mechanical stability and chemical nature of the solids. The elastic constants C$_{ij}$ are calculated from the strain$-$stress relationship~\cite{kachanov2003handbook}. From the Voigt$-$Reuss$-$Hill (VRH) theory ~\cite{den1999relation}. one can calculate the elastic properties, such as bulk and shear modulus from the elastic constants. Table. \ref{tbl:2} shows the calculated elastic constants C$_{ij}$ and elastic moduli. From the calculated C$_{ij}$ it is found that all the currently investigated compounds in the present study are mechanically stable as they obey the Cauchy$-$Born rule~\cite{ericksen1984phase, born1940stability} about the stability criteria for the cubic and tetragonal structures. The calculated elastic constants and moduli show an increasing trend as Ti is substituted with the La/V, Hf or Zr in the TiNiSn system. The Hf substituted TiNiSn have large value of elastic constants and moduli and hence exhibit superior mechanical properties compared to other substituted systems. The calculated bulk modulus, shear modulus, Young's modulus  and  Poisson's ratio  are listed in Table. \ref{tbl:2}. 
\par 
Finally the lattice thermal conductivity is calculated from Slack's equation using the Debye temperature and Gr\"{u}neisen parameter calculated from elastic moduli such as B and G. (see Fig. \ref{fgr:8}. (a)). We have also calculated the $\kappa_{L}$ from the phonon band structure using Phono3py code. Fig. \ref{fgr:8}. (a) shows the comparison between the calculated $\kappa_{L}$ from Phono3py using phonon dispersion and Slack's equation using the calculated elastic constants. From the Fig. \ref{fgr:8}. (a) we can see that at lower temperatures i.e below 300\,K the difference between the calculated $\kappa_{L}$ values from Slack and Phono3py are larger than the values at higher temperature for parent TiNiSn and the pentanary systems. However at higher temperatures we observed smaller difference between them. In the case of TiNiSn (see the inset Fig. \ref{fgr:8}. (a) the difference between the $\kappa_{L}$ values obtained from these two approaches is very small and it overlap at 500\,K. The calculated $\kappa_{L}$ value for TiNiSn from the phonon dispersion relation and Slack's approach are 6.87 and 6.02 W/mK at 550\,K. Also we observed that our calculated $\kappa_{L}$ value for TiNiSn from Phono3py code is well matched with the  reported experimental studies~\cite{bhattacharya2002grain, katayama2003effects,downie2013enhanced}.
\par 
From the Fig. \ref{fgr:8}. (a) we observed that the multinary substituted systems have lower $\kappa_{L}$ values than those from the parent TiNiSn as we have expected due to increase in phonon scattering. In the case of La$_{0.25}$Zr$_{0.5}$V$_{0.25}$NiSn and La$_{0.25}$Hf$_{0.5}$V$_{0.25}$NiSn the calculated $\kappa_{L}$ from Slack's equation at 300\,K are 1.91 and 1.74 W/mK, respectively. However, the calculated $\kappa_{L}$ value from phonon dispersion curve for these compounds are much smaller (only 0.66 and 0.31 W/mK, respectively) than that from Slack's approach. It may be noted  that our calculated lattice thermal conductivity from both the methods follows the same trend i.e reduces with increase of temperature. The La$_{0.25}$Hf$_{0.5}$V$_{0.25}$NiSn system shows small value of $\kappa_{L}$ and this can be understood from the difference in Debye temperature, Gr\"{u}neisen parameter. The La$_{0.25}$Hf$_{0.5}$V$_{0.25}$NiSn has low value of  Debye temperature and high value of Gr\"{u}neisen parameter which is favorable for low $\kappa_{L}$. 
\par 
The \textit{ZT} of parent and the pentanary systems are evaluated by substituting the corresponding calculated value of S, $\sigma$, T, $\kappa_{e}$ and the $\kappa_{L}$ into the \textit{ ZT} equation \ref{eqn1}. Figure. \ref{fgr:8}. (b) shows the calculated \textit{ZT} value as a function of temperature. We have also calculated the \textit{ZT} value by including the $\kappa_{L}$  calculated from Slack's equation and from phonon dispersion relation using the Phono3py code. From the Fig.  \ref{fgr:8}. (b) one can see that the \textit{ZT} value for TiNiSn calculated from both the methods increases with temperature and reaches maximum value at around 800\,K. Also at 500\,K the \textit{ ZT} values obtained based on $\kappa_{L}$ values obtained from both the approach overlap with each other which follow the same trend as $\kappa_{L}$ . The calculated \textit{ZT} for TiNiSn including $\kappa_{L}$ value calculated from Slack and Phono3py are 0.27 and 0.22 at 800\,K, respectively. 
In the case of La$_{0.25}$Zr$_{0.5}$V$_{0.25}$NiSn and La$_{0.25}$Hf$_{0.5}$V$_{0.25}$NiSn we find the \textit{ZT} value increases with temperature and reaches maximum value at 550\,K and then reduces at higher temperatures. We can observe the significant difference in the \textit{ZT} values calculated from both the methods. The calculated \textit{ZT} value for La$_{0.25}$Zr$_{0.5}$V$_{0.25}$NiSn  from Slack's equation and from phonon dispersion relation using the Phono3py code are 0.54 and 0.4 at 550\,K, respectively. However, the difference between the \textit{ZT} values obtained from these two approaches are minimum at 800\,K. In the case of La$_{0.25}$Hf$_{0.5}$V$_{0.25}$NiSn also  we found the same trend in the \textit{ZT} as La$_{0.25}$Zr$_{0.5}$V$_{0.25}$NiSn  and the calculated values for \textit{ZT} from these two approaches are 0.77 and 0.53 at 550\,K, respectively.
\par

We have also compared our calculated $\kappa_{L}$  and \textit{ ZT} values with the available experimental results (see inset figure in Fig. \ref{fgr:8}. (a)) and (b), respectively. Our calculated $\kappa_{L}$ from phonon dispersion relation using the Phono3py code shows slightly larger value as compared to experiment. However, the $\kappa_{L}$ calculated from Slack's approach shows the large difference at lower temperature (below 300\,K), but shows very good agreement at high temperature (above 400\,K). However, the experimentally estimated $\kappa_{L}$ values are found in the range between 2.4$-$6.08 W/mK. Our reported $\kappa_{L}$ provide reasonable agreement with the reported experimental studies. The experimentally estimated \textit{ZT} show the little large value compared with our \textit{ZT} calculated from phonon dispersion relation using the Phono3py code and Slack's equation. The discrepancy between the theoretical and experimental calculated $\kappa_{L}$  and \textit{ ZT}  may be attributed to the fact that the theoretical results are applicable to defect free single crystal and experimetally synthesised samples are usually poly crytal with various defects that could influence the transport properties. Moreover, synthesizing the pure TiNiSn is very challenging, small amounts of impurity phases and interstitial Ni defects change the majority and minority charge carrier concentration and therefore the TE transport properties~\cite{douglas2012enhanced, kirievsky2013phase,berche2018oxidation,young2019processing}. Unfortunately there are no experimental investigations on the TE properties of La$_{0.25}$Zr$_{0.5}$V$_{0.25}$NiSn and La$_{0.25}$Hf$_{0.5}$V$_{0.25}$NiSn  to compare with the present results. Therefore, we hope that there will be more theoretical and experimental work on these materials in  future to develop high efficiency TE materials based on HH alloys. 

\subsection{Lattice dynamic calculation of pentanary substituted TiNiSn}
The thermodynamic properties of selected systems are calculated with the Phonopy and Phono3py codes using the finite displacement and supercell approach. The generation of the supercell structures with displacements are made using the Phonopy code. The supercell structure contains the information about the atomic displacements, then  the forces acting on the atoms corresponding to each displacement set are calculated by VASP code. All the force sets calculated by VASP code are then used to calculate the force constants. The dynamical matrix is built from the force constants and from the matrix one can generate the phonon frequencies and eigen vectors. Figure. \ref{fgr:9} shows the phonon band structure for some of the selected pentanary systems. 
\par
The phonon dispersion curves are plotted along the high symmetry direction in the first BZ. The number of vibrational modes of the system depends on the number of atoms present in the unit cell. The 12 atoms present in the selected systems gives a total of 36 phonon branches, including one longitudinal acoustic (LA) mode, two transverse acoustic (TA) modes, and 33 optical modes. From the phonon dispersion curves shown in Fig. \ref{fgr:9}. (a), (b) and (c) for La$_{0.25}$X$^{IV}_{0.5}$V$_{0.25}$NiSn where (X$^{IV}$ $=$ Ti, Zr, Hf), it  can be seen that for all these selected pentanary systems both the acoustic and optical bands are well mixed. By comparing the shape of the acoustic and optical bands of these systems we found that there is a frequency shift from La$_{0.25}$Ti$_{0.5}$V$_{0.25}$NiSn to La$_{0.25}$Hf$_{0.5}$V$_{0.25}$NiSn and this is due to the mass difference between the Ti, Zr and Hf atoms. 

\begin{figure}
\centering
\includegraphics[height=7cm]{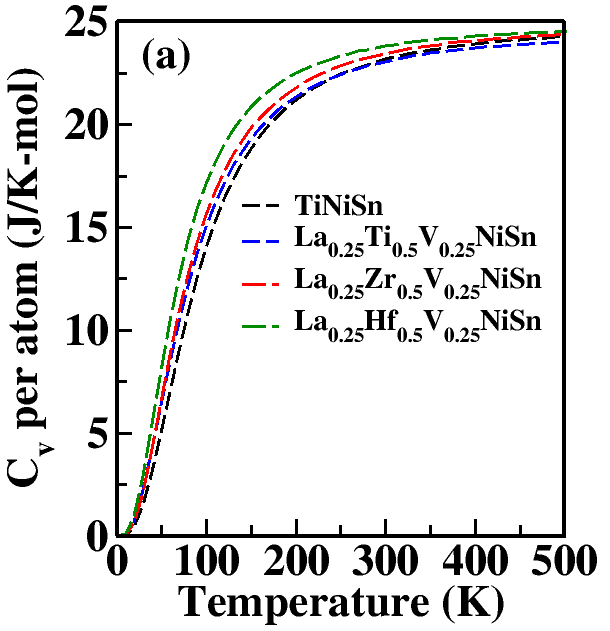}
\caption{The calculated heat capacity as a function of temperature for  pure, quaternary and pentanary substituted TiNiSn series.}
\label{fgr:10}
\end{figure}

In the case of La$_{0.25}$Zr$_{0.5}$V$_{0.25}$NiSn and La$_{0.25}$Hf$_{0.5}$V$_{0.25}$NiSn we can find a strong optical$-$acoustic band mixing and therefore these systems are expected to have more optical$-$acoustic phonon$-$phonon scattering. We have also noticed the frequency shift towards lower energy for the Hf containing system in our calculated phonon dispersion curves compared to that of Ti and Zr containing systems.  To know the contribution of various atoms to the heat capacity (C$_{v}$) we have calculated the heat capacity at constant volume (C$_{v}$).
Figure. \ref{fgr:10}. shows the (C$_{v}$) at temperatures from 0 to 500\,K, for all selected systems. The calculated C$_{v}$ values for La$_{0.25}$Ti$_{0.5}$V$_{0.25}$NiSn, La$_{0.25}$Zr$_{0.5}$V$_{0.25}$NiSn and La$_{0.25}$Hf$_{0.5}$V$_{0.25}$NiSn at 100\,K are 15.2, 15.9 and 17.25 $(J.K^{-1}.mol^{-1})$, respectively. The shift in frequency towards lower frequency in the phonon band structure for La$_{0.25}$Hf$_{0.5}$V$_{0.25}$NiSn over Ti or Zr containing system is due to the large average atomic mass in this system that results in higher C$_{v}$ value compared to the other systems.   

\section{Conclusion}
This work focuses on designing the new series of pentanary substituted TiNiSn HH alloys with the aliovalent substitution retaining the 18 valence electrons count using VASP$-$PAW method and studied the electronic structure of nine pentanary systems.  The thermoelectric transport properties have been studied by combining Boltzmann transport theory with the electronic structure obtained from full potential Wien2k code. Our calculated band structure analysis shows that all pentanary substituted TiNiSn systems has semiconducting behavior and small band gap compared to pure TiNiSn. The calculated band gap values for the pentanary systems in the present work ranges from 0.17 to 0.32 eV. The calculated lattice part of thermal conductivity  $\kappa_{L}$ using finite difference method implemented in the Phono3py code and the calculated elastic constants with Slack's equation show good agreement at high temperatures. From the calculated  $\kappa_{L}$ we show that the pentanary substituted systems have a lower value compared to pure TiNiSn due to mass fluctuation. Moreover, we have obtained comparable $\kappa_{L}$ values from Phono3py code and Slack's equation for the pentanary systems  La$_{0.25}$Hf$_{0.5}$V$_{0.25}$NiSn and La$_{0.25}$Zr$_{0.5}$V$_{0.25}$NiSn which are found to be 0.37 (1.04) and 0.16 (0.95) W/mK at 550\,K, respectively and the corresponding maximum \textit{ZT} value are found to be 0.54 (0.4) and 0.77 (0.53) at 550\,K, respectively. These results show that the pentanary substituted TiNiSn is an excellent example where we can increase the \textit{ZT} value by lowering the band gap through electronic band structure engineering thus increase electrical conductivity and decrease the thermal conductivity  by mass fluctuation. The improved thermoelectric efficiency in these pentanary based systems suggests a novel means to further improve thermoelectric performance through band engineering and lowering the thermal conductivity by multinary addition thus  plays an important role in the development of  efficient TE materials for high efficiency  TE devices.

\section*{Acknowledgment}
The authors are grateful to the Science and Engineering Research Board (SERB) a stationary body of  Department of Science and Technology, Government of India, for the funding support  under the scheme SERB$-$Overseas Visiting Doctoral Fellowship(OVDF) via award  no. ODF/2018/000845  and the Research Council of Norway for providing the computer time (under the project number NN2875k) at the Norwegian supercomputer facility. The authors would also like to acknowledge the SERB$-$Core Research Grant (CRG) vide file no.CRG/2020/001399.



\bibliography{bib}

\begin{thebibliography}{121}%
\makeatletter
\providecommand \@ifxundefined [1]{%
 \@ifx{#1\undefined}
}%
\providecommand \@ifnum [1]{%
 \ifnum #1\expandafter \@firstoftwo
 \else \expandafter \@secondoftwo
 \fi
}%
\providecommand \@ifx [1]{%
 \ifx #1\expandafter \@firstoftwo
 \else \expandafter \@secondoftwo
 \fi
}%
\providecommand \natexlab [1]{#1}%
\providecommand \enquote  [1]{``#1''}%
\providecommand \bibnamefont  [1]{#1}%
\providecommand \bibfnamefont [1]{#1}%
\providecommand \citenamefont [1]{#1}%
\providecommand \href@noop [0]{\@secondoftwo}%
\providecommand \href [0]{\begingroup \@sanitize@url \@href}%
\providecommand \@href[1]{\@@startlink{#1}\@@href}%
\providecommand \@@href[1]{\endgroup#1\@@endlink}%
\providecommand \@sanitize@url [0]{\catcode `\\12\catcode `\$12\catcode
  `\&12\catcode `\#12\catcode `\^12\catcode `\_12\catcode `\%12\relax}%
\providecommand \@@startlink[1]{}%
\providecommand \@@endlink[0]{}%
\providecommand \url  [0]{\begingroup\@sanitize@url \@url }%
\providecommand \@url [1]{\endgroup\@href {#1}{\urlprefix }}%
\providecommand \urlprefix  [0]{URL }%
\providecommand \Eprint [0]{\href }%
\providecommand \doibase [0]{http://dx.doi.org/}%
\providecommand \selectlanguage [0]{\@gobble}%
\providecommand \bibinfo  [0]{\@secondoftwo}%
\providecommand \bibfield  [0]{\@secondoftwo}%
\providecommand \translation [1]{[#1]}%
\providecommand \BibitemOpen [0]{}%
\providecommand \bibitemStop [0]{}%
\providecommand \bibitemNoStop [0]{.\EOS\space}%
\providecommand \EOS [0]{\spacefactor3000\relax}%
\providecommand \BibitemShut  [1]{\csname bibitem#1\endcsname}%
\let\auto@bib@innerbib\@empty
\bibitem [{\citenamefont {Bell}(2008)}]{bell2008cooling}%
  \BibitemOpen
  \bibfield  {author} {\bibinfo {author} {\bibfnamefont {L.~E.}\ \bibnamefont
  {Bell}},\ }\href@noop {} {\bibfield  {journal} {\bibinfo  {journal}
  {Science}\ }\textbf {\bibinfo {volume} {321}},\ \bibinfo {pages} {1457}
  (\bibinfo {year} {2008})}\BibitemShut {NoStop}%
\bibitem [{\citenamefont {Jaumot}(1958)}]{jaumot1958thermoelectric}%
  \BibitemOpen
  \bibfield  {author} {\bibinfo {author} {\bibfnamefont {F.~E.}\ \bibnamefont
  {Jaumot}},\ }\href@noop {} {\bibfield  {journal} {\bibinfo  {journal}
  {Proceedings of the IRE}\ }\textbf {\bibinfo {volume} {46}},\ \bibinfo
  {pages} {538} (\bibinfo {year} {1958})}\BibitemShut {NoStop}%
\bibitem [{\citenamefont {Zhao}\ and\ \citenamefont
  {Tan}(2014)}]{zhao2014review}%
  \BibitemOpen
  \bibfield  {author} {\bibinfo {author} {\bibfnamefont {D.}~\bibnamefont
  {Zhao}}\ and\ \bibinfo {author} {\bibfnamefont {G.}~\bibnamefont {Tan}},\
  }\href@noop {} {\bibfield  {journal} {\bibinfo  {journal} {Applied Thermal
  Engineering}\ }\textbf {\bibinfo {volume} {66}},\ \bibinfo {pages} {15}
  (\bibinfo {year} {2014})}\BibitemShut {NoStop}%
\bibitem [{\citenamefont {Joshi}\ \emph {et~al.}(2011)\citenamefont {Joshi},
  \citenamefont {Yan}, \citenamefont {Wang}, \citenamefont {Liu}, \citenamefont
  {Chen},\ and\ \citenamefont {Ren}}]{joshi2011enhancement}%
  \BibitemOpen
  \bibfield  {author} {\bibinfo {author} {\bibfnamefont {G.}~\bibnamefont
  {Joshi}}, \bibinfo {author} {\bibfnamefont {X.}~\bibnamefont {Yan}}, \bibinfo
  {author} {\bibfnamefont {H.}~\bibnamefont {Wang}}, \bibinfo {author}
  {\bibfnamefont {W.}~\bibnamefont {Liu}}, \bibinfo {author} {\bibfnamefont
  {G.}~\bibnamefont {Chen}}, \ and\ \bibinfo {author} {\bibfnamefont
  {Z.}~\bibnamefont {Ren}},\ }\href@noop {} {\bibfield  {journal} {\bibinfo
  {journal} {Advanced Energy Materials}\ }\textbf {\bibinfo {volume} {1}},\
  \bibinfo {pages} {643} (\bibinfo {year} {2011})}\BibitemShut {NoStop}%
\bibitem [{\citenamefont {Kanatzidis}(2009)}]{kanatzidis2009nanostructured}%
  \BibitemOpen
  \bibfield  {author} {\bibinfo {author} {\bibfnamefont {M.~G.}\ \bibnamefont
  {Kanatzidis}},\ }\href@noop {} {\bibfield  {journal} {\bibinfo  {journal}
  {Chemistry of materials}\ }\textbf {\bibinfo {volume} {22}},\ \bibinfo
  {pages} {648} (\bibinfo {year} {2009})}\BibitemShut {NoStop}%
\bibitem [{\citenamefont {Pei}\ \emph {et~al.}(2011{\natexlab{a}})\citenamefont
  {Pei}, \citenamefont {Lensch-Falk}, \citenamefont {Toberer}, \citenamefont
  {Medlin},\ and\ \citenamefont {Snyder}}]{pei2011high}%
  \BibitemOpen
  \bibfield  {author} {\bibinfo {author} {\bibfnamefont {Y.}~\bibnamefont
  {Pei}}, \bibinfo {author} {\bibfnamefont {J.}~\bibnamefont {Lensch-Falk}},
  \bibinfo {author} {\bibfnamefont {E.~S.}\ \bibnamefont {Toberer}}, \bibinfo
  {author} {\bibfnamefont {D.~L.}\ \bibnamefont {Medlin}}, \ and\ \bibinfo
  {author} {\bibfnamefont {G.~J.}\ \bibnamefont {Snyder}},\ }\href@noop {}
  {\bibfield  {journal} {\bibinfo  {journal} {Advanced Functional Materials}\
  }\textbf {\bibinfo {volume} {21}},\ \bibinfo {pages} {241} (\bibinfo {year}
  {2011}{\natexlab{a}})}\BibitemShut {NoStop}%
\bibitem [{\citenamefont {Xie}\ \emph {et~al.}(2010)\citenamefont {Xie},
  \citenamefont {He}, \citenamefont {Kang}, \citenamefont {Tang}, \citenamefont
  {Zhu}, \citenamefont {Laver}, \citenamefont {Wang}, \citenamefont {Copley},
  \citenamefont {Brown}, \citenamefont {Zhang} \emph
  {et~al.}}]{xie2010identifying}%
  \BibitemOpen
  \bibfield  {author} {\bibinfo {author} {\bibfnamefont {W.}~\bibnamefont
  {Xie}}, \bibinfo {author} {\bibfnamefont {J.}~\bibnamefont {He}}, \bibinfo
  {author} {\bibfnamefont {H.~J.}\ \bibnamefont {Kang}}, \bibinfo {author}
  {\bibfnamefont {X.}~\bibnamefont {Tang}}, \bibinfo {author} {\bibfnamefont
  {S.}~\bibnamefont {Zhu}}, \bibinfo {author} {\bibfnamefont {M.}~\bibnamefont
  {Laver}}, \bibinfo {author} {\bibfnamefont {S.}~\bibnamefont {Wang}},
  \bibinfo {author} {\bibfnamefont {J.~R.}\ \bibnamefont {Copley}}, \bibinfo
  {author} {\bibfnamefont {C.~M.}\ \bibnamefont {Brown}}, \bibinfo {author}
  {\bibfnamefont {Q.}~\bibnamefont {Zhang}},  \emph {et~al.},\ }\href@noop {}
  {\bibfield  {journal} {\bibinfo  {journal} {Nano letters}\ }\textbf {\bibinfo
  {volume} {10}},\ \bibinfo {pages} {3283} (\bibinfo {year}
  {2010})}\BibitemShut {NoStop}%
\bibitem [{\citenamefont {Poudeu}\ \emph {et~al.}(2009)\citenamefont {Poudeu},
  \citenamefont {Gu{\'e}guen}, \citenamefont {Wu}, \citenamefont {Hogan},\ and\
  \citenamefont {Kanatzidis}}]{poudeu2009high}%
  \BibitemOpen
  \bibfield  {author} {\bibinfo {author} {\bibfnamefont {P.~F.}\ \bibnamefont
  {Poudeu}}, \bibinfo {author} {\bibfnamefont {A.}~\bibnamefont {Gu{\'e}guen}},
  \bibinfo {author} {\bibfnamefont {C.-I.}\ \bibnamefont {Wu}}, \bibinfo
  {author} {\bibfnamefont {T.}~\bibnamefont {Hogan}}, \ and\ \bibinfo {author}
  {\bibfnamefont {M.~G.}\ \bibnamefont {Kanatzidis}},\ }\href@noop {}
  {\bibfield  {journal} {\bibinfo  {journal} {Chemistry of materials}\ }\textbf
  {\bibinfo {volume} {22}},\ \bibinfo {pages} {1046} (\bibinfo {year}
  {2009})}\BibitemShut {NoStop}%
\bibitem [{\citenamefont {Joshi}\ \emph {et~al.}(2008)\citenamefont {Joshi},
  \citenamefont {Lee}, \citenamefont {Lan}, \citenamefont {Wang}, \citenamefont
  {Zhu}, \citenamefont {Wang}, \citenamefont {Gould}, \citenamefont {Cuff},
  \citenamefont {Tang}, \citenamefont {Dresselhaus} \emph
  {et~al.}}]{joshi2008enhanced}%
  \BibitemOpen
  \bibfield  {author} {\bibinfo {author} {\bibfnamefont {G.}~\bibnamefont
  {Joshi}}, \bibinfo {author} {\bibfnamefont {H.}~\bibnamefont {Lee}}, \bibinfo
  {author} {\bibfnamefont {Y.}~\bibnamefont {Lan}}, \bibinfo {author}
  {\bibfnamefont {X.}~\bibnamefont {Wang}}, \bibinfo {author} {\bibfnamefont
  {G.}~\bibnamefont {Zhu}}, \bibinfo {author} {\bibfnamefont {D.}~\bibnamefont
  {Wang}}, \bibinfo {author} {\bibfnamefont {R.~W.}\ \bibnamefont {Gould}},
  \bibinfo {author} {\bibfnamefont {D.~C.}\ \bibnamefont {Cuff}}, \bibinfo
  {author} {\bibfnamefont {M.~Y.}\ \bibnamefont {Tang}}, \bibinfo {author}
  {\bibfnamefont {M.~S.}\ \bibnamefont {Dresselhaus}},  \emph {et~al.},\
  }\href@noop {} {\bibfield  {journal} {\bibinfo  {journal} {Nano letters}\
  }\textbf {\bibinfo {volume} {8}},\ \bibinfo {pages} {4670} (\bibinfo {year}
  {2008})}\BibitemShut {NoStop}%
\bibitem [{\citenamefont {Liu}\ \emph {et~al.}(2012{\natexlab{a}})\citenamefont
  {Liu}, \citenamefont {Yan}, \citenamefont {Chen},\ and\ \citenamefont
  {Ren}}]{liu2012recent}%
  \BibitemOpen
  \bibfield  {author} {\bibinfo {author} {\bibfnamefont {W.}~\bibnamefont
  {Liu}}, \bibinfo {author} {\bibfnamefont {X.}~\bibnamefont {Yan}}, \bibinfo
  {author} {\bibfnamefont {G.}~\bibnamefont {Chen}}, \ and\ \bibinfo {author}
  {\bibfnamefont {Z.}~\bibnamefont {Ren}},\ }\href@noop {} {\bibfield
  {journal} {\bibinfo  {journal} {Nano Energy}\ }\textbf {\bibinfo {volume}
  {1}},\ \bibinfo {pages} {42} (\bibinfo {year}
  {2012}{\natexlab{a}})}\BibitemShut {NoStop}%
\bibitem [{\citenamefont {Alam}\ and\ \citenamefont
  {Ramakrishna}(2013)}]{alam2013review}%
  \BibitemOpen
  \bibfield  {author} {\bibinfo {author} {\bibfnamefont {H.}~\bibnamefont
  {Alam}}\ and\ \bibinfo {author} {\bibfnamefont {S.}~\bibnamefont
  {Ramakrishna}},\ }\href@noop {} {\bibfield  {journal} {\bibinfo  {journal}
  {Nano energy}\ }\textbf {\bibinfo {volume} {2}},\ \bibinfo {pages} {190}
  (\bibinfo {year} {2013})}\BibitemShut {NoStop}%
\bibitem [{\citenamefont {Slack}(1995)}]{slack1995new}%
  \BibitemOpen
  \bibfield  {author} {\bibinfo {author} {\bibfnamefont {G.~A.}\ \bibnamefont
  {Slack}},\ }\href@noop {} {\bibfield  {journal} {\bibinfo  {journal} {CRC
  handbook of thermoelectrics}\ ,\ \bibinfo {pages} {407}} (\bibinfo {year}
  {1995})}\BibitemShut {NoStop}%
\bibitem [{\citenamefont {Toberer}\ \emph {et~al.}(2011)\citenamefont
  {Toberer}, \citenamefont {Zevalkink},\ and\ \citenamefont
  {Snyder}}]{toberer2011phonon}%
  \BibitemOpen
  \bibfield  {author} {\bibinfo {author} {\bibfnamefont {E.~S.}\ \bibnamefont
  {Toberer}}, \bibinfo {author} {\bibfnamefont {A.}~\bibnamefont {Zevalkink}},
  \ and\ \bibinfo {author} {\bibfnamefont {G.~J.}\ \bibnamefont {Snyder}},\
  }\href@noop {} {\bibfield  {journal} {\bibinfo  {journal} {Journal of
  Materials Chemistry}\ }\textbf {\bibinfo {volume} {21}},\ \bibinfo {pages}
  {15843} (\bibinfo {year} {2011})}\BibitemShut {NoStop}%
\bibitem [{\citenamefont {Cohn}\ \emph {et~al.}(1999)\citenamefont {Cohn},
  \citenamefont {Nolas}, \citenamefont {Fessatidis}, \citenamefont {Metcalf},\
  and\ \citenamefont {Slack}}]{cohn1999glasslike}%
  \BibitemOpen
  \bibfield  {author} {\bibinfo {author} {\bibfnamefont {J.}~\bibnamefont
  {Cohn}}, \bibinfo {author} {\bibfnamefont {G.}~\bibnamefont {Nolas}},
  \bibinfo {author} {\bibfnamefont {V.}~\bibnamefont {Fessatidis}}, \bibinfo
  {author} {\bibfnamefont {T.}~\bibnamefont {Metcalf}}, \ and\ \bibinfo
  {author} {\bibfnamefont {G.}~\bibnamefont {Slack}},\ }\href@noop {}
  {\bibfield  {journal} {\bibinfo  {journal} {Physical Review Letters}\
  }\textbf {\bibinfo {volume} {82}},\ \bibinfo {pages} {779} (\bibinfo {year}
  {1999})}\BibitemShut {NoStop}%
\bibitem [{\citenamefont {Nolas}\ \emph {et~al.}(1998)\citenamefont {Nolas},
  \citenamefont {Cohn},\ and\ \citenamefont {Slack}}]{nolas1998effect}%
  \BibitemOpen
  \bibfield  {author} {\bibinfo {author} {\bibfnamefont {G.}~\bibnamefont
  {Nolas}}, \bibinfo {author} {\bibfnamefont {J.}~\bibnamefont {Cohn}}, \ and\
  \bibinfo {author} {\bibfnamefont {G.}~\bibnamefont {Slack}},\ }\href@noop {}
  {\bibfield  {journal} {\bibinfo  {journal} {Physical Review B}\ }\textbf
  {\bibinfo {volume} {58}},\ \bibinfo {pages} {164} (\bibinfo {year}
  {1998})}\BibitemShut {NoStop}%
\bibitem [{\citenamefont {Caillat}\ \emph {et~al.}(1996)\citenamefont
  {Caillat}, \citenamefont {Borshchevsky},\ and\ \citenamefont
  {Fleurial}}]{caillat1996properties}%
  \BibitemOpen
  \bibfield  {author} {\bibinfo {author} {\bibfnamefont {T.}~\bibnamefont
  {Caillat}}, \bibinfo {author} {\bibfnamefont {A.}~\bibnamefont
  {Borshchevsky}}, \ and\ \bibinfo {author} {\bibfnamefont {J.-P.}\
  \bibnamefont {Fleurial}},\ }\href@noop {} {\bibfield  {journal} {\bibinfo
  {journal} {Journal of Applied Physics}\ }\textbf {\bibinfo {volume} {80}},\
  \bibinfo {pages} {4442} (\bibinfo {year} {1996})}\BibitemShut {NoStop}%
\bibitem [{\citenamefont {Toberer}\ \emph {et~al.}(2009)\citenamefont
  {Toberer}, \citenamefont {May},\ and\ \citenamefont
  {Snyder}}]{toberer2009zintl}%
  \BibitemOpen
  \bibfield  {author} {\bibinfo {author} {\bibfnamefont {E.~S.}\ \bibnamefont
  {Toberer}}, \bibinfo {author} {\bibfnamefont {A.~F.}\ \bibnamefont {May}}, \
  and\ \bibinfo {author} {\bibfnamefont {G.~J.}\ \bibnamefont {Snyder}},\
  }\href@noop {} {\bibfield  {journal} {\bibinfo  {journal} {Chemistry of
  Materials}\ }\textbf {\bibinfo {volume} {22}},\ \bibinfo {pages} {624}
  (\bibinfo {year} {2009})}\BibitemShut {NoStop}%
\bibitem [{\citenamefont {Kauzlarich}\ \emph {et~al.}(2007)\citenamefont
  {Kauzlarich}, \citenamefont {Brown},\ and\ \citenamefont
  {Snyder}}]{kauzlarich2007zintl}%
  \BibitemOpen
  \bibfield  {author} {\bibinfo {author} {\bibfnamefont {S.~M.}\ \bibnamefont
  {Kauzlarich}}, \bibinfo {author} {\bibfnamefont {S.~R.}\ \bibnamefont
  {Brown}}, \ and\ \bibinfo {author} {\bibfnamefont {G.~J.}\ \bibnamefont
  {Snyder}},\ }\href@noop {} {\bibfield  {journal} {\bibinfo  {journal} {Dalton
  Transactions}\ ,\ \bibinfo {pages} {2099}} (\bibinfo {year}
  {2007})}\BibitemShut {NoStop}%
\bibitem [{\citenamefont {Toberer}\ \emph {et~al.}(2010)\citenamefont
  {Toberer}, \citenamefont {May}, \citenamefont {Melot}, \citenamefont
  {Flage-Larsen},\ and\ \citenamefont {Snyder}}]{toberer2010electronic}%
  \BibitemOpen
  \bibfield  {author} {\bibinfo {author} {\bibfnamefont {E.~S.}\ \bibnamefont
  {Toberer}}, \bibinfo {author} {\bibfnamefont {A.~F.}\ \bibnamefont {May}},
  \bibinfo {author} {\bibfnamefont {B.~C.}\ \bibnamefont {Melot}}, \bibinfo
  {author} {\bibfnamefont {E.}~\bibnamefont {Flage-Larsen}}, \ and\ \bibinfo
  {author} {\bibfnamefont {G.~J.}\ \bibnamefont {Snyder}},\ }\href@noop {}
  {\bibfield  {journal} {\bibinfo  {journal} {Dalton transactions}\ }\textbf
  {\bibinfo {volume} {39}},\ \bibinfo {pages} {1046} (\bibinfo {year}
  {2010})}\BibitemShut {NoStop}%
\bibitem [{\citenamefont {Zhang}\ \emph {et~al.}(2010)\citenamefont {Zhang},
  \citenamefont {Yu}, \citenamefont {Wang},\ and\ \citenamefont
  {Li}}]{zhang2010nanostructures}%
  \BibitemOpen
  \bibfield  {author} {\bibinfo {author} {\bibfnamefont {G.}~\bibnamefont
  {Zhang}}, \bibinfo {author} {\bibfnamefont {Q.}~\bibnamefont {Yu}}, \bibinfo
  {author} {\bibfnamefont {W.}~\bibnamefont {Wang}}, \ and\ \bibinfo {author}
  {\bibfnamefont {X.}~\bibnamefont {Li}},\ }\href@noop {} {\bibfield  {journal}
  {\bibinfo  {journal} {Advanced Materials}\ }\textbf {\bibinfo {volume}
  {22}},\ \bibinfo {pages} {1959} (\bibinfo {year} {2010})}\BibitemShut
  {NoStop}%
\bibitem [{\citenamefont {Hicks}\ and\ \citenamefont
  {Dresselhaus}(1993)}]{hicks1993effect}%
  \BibitemOpen
  \bibfield  {author} {\bibinfo {author} {\bibfnamefont {L.~D.}\ \bibnamefont
  {Hicks}}\ and\ \bibinfo {author} {\bibfnamefont {M.~S.}\ \bibnamefont
  {Dresselhaus}},\ }\href@noop {} {\bibfield  {journal} {\bibinfo  {journal}
  {Physical Review B}\ }\textbf {\bibinfo {volume} {47}},\ \bibinfo {pages}
  {12727} (\bibinfo {year} {1993})}\BibitemShut {NoStop}%
\bibitem [{\citenamefont {Venkatasubramanian}\ \emph
  {et~al.}(2001)\citenamefont {Venkatasubramanian}, \citenamefont {Siivola},
  \citenamefont {Colpitts},\ and\ \citenamefont
  {O'quinn}}]{venkatasubramanian2001thin}%
  \BibitemOpen
  \bibfield  {author} {\bibinfo {author} {\bibfnamefont {R.}~\bibnamefont
  {Venkatasubramanian}}, \bibinfo {author} {\bibfnamefont {E.}~\bibnamefont
  {Siivola}}, \bibinfo {author} {\bibfnamefont {T.}~\bibnamefont {Colpitts}}, \
  and\ \bibinfo {author} {\bibfnamefont {B.}~\bibnamefont {O'quinn}},\
  }\href@noop {} {\bibfield  {journal} {\bibinfo  {journal} {Nature}\ }\textbf
  {\bibinfo {volume} {413}},\ \bibinfo {pages} {597} (\bibinfo {year}
  {2001})}\BibitemShut {NoStop}%
\bibitem [{\citenamefont {Xie}\ \emph {et~al.}(2009)\citenamefont {Xie},
  \citenamefont {Tang}, \citenamefont {Yan}, \citenamefont {Zhang},\ and\
  \citenamefont {Tritt}}]{xie2009high}%
  \BibitemOpen
  \bibfield  {author} {\bibinfo {author} {\bibfnamefont {W.}~\bibnamefont
  {Xie}}, \bibinfo {author} {\bibfnamefont {X.}~\bibnamefont {Tang}}, \bibinfo
  {author} {\bibfnamefont {Y.}~\bibnamefont {Yan}}, \bibinfo {author}
  {\bibfnamefont {Q.}~\bibnamefont {Zhang}}, \ and\ \bibinfo {author}
  {\bibfnamefont {T.~M.}\ \bibnamefont {Tritt}},\ }\href@noop {} {\bibfield
  {journal} {\bibinfo  {journal} {Journal of Applied Physics}\ }\textbf
  {\bibinfo {volume} {105}},\ \bibinfo {pages} {113713} (\bibinfo {year}
  {2009})}\BibitemShut {NoStop}%
\bibitem [{\citenamefont {Kuentzler}\ \emph {et~al.}(1992)\citenamefont
  {Kuentzler}, \citenamefont {Clad}, \citenamefont {Schmerber},\ and\
  \citenamefont {Dossmann}}]{kuentzler1992gap}%
  \BibitemOpen
  \bibfield  {author} {\bibinfo {author} {\bibfnamefont {R.}~\bibnamefont
  {Kuentzler}}, \bibinfo {author} {\bibfnamefont {R.}~\bibnamefont {Clad}},
  \bibinfo {author} {\bibfnamefont {G.}~\bibnamefont {Schmerber}}, \ and\
  \bibinfo {author} {\bibfnamefont {Y.}~\bibnamefont {Dossmann}},\ }\href@noop
  {} {\bibfield  {journal} {\bibinfo  {journal} {Journal of Magnetism and
  Magnetic Materials}\ }\textbf {\bibinfo {volume} {104}},\ \bibinfo {pages}
  {1976} (\bibinfo {year} {1992})}\BibitemShut {NoStop}%
\bibitem [{\citenamefont {Tobola}\ \emph {et~al.}(1998)\citenamefont {Tobola},
  \citenamefont {Pierre}, \citenamefont {Kaprzyk}, \citenamefont {Skolozdra},\
  and\ \citenamefont {Kouacou}}]{tobola1998crossover}%
  \BibitemOpen
  \bibfield  {author} {\bibinfo {author} {\bibfnamefont {J.}~\bibnamefont
  {Tobola}}, \bibinfo {author} {\bibfnamefont {J.}~\bibnamefont {Pierre}},
  \bibinfo {author} {\bibfnamefont {S.}~\bibnamefont {Kaprzyk}}, \bibinfo
  {author} {\bibfnamefont {R.}~\bibnamefont {Skolozdra}}, \ and\ \bibinfo
  {author} {\bibfnamefont {M.}~\bibnamefont {Kouacou}},\ }\href@noop {}
  {\bibfield  {journal} {\bibinfo  {journal} {Journal of Physics: Condensed
  Matter}\ }\textbf {\bibinfo {volume} {10}},\ \bibinfo {pages} {1013}
  (\bibinfo {year} {1998})}\BibitemShut {NoStop}%
\bibitem [{\citenamefont {Ren}\ \emph {et~al.}(2020)\citenamefont {Ren},
  \citenamefont {Fu}, \citenamefont {Qiu}, \citenamefont {Dai}, \citenamefont
  {Liu}, \citenamefont {Masuda}, \citenamefont {Asai}, \citenamefont
  {Hagihala}, \citenamefont {Lee}, \citenamefont {Torri} \emph
  {et~al.}}]{ren2020establishing}%
  \BibitemOpen
  \bibfield  {author} {\bibinfo {author} {\bibfnamefont {Q.}~\bibnamefont
  {Ren}}, \bibinfo {author} {\bibfnamefont {C.}~\bibnamefont {Fu}}, \bibinfo
  {author} {\bibfnamefont {Q.}~\bibnamefont {Qiu}}, \bibinfo {author}
  {\bibfnamefont {S.}~\bibnamefont {Dai}}, \bibinfo {author} {\bibfnamefont
  {Z.}~\bibnamefont {Liu}}, \bibinfo {author} {\bibfnamefont {T.}~\bibnamefont
  {Masuda}}, \bibinfo {author} {\bibfnamefont {S.}~\bibnamefont {Asai}},
  \bibinfo {author} {\bibfnamefont {M.}~\bibnamefont {Hagihala}}, \bibinfo
  {author} {\bibfnamefont {S.}~\bibnamefont {Lee}}, \bibinfo {author}
  {\bibfnamefont {S.}~\bibnamefont {Torri}},  \emph {et~al.},\ }\href@noop {}
  {\bibfield  {journal} {\bibinfo  {journal} {Nature communications}\ }\textbf
  {\bibinfo {volume} {11}},\ \bibinfo {pages} {1} (\bibinfo {year}
  {2020})}\BibitemShut {NoStop}%
\bibitem [{\citenamefont {Ioffe}(1957)}]{ioffe1957semiconductor}%
  \BibitemOpen
  \bibfield  {author} {\bibinfo {author} {\bibfnamefont {A.~F.}\ \bibnamefont
  {Ioffe}},\ }\href@noop {} {\emph {\bibinfo {title} {Semiconductor
  Thermoelements, and}}}\ (\bibinfo  {publisher} {Info-search, Limited},\
  \bibinfo {year} {1957})\BibitemShut {NoStop}%
\bibitem [{\citenamefont {Jaldurgam}\ \emph {et~al.}(2021)\citenamefont
  {Jaldurgam}, \citenamefont {Ahmad},\ and\ \citenamefont
  {Touati}}]{jaldurgam2021low}%
  \BibitemOpen
  \bibfield  {author} {\bibinfo {author} {\bibfnamefont {F.~F.}\ \bibnamefont
  {Jaldurgam}}, \bibinfo {author} {\bibfnamefont {Z.}~\bibnamefont {Ahmad}}, \
  and\ \bibinfo {author} {\bibfnamefont {F.}~\bibnamefont {Touati}},\
  }\href@noop {} {\bibfield  {journal} {\bibinfo  {journal} {Nanomaterials}\
  }\textbf {\bibinfo {volume} {11}},\ \bibinfo {pages} {895} (\bibinfo {year}
  {2021})}\BibitemShut {NoStop}%
\bibitem [{\citenamefont {Wang}\ \emph {et~al.}(2013)\citenamefont {Wang},
  \citenamefont {Pei}, \citenamefont {LaLonde},\ and\ \citenamefont
  {Snyder}}]{wang2013material}%
  \BibitemOpen
  \bibfield  {author} {\bibinfo {author} {\bibfnamefont {H.}~\bibnamefont
  {Wang}}, \bibinfo {author} {\bibfnamefont {Y.}~\bibnamefont {Pei}}, \bibinfo
  {author} {\bibfnamefont {A.~D.}\ \bibnamefont {LaLonde}}, \ and\ \bibinfo
  {author} {\bibfnamefont {G.~J.}\ \bibnamefont {Snyder}},\ }in\ \href@noop {}
  {\emph {\bibinfo {booktitle} {Thermoelectric Nanomaterials}}}\ (\bibinfo
  {publisher} {Springer},\ \bibinfo {year} {2013})\ pp.\ \bibinfo {pages}
  {3--32}\BibitemShut {NoStop}%
\bibitem [{\citenamefont {Zhu}\ \emph {et~al.}(2015)\citenamefont {Zhu},
  \citenamefont {Fu}, \citenamefont {Xie}, \citenamefont {Liu},\ and\
  \citenamefont {Zhao}}]{zhu2015high}%
  \BibitemOpen
  \bibfield  {author} {\bibinfo {author} {\bibfnamefont {T.}~\bibnamefont
  {Zhu}}, \bibinfo {author} {\bibfnamefont {C.}~\bibnamefont {Fu}}, \bibinfo
  {author} {\bibfnamefont {H.}~\bibnamefont {Xie}}, \bibinfo {author}
  {\bibfnamefont {Y.}~\bibnamefont {Liu}}, \ and\ \bibinfo {author}
  {\bibfnamefont {X.}~\bibnamefont {Zhao}},\ }\href@noop {} {\bibfield
  {journal} {\bibinfo  {journal} {Advanced Energy Materials}\ }\textbf
  {\bibinfo {volume} {5}},\ \bibinfo {pages} {1500588} (\bibinfo {year}
  {2015})}\BibitemShut {NoStop}%
\bibitem [{\citenamefont {Bulman}\ and\ \citenamefont
  {Cook}(2014)}]{bulman2014high}%
  \BibitemOpen
  \bibfield  {author} {\bibinfo {author} {\bibfnamefont {G.}~\bibnamefont
  {Bulman}}\ and\ \bibinfo {author} {\bibfnamefont {B.}~\bibnamefont {Cook}},\
  }in\ \href@noop {} {\emph {\bibinfo {booktitle} {Energy Harvesting and
  Storage: Materials, Devices, and Applications V}}},\ Vol.\ \bibinfo {volume}
  {9115}\ (\bibinfo {organization} {International Society for Optics and
  Photonics},\ \bibinfo {year} {2014})\ p.\ \bibinfo {pages}
  {911507}\BibitemShut {NoStop}%
\bibitem [{\citenamefont {Xie}\ \emph {et~al.}(2014)\citenamefont {Xie},
  \citenamefont {Wang}, \citenamefont {Fu}, \citenamefont {Liu}, \citenamefont
  {Snyder}, \citenamefont {Zhao},\ and\ \citenamefont
  {Zhu}}]{xie2014intrinsic}%
  \BibitemOpen
  \bibfield  {author} {\bibinfo {author} {\bibfnamefont {H.}~\bibnamefont
  {Xie}}, \bibinfo {author} {\bibfnamefont {H.}~\bibnamefont {Wang}}, \bibinfo
  {author} {\bibfnamefont {C.}~\bibnamefont {Fu}}, \bibinfo {author}
  {\bibfnamefont {Y.}~\bibnamefont {Liu}}, \bibinfo {author} {\bibfnamefont
  {G.~J.}\ \bibnamefont {Snyder}}, \bibinfo {author} {\bibfnamefont
  {X.}~\bibnamefont {Zhao}}, \ and\ \bibinfo {author} {\bibfnamefont
  {T.}~\bibnamefont {Zhu}},\ }\href@noop {} {\bibfield  {journal} {\bibinfo
  {journal} {Scientific reports}\ }\textbf {\bibinfo {volume} {4}},\ \bibinfo
  {pages} {6888} (\bibinfo {year} {2014})}\BibitemShut {NoStop}%
\bibitem [{\citenamefont {Kutorasinski}\ \emph {et~al.}(2014)\citenamefont
  {Kutorasinski}, \citenamefont {Tobola},\ and\ \citenamefont
  {Kaprzyk}}]{kutorasinski2014application}%
  \BibitemOpen
  \bibfield  {author} {\bibinfo {author} {\bibfnamefont {K.}~\bibnamefont
  {Kutorasinski}}, \bibinfo {author} {\bibfnamefont {J.}~\bibnamefont
  {Tobola}}, \ and\ \bibinfo {author} {\bibfnamefont {S.}~\bibnamefont
  {Kaprzyk}},\ }\href@noop {} {\bibfield  {journal} {\bibinfo  {journal}
  {physica status solidi (a)}\ }\textbf {\bibinfo {volume} {211}},\ \bibinfo
  {pages} {1229} (\bibinfo {year} {2014})}\BibitemShut {NoStop}%
\bibitem [{\citenamefont {Chaput}\ \emph {et~al.}(2006)\citenamefont {Chaput},
  \citenamefont {Tobola}, \citenamefont {P{\'e}cheur},\ and\ \citenamefont
  {Scherrer}}]{chaput2006electronic}%
  \BibitemOpen
  \bibfield  {author} {\bibinfo {author} {\bibfnamefont {L.}~\bibnamefont
  {Chaput}}, \bibinfo {author} {\bibfnamefont {J.}~\bibnamefont {Tobola}},
  \bibinfo {author} {\bibfnamefont {P.}~\bibnamefont {P{\'e}cheur}}, \ and\
  \bibinfo {author} {\bibfnamefont {H.}~\bibnamefont {Scherrer}},\ }\href@noop
  {} {\bibfield  {journal} {\bibinfo  {journal} {Physical Review B}\ }\textbf
  {\bibinfo {volume} {73}},\ \bibinfo {pages} {045121} (\bibinfo {year}
  {2006})}\BibitemShut {NoStop}%
\bibitem [{\citenamefont {Xie}\ \emph {et~al.}(2013)\citenamefont {Xie},
  \citenamefont {Wang}, \citenamefont {Pei}, \citenamefont {Fu}, \citenamefont
  {Liu}, \citenamefont {Snyder}, \citenamefont {Zhao},\ and\ \citenamefont
  {Zhu}}]{xie2013beneficial}%
  \BibitemOpen
  \bibfield  {author} {\bibinfo {author} {\bibfnamefont {H.}~\bibnamefont
  {Xie}}, \bibinfo {author} {\bibfnamefont {H.}~\bibnamefont {Wang}}, \bibinfo
  {author} {\bibfnamefont {Y.}~\bibnamefont {Pei}}, \bibinfo {author}
  {\bibfnamefont {C.}~\bibnamefont {Fu}}, \bibinfo {author} {\bibfnamefont
  {X.}~\bibnamefont {Liu}}, \bibinfo {author} {\bibfnamefont {G.~J.}\
  \bibnamefont {Snyder}}, \bibinfo {author} {\bibfnamefont {X.}~\bibnamefont
  {Zhao}}, \ and\ \bibinfo {author} {\bibfnamefont {T.}~\bibnamefont {Zhu}},\
  }\href@noop {} {\bibfield  {journal} {\bibinfo  {journal} {Advanced
  Functional Materials}\ }\textbf {\bibinfo {volume} {23}},\ \bibinfo {pages}
  {5123} (\bibinfo {year} {2013})}\BibitemShut {NoStop}%
\bibitem [{\citenamefont {Lee}\ and\ \citenamefont {Chao}(2010)}]{lee2010high}%
  \BibitemOpen
  \bibfield  {author} {\bibinfo {author} {\bibfnamefont {P.-J.}\ \bibnamefont
  {Lee}}\ and\ \bibinfo {author} {\bibfnamefont {L.-S.}\ \bibnamefont {Chao}},\
  }\href@noop {} {\bibfield  {journal} {\bibinfo  {journal} {Journal of Alloys
  and Compounds}\ }\textbf {\bibinfo {volume} {504}},\ \bibinfo {pages} {192}
  (\bibinfo {year} {2010})}\BibitemShut {NoStop}%
\bibitem [{\citenamefont {Simonson}\ \emph {et~al.}(2011)\citenamefont
  {Simonson}, \citenamefont {Wu}, \citenamefont {Xie}, \citenamefont {Tritt},\
  and\ \citenamefont {Poon}}]{simonson2011introduction}%
  \BibitemOpen
  \bibfield  {author} {\bibinfo {author} {\bibfnamefont {J.}~\bibnamefont
  {Simonson}}, \bibinfo {author} {\bibfnamefont {D.}~\bibnamefont {Wu}},
  \bibinfo {author} {\bibfnamefont {W.}~\bibnamefont {Xie}}, \bibinfo {author}
  {\bibfnamefont {T.}~\bibnamefont {Tritt}}, \ and\ \bibinfo {author}
  {\bibfnamefont {S.}~\bibnamefont {Poon}},\ }\href@noop {} {\bibfield
  {journal} {\bibinfo  {journal} {Physical Review B}\ }\textbf {\bibinfo
  {volume} {83}},\ \bibinfo {pages} {235211} (\bibinfo {year}
  {2011})}\BibitemShut {NoStop}%
\bibitem [{\citenamefont {Xie}\ \emph {et~al.}(2012)\citenamefont {Xie},
  \citenamefont {Weidenkaff}, \citenamefont {Tang}, \citenamefont {Zhang},
  \citenamefont {Poon},\ and\ \citenamefont {Tritt}}]{xie2012recent}%
  \BibitemOpen
  \bibfield  {author} {\bibinfo {author} {\bibfnamefont {W.}~\bibnamefont
  {Xie}}, \bibinfo {author} {\bibfnamefont {A.}~\bibnamefont {Weidenkaff}},
  \bibinfo {author} {\bibfnamefont {X.}~\bibnamefont {Tang}}, \bibinfo {author}
  {\bibfnamefont {Q.}~\bibnamefont {Zhang}}, \bibinfo {author} {\bibfnamefont
  {J.}~\bibnamefont {Poon}}, \ and\ \bibinfo {author} {\bibfnamefont
  {T.}~\bibnamefont {Tritt}},\ }\href@noop {} {\bibfield  {journal} {\bibinfo
  {journal} {Nanomaterials}\ }\textbf {\bibinfo {volume} {2}},\ \bibinfo
  {pages} {379} (\bibinfo {year} {2012})}\BibitemShut {NoStop}%
\bibitem [{\citenamefont {Xie}\ \emph {et~al.}(2008)\citenamefont {Xie},
  \citenamefont {Jin},\ and\ \citenamefont {Tang}}]{xie2008preparation}%
  \BibitemOpen
  \bibfield  {author} {\bibinfo {author} {\bibfnamefont {W.}~\bibnamefont
  {Xie}}, \bibinfo {author} {\bibfnamefont {Q.}~\bibnamefont {Jin}}, \ and\
  \bibinfo {author} {\bibfnamefont {X.}~\bibnamefont {Tang}},\ }\href@noop {}
  {\bibfield  {journal} {\bibinfo  {journal} {Journal of Applied Physics}\
  }\textbf {\bibinfo {volume} {103}},\ \bibinfo {pages} {043711} (\bibinfo
  {year} {2008})}\BibitemShut {NoStop}%
\bibitem [{\citenamefont {Sekimoto}\ \emph {et~al.}(2007)\citenamefont
  {Sekimoto}, \citenamefont {Kurosaki}, \citenamefont {Muta},\ and\
  \citenamefont {Yamanaka}}]{sekimoto2007high}%
  \BibitemOpen
  \bibfield  {author} {\bibinfo {author} {\bibfnamefont {T.}~\bibnamefont
  {Sekimoto}}, \bibinfo {author} {\bibfnamefont {K.}~\bibnamefont {Kurosaki}},
  \bibinfo {author} {\bibfnamefont {H.}~\bibnamefont {Muta}}, \ and\ \bibinfo
  {author} {\bibfnamefont {S.}~\bibnamefont {Yamanaka}},\ }\href@noop {}
  {\bibfield  {journal} {\bibinfo  {journal} {Japanese Journal of Applied
  Physics}\ }\textbf {\bibinfo {volume} {46}},\ \bibinfo {pages} {L673}
  (\bibinfo {year} {2007})}\BibitemShut {NoStop}%
\bibitem [{\citenamefont {Yang}\ \emph {et~al.}(2004)\citenamefont {Yang},
  \citenamefont {Meisner},\ and\ \citenamefont {Chen}}]{yang2004strain}%
  \BibitemOpen
  \bibfield  {author} {\bibinfo {author} {\bibfnamefont {J.}~\bibnamefont
  {Yang}}, \bibinfo {author} {\bibfnamefont {G.}~\bibnamefont {Meisner}}, \
  and\ \bibinfo {author} {\bibfnamefont {L.}~\bibnamefont {Chen}},\ }\href@noop
  {} {\bibfield  {journal} {\bibinfo  {journal} {Applied physics letters}\
  }\textbf {\bibinfo {volume} {85}},\ \bibinfo {pages} {1140} (\bibinfo {year}
  {2004})}\BibitemShut {NoStop}%
\bibitem [{\citenamefont {Sakurada}\ and\ \citenamefont
  {Shutoh}(2005)}]{sakurada2005effect}%
  \BibitemOpen
  \bibfield  {author} {\bibinfo {author} {\bibfnamefont {S.}~\bibnamefont
  {Sakurada}}\ and\ \bibinfo {author} {\bibfnamefont {N.}~\bibnamefont
  {Shutoh}},\ }\href@noop {} {\bibfield  {journal} {\bibinfo  {journal}
  {Applied Physics Letters}\ }\textbf {\bibinfo {volume} {86}},\ \bibinfo
  {pages} {082105} (\bibinfo {year} {2005})}\BibitemShut {NoStop}%
\bibitem [{\citenamefont {Yan}\ \emph {et~al.}(2013)\citenamefont {Yan},
  \citenamefont {Liu}, \citenamefont {Chen}, \citenamefont {Wang},
  \citenamefont {Zhang}, \citenamefont {Chen},\ and\ \citenamefont
  {Ren}}]{yan2013thermoelectric}%
  \BibitemOpen
  \bibfield  {author} {\bibinfo {author} {\bibfnamefont {X.}~\bibnamefont
  {Yan}}, \bibinfo {author} {\bibfnamefont {W.}~\bibnamefont {Liu}}, \bibinfo
  {author} {\bibfnamefont {S.}~\bibnamefont {Chen}}, \bibinfo {author}
  {\bibfnamefont {H.}~\bibnamefont {Wang}}, \bibinfo {author} {\bibfnamefont
  {Q.}~\bibnamefont {Zhang}}, \bibinfo {author} {\bibfnamefont
  {G.}~\bibnamefont {Chen}}, \ and\ \bibinfo {author} {\bibfnamefont
  {Z.}~\bibnamefont {Ren}},\ }\href@noop {} {\bibfield  {journal} {\bibinfo
  {journal} {Advanced Energy Materials}\ }\textbf {\bibinfo {volume} {3}},\
  \bibinfo {pages} {1195} (\bibinfo {year} {2013})}\BibitemShut {NoStop}%
\bibitem [{\citenamefont {Katayama}\ \emph {et~al.}(2003)\citenamefont
  {Katayama}, \citenamefont {Kim}, \citenamefont {Kimura},\ and\ \citenamefont
  {Mishima}}]{katayama2003effects}%
  \BibitemOpen
  \bibfield  {author} {\bibinfo {author} {\bibfnamefont {T.}~\bibnamefont
  {Katayama}}, \bibinfo {author} {\bibfnamefont {S.~W.}\ \bibnamefont {Kim}},
  \bibinfo {author} {\bibfnamefont {Y.}~\bibnamefont {Kimura}}, \ and\ \bibinfo
  {author} {\bibfnamefont {Y.}~\bibnamefont {Mishima}},\ }\href@noop {}
  {\bibfield  {journal} {\bibinfo  {journal} {Journal of electronic materials}\
  }\textbf {\bibinfo {volume} {32}},\ \bibinfo {pages} {1160} (\bibinfo {year}
  {2003})}\BibitemShut {NoStop}%
\bibitem [{\citenamefont {Gelbstein}\ \emph {et~al.}(2011)\citenamefont
  {Gelbstein}, \citenamefont {Tal}, \citenamefont {Yarmek}, \citenamefont
  {Rosenberg}, \citenamefont {Dariel}, \citenamefont {Ouardi}, \citenamefont
  {Balke}, \citenamefont {Felser},\ and\ \citenamefont
  {K{\"o}hne}}]{gelbstein2011thermoelectric}%
  \BibitemOpen
  \bibfield  {author} {\bibinfo {author} {\bibfnamefont {Y.}~\bibnamefont
  {Gelbstein}}, \bibinfo {author} {\bibfnamefont {N.}~\bibnamefont {Tal}},
  \bibinfo {author} {\bibfnamefont {A.}~\bibnamefont {Yarmek}}, \bibinfo
  {author} {\bibfnamefont {Y.}~\bibnamefont {Rosenberg}}, \bibinfo {author}
  {\bibfnamefont {M.~P.}\ \bibnamefont {Dariel}}, \bibinfo {author}
  {\bibfnamefont {S.}~\bibnamefont {Ouardi}}, \bibinfo {author} {\bibfnamefont
  {B.}~\bibnamefont {Balke}}, \bibinfo {author} {\bibfnamefont
  {C.}~\bibnamefont {Felser}}, \ and\ \bibinfo {author} {\bibfnamefont
  {M.}~\bibnamefont {K{\"o}hne}},\ }\href@noop {} {\bibfield  {journal}
  {\bibinfo  {journal} {Journal of Materials Research}\ }\textbf {\bibinfo
  {volume} {26}},\ \bibinfo {pages} {1919} (\bibinfo {year}
  {2011})}\BibitemShut {NoStop}%
\bibitem [{\citenamefont {Birkel}\ \emph {et~al.}(2012)\citenamefont {Birkel},
  \citenamefont {Zeier}, \citenamefont {Douglas}, \citenamefont {Lettiere},
  \citenamefont {Mills}, \citenamefont {Seward}, \citenamefont {Birkel},
  \citenamefont {Snedaker}, \citenamefont {Zhang}, \citenamefont {Snyder} \emph
  {et~al.}}]{birkel2012rapid}%
  \BibitemOpen
  \bibfield  {author} {\bibinfo {author} {\bibfnamefont {C.~S.}\ \bibnamefont
  {Birkel}}, \bibinfo {author} {\bibfnamefont {W.~G.}\ \bibnamefont {Zeier}},
  \bibinfo {author} {\bibfnamefont {J.~E.}\ \bibnamefont {Douglas}}, \bibinfo
  {author} {\bibfnamefont {B.~R.}\ \bibnamefont {Lettiere}}, \bibinfo {author}
  {\bibfnamefont {C.~E.}\ \bibnamefont {Mills}}, \bibinfo {author}
  {\bibfnamefont {G.}~\bibnamefont {Seward}}, \bibinfo {author} {\bibfnamefont
  {A.}~\bibnamefont {Birkel}}, \bibinfo {author} {\bibfnamefont {M.~L.}\
  \bibnamefont {Snedaker}}, \bibinfo {author} {\bibfnamefont {Y.}~\bibnamefont
  {Zhang}}, \bibinfo {author} {\bibfnamefont {G.~J.}\ \bibnamefont {Snyder}},
  \emph {et~al.},\ }\href@noop {} {\bibfield  {journal} {\bibinfo  {journal}
  {Chemistry of Materials}\ }\textbf {\bibinfo {volume} {24}},\ \bibinfo
  {pages} {2558} (\bibinfo {year} {2012})}\BibitemShut {NoStop}%
\bibitem [{\citenamefont {Douglas}\ \emph {et~al.}(2012)\citenamefont
  {Douglas}, \citenamefont {Birkel}, \citenamefont {Miao}, \citenamefont
  {Torbet}, \citenamefont {Stucky}, \citenamefont {Pollock},\ and\
  \citenamefont {Seshadri}}]{douglas2012enhanced}%
  \BibitemOpen
  \bibfield  {author} {\bibinfo {author} {\bibfnamefont {J.~E.}\ \bibnamefont
  {Douglas}}, \bibinfo {author} {\bibfnamefont {C.~S.}\ \bibnamefont {Birkel}},
  \bibinfo {author} {\bibfnamefont {M.-S.}\ \bibnamefont {Miao}}, \bibinfo
  {author} {\bibfnamefont {C.~J.}\ \bibnamefont {Torbet}}, \bibinfo {author}
  {\bibfnamefont {G.~D.}\ \bibnamefont {Stucky}}, \bibinfo {author}
  {\bibfnamefont {T.~M.}\ \bibnamefont {Pollock}}, \ and\ \bibinfo {author}
  {\bibfnamefont {R.}~\bibnamefont {Seshadri}},\ }\href@noop {} {\bibfield
  {journal} {\bibinfo  {journal} {Applied Physics Letters}\ }\textbf {\bibinfo
  {volume} {101}},\ \bibinfo {pages} {183902} (\bibinfo {year}
  {2012})}\BibitemShut {NoStop}%
\bibitem [{\citenamefont {G{\"u}rth}\ \emph {et~al.}(2016)\citenamefont
  {G{\"u}rth}, \citenamefont {Rogl}, \citenamefont {Romaka}, \citenamefont
  {Grytsiv}, \citenamefont {Bauer},\ and\ \citenamefont
  {Rogl}}]{gurth2016thermoelectric}%
  \BibitemOpen
  \bibfield  {author} {\bibinfo {author} {\bibfnamefont {M.}~\bibnamefont
  {G{\"u}rth}}, \bibinfo {author} {\bibfnamefont {G.}~\bibnamefont {Rogl}},
  \bibinfo {author} {\bibfnamefont {V.}~\bibnamefont {Romaka}}, \bibinfo
  {author} {\bibfnamefont {A.}~\bibnamefont {Grytsiv}}, \bibinfo {author}
  {\bibfnamefont {E.}~\bibnamefont {Bauer}}, \ and\ \bibinfo {author}
  {\bibfnamefont {P.}~\bibnamefont {Rogl}},\ }\href@noop {} {\bibfield
  {journal} {\bibinfo  {journal} {Acta Materialia}\ }\textbf {\bibinfo {volume}
  {104}},\ \bibinfo {pages} {210} (\bibinfo {year} {2016})}\BibitemShut
  {NoStop}%
\bibitem [{\citenamefont {Fu}\ \emph {et~al.}(2015)\citenamefont {Fu},
  \citenamefont {Bai}, \citenamefont {Liu}, \citenamefont {Tang}, \citenamefont
  {Chen}, \citenamefont {Zhao},\ and\ \citenamefont {Zhu}}]{fu2015realizing}%
  \BibitemOpen
  \bibfield  {author} {\bibinfo {author} {\bibfnamefont {C.}~\bibnamefont
  {Fu}}, \bibinfo {author} {\bibfnamefont {S.}~\bibnamefont {Bai}}, \bibinfo
  {author} {\bibfnamefont {Y.}~\bibnamefont {Liu}}, \bibinfo {author}
  {\bibfnamefont {Y.}~\bibnamefont {Tang}}, \bibinfo {author} {\bibfnamefont
  {L.}~\bibnamefont {Chen}}, \bibinfo {author} {\bibfnamefont {X.}~\bibnamefont
  {Zhao}}, \ and\ \bibinfo {author} {\bibfnamefont {T.}~\bibnamefont {Zhu}},\
  }\href@noop {} {\bibfield  {journal} {\bibinfo  {journal} {Nature
  communications}\ }\textbf {\bibinfo {volume} {6}},\ \bibinfo {pages} {1}
  (\bibinfo {year} {2015})}\BibitemShut {NoStop}%
\bibitem [{\citenamefont {Kresse}\ and\ \citenamefont
  {Joubert}(1999)}]{kresse1999ultrasoft}%
  \BibitemOpen
  \bibfield  {author} {\bibinfo {author} {\bibfnamefont {G.}~\bibnamefont
  {Kresse}}\ and\ \bibinfo {author} {\bibfnamefont {D.}~\bibnamefont
  {Joubert}},\ }\href@noop {} {\bibfield  {journal} {\bibinfo  {journal} {Phys.
  Rev. B}\ }\textbf {\bibinfo {volume} {59}},\ \bibinfo {pages} {1758}
  (\bibinfo {year} {1999})}\BibitemShut {NoStop}%
\bibitem [{\citenamefont {Kresse}\ and\ \citenamefont
  {Furthm{\"u}ller}(1996)}]{kresse1996efficiency}%
  \BibitemOpen
  \bibfield  {author} {\bibinfo {author} {\bibfnamefont {G.}~\bibnamefont
  {Kresse}}\ and\ \bibinfo {author} {\bibfnamefont {J.}~\bibnamefont
  {Furthm{\"u}ller}},\ }\href@noop {} {\bibfield  {journal} {\bibinfo
  {journal} {Comput. Mater. Sci.}\ }\textbf {\bibinfo {volume} {6}},\ \bibinfo
  {pages} {15} (\bibinfo {year} {1996})}\BibitemShut {NoStop}%
\bibitem [{\citenamefont {Perdew}\ \emph {et~al.}(1996)\citenamefont {Perdew},
  \citenamefont {Burke},\ and\ \citenamefont
  {Ernzerhof}}]{perdew1996generalized}%
  \BibitemOpen
  \bibfield  {author} {\bibinfo {author} {\bibfnamefont {J.~P.}\ \bibnamefont
  {Perdew}}, \bibinfo {author} {\bibfnamefont {K.}~\bibnamefont {Burke}}, \
  and\ \bibinfo {author} {\bibfnamefont {M.}~\bibnamefont {Ernzerhof}},\
  }\href@noop {} {\bibfield  {journal} {\bibinfo  {journal} {Phys. Rev. Lett}\
  }\textbf {\bibinfo {volume} {77}},\ \bibinfo {pages} {3865} (\bibinfo {year}
  {1996})}\BibitemShut {NoStop}%
\bibitem [{\citenamefont {Monkhorst}\ and\ \citenamefont
  {Pack}(1976)}]{monkhorst1976special}%
  \BibitemOpen
  \bibfield  {author} {\bibinfo {author} {\bibfnamefont {H.~J.}\ \bibnamefont
  {Monkhorst}}\ and\ \bibinfo {author} {\bibfnamefont {J.~D.}\ \bibnamefont
  {Pack}},\ }\href@noop {} {\bibfield  {journal} {\bibinfo  {journal} {Phys.
  Rev. B}\ }\textbf {\bibinfo {volume} {13}},\ \bibinfo {pages} {5188}
  (\bibinfo {year} {1976})}\BibitemShut {NoStop}%
\bibitem [{\citenamefont {Bl{\"o}chl}\ \emph {et~al.}(1994)\citenamefont
  {Bl{\"o}chl}, \citenamefont {Jepsen},\ and\ \citenamefont
  {Andersen}}]{blochl1994improved}%
  \BibitemOpen
  \bibfield  {author} {\bibinfo {author} {\bibfnamefont {P.~E.}\ \bibnamefont
  {Bl{\"o}chl}}, \bibinfo {author} {\bibfnamefont {O.}~\bibnamefont {Jepsen}},
  \ and\ \bibinfo {author} {\bibfnamefont {O.~K.}\ \bibnamefont {Andersen}},\
  }\href@noop {} {\bibfield  {journal} {\bibinfo  {journal} {Physical Review
  B}\ }\textbf {\bibinfo {volume} {49}},\ \bibinfo {pages} {16223} (\bibinfo
  {year} {1994})}\BibitemShut {NoStop}%
\bibitem [{\citenamefont {Choudhary}\ and\ \citenamefont
  {Ravindran}(2020)}]{choudhary2020thermal}%
  \BibitemOpen
  \bibfield  {author} {\bibinfo {author} {\bibfnamefont {M.~K.}\ \bibnamefont
  {Choudhary}}\ and\ \bibinfo {author} {\bibfnamefont {P.}~\bibnamefont
  {Ravindran}},\ }\href@noop {} {\bibfield  {journal} {\bibinfo  {journal}
  {Sustainable Energy \& Fuels}\ }\textbf {\bibinfo {volume} {4}},\ \bibinfo
  {pages} {895} (\bibinfo {year} {2020})}\BibitemShut {NoStop}%
\bibitem [{\citenamefont {Blaha}\ \emph {et~al.}(2001)\citenamefont {Blaha},
  \citenamefont {Schwarz}, \citenamefont {Madsen}, \citenamefont {Kvasnicka},\
  and\ \citenamefont {Luitz}}]{blaha2001wien2k}%
  \BibitemOpen
  \bibfield  {author} {\bibinfo {author} {\bibfnamefont {P.}~\bibnamefont
  {Blaha}}, \bibinfo {author} {\bibfnamefont {K.}~\bibnamefont {Schwarz}},
  \bibinfo {author} {\bibfnamefont {G.}~\bibnamefont {Madsen}}, \bibinfo
  {author} {\bibfnamefont {D.}~\bibnamefont {Kvasnicka}}, \ and\ \bibinfo
  {author} {\bibfnamefont {J.}~\bibnamefont {Luitz}},\ }\href@noop {}
  {\bibfield  {journal} {\bibinfo  {journal} {Austria ISBN}\ ,\ \bibinfo
  {pages} {3}} (\bibinfo {year} {2001})}\BibitemShut {NoStop}%
\bibitem [{\citenamefont {Schwarz}\ and\ \citenamefont
  {Blaha}(2003)}]{schwarz2003solid}%
  \BibitemOpen
  \bibfield  {author} {\bibinfo {author} {\bibfnamefont {K.}~\bibnamefont
  {Schwarz}}\ and\ \bibinfo {author} {\bibfnamefont {P.}~\bibnamefont
  {Blaha}},\ }\href@noop {} {\bibfield  {journal} {\bibinfo  {journal}
  {Computational Materials Science}\ }\textbf {\bibinfo {volume} {28}},\
  \bibinfo {pages} {259} (\bibinfo {year} {2003})}\BibitemShut {NoStop}%
\bibitem [{\citenamefont {Madsen}\ and\ \citenamefont
  {Singh}(2006)}]{madsen2006boltZTrap}%
  \BibitemOpen
  \bibfield  {author} {\bibinfo {author} {\bibfnamefont {G.~K.}\ \bibnamefont
  {Madsen}}\ and\ \bibinfo {author} {\bibfnamefont {D.~J.}\ \bibnamefont
  {Singh}},\ }\href@noop {} {\bibfield  {journal} {\bibinfo  {journal}
  {Computer Physics Communications}\ }\textbf {\bibinfo {volume} {175}},\
  \bibinfo {pages} {67} (\bibinfo {year} {2006})}\BibitemShut {NoStop}%
\bibitem [{\citenamefont {Togo}\ and\ \citenamefont
  {Tanaka}(2015)}]{togo2015first}%
  \BibitemOpen
  \bibfield  {author} {\bibinfo {author} {\bibfnamefont {A.}~\bibnamefont
  {Togo}}\ and\ \bibinfo {author} {\bibfnamefont {I.}~\bibnamefont {Tanaka}},\
  }\href@noop {} {\bibfield  {journal} {\bibinfo  {journal} {Scripta
  Materialia}\ }\textbf {\bibinfo {volume} {108}},\ \bibinfo {pages} {1}
  (\bibinfo {year} {2015})}\BibitemShut {NoStop}%
\bibitem [{\citenamefont {Togo}\ \emph {et~al.}(2015)\citenamefont {Togo},
  \citenamefont {Chaput},\ and\ \citenamefont
  {Tanaka}}]{togo2015distributions}%
  \BibitemOpen
  \bibfield  {author} {\bibinfo {author} {\bibfnamefont {A.}~\bibnamefont
  {Togo}}, \bibinfo {author} {\bibfnamefont {L.}~\bibnamefont {Chaput}}, \ and\
  \bibinfo {author} {\bibfnamefont {I.}~\bibnamefont {Tanaka}},\ }\href@noop {}
  {\bibfield  {journal} {\bibinfo  {journal} {Physical Review B}\ }\textbf
  {\bibinfo {volume} {91}},\ \bibinfo {pages} {094306} (\bibinfo {year}
  {2015})}\BibitemShut {NoStop}%
\bibitem [{\citenamefont {K{\"u}bler}\ \emph {et~al.}(1983)\citenamefont
  {K{\"u}bler}, \citenamefont {William},\ and\ \citenamefont
  {Sommers}}]{kubler1983formation}%
  \BibitemOpen
  \bibfield  {author} {\bibinfo {author} {\bibfnamefont {J.}~\bibnamefont
  {K{\"u}bler}}, \bibinfo {author} {\bibfnamefont {A.}~\bibnamefont {William}},
  \ and\ \bibinfo {author} {\bibfnamefont {C.}~\bibnamefont {Sommers}},\
  }\href@noop {} {\bibfield  {journal} {\bibinfo  {journal} {Physical Review
  B}\ }\textbf {\bibinfo {volume} {28}},\ \bibinfo {pages} {1745} (\bibinfo
  {year} {1983})}\BibitemShut {NoStop}%
\bibitem [{\citenamefont {Pierre}\ \emph {et~al.}(1997)\citenamefont {Pierre},
  \citenamefont {Skolozdra}, \citenamefont {Tobola}, \citenamefont {Kaprzyk},
  \citenamefont {Hordequin}, \citenamefont {Kouacou}, \citenamefont {Karla},
  \citenamefont {Currat},\ and\ \citenamefont
  {Lelievre-Berna}}]{pierre1997properties}%
  \BibitemOpen
  \bibfield  {author} {\bibinfo {author} {\bibfnamefont {J.}~\bibnamefont
  {Pierre}}, \bibinfo {author} {\bibfnamefont {R.}~\bibnamefont {Skolozdra}},
  \bibinfo {author} {\bibfnamefont {J.}~\bibnamefont {Tobola}}, \bibinfo
  {author} {\bibfnamefont {S.}~\bibnamefont {Kaprzyk}}, \bibinfo {author}
  {\bibfnamefont {C.}~\bibnamefont {Hordequin}}, \bibinfo {author}
  {\bibfnamefont {M.}~\bibnamefont {Kouacou}}, \bibinfo {author} {\bibfnamefont
  {I.}~\bibnamefont {Karla}}, \bibinfo {author} {\bibfnamefont
  {R.}~\bibnamefont {Currat}}, \ and\ \bibinfo {author} {\bibfnamefont
  {E.}~\bibnamefont {Lelievre-Berna}},\ }\href@noop {} {\bibfield  {journal}
  {\bibinfo  {journal} {Journal of alloys and compounds}\ }\textbf {\bibinfo
  {volume} {262}},\ \bibinfo {pages} {101} (\bibinfo {year}
  {1997})}\BibitemShut {NoStop}%
\bibitem [{\citenamefont {Tobo~{\ l} a}\ and\ \citenamefont
  {Pierre}(2000)}]{tobola2000electronic}%
  \BibitemOpen
  \bibfield  {author} {\bibinfo {author} {\bibfnamefont {J.}~\bibnamefont
  {Tobo~{\ l} a}}\ and\ \bibinfo {author} {\bibfnamefont {J.}~\bibnamefont
  {Pierre}},\ }\href@noop {} {\bibfield  {journal} {\bibinfo  {journal}
  {Journal of alloys and compounds}\ }\textbf {\bibinfo {volume} {296}},\
  \bibinfo {pages} {243} (\bibinfo {year} {2000})}\BibitemShut {NoStop}%
\bibitem [{\citenamefont {Offernes}\ \emph {et~al.}(2007)\citenamefont
  {Offernes}, \citenamefont {Ravindran},\ and\ \citenamefont
  {Kjekshus}}]{offernes2007electronic}%
  \BibitemOpen
  \bibfield  {author} {\bibinfo {author} {\bibfnamefont {L.}~\bibnamefont
  {Offernes}}, \bibinfo {author} {\bibfnamefont {P.}~\bibnamefont {Ravindran}},
  \ and\ \bibinfo {author} {\bibfnamefont {A.}~\bibnamefont {Kjekshus}},\
  }\href@noop {} {\bibfield  {journal} {\bibinfo  {journal} {Journal of alloys
  and compounds}\ }\textbf {\bibinfo {volume} {439}},\ \bibinfo {pages} {37}
  (\bibinfo {year} {2007})}\BibitemShut {NoStop}%
\bibitem [{\citenamefont {Graf}\ \emph {et~al.}(2011)\citenamefont {Graf},
  \citenamefont {Felser},\ and\ \citenamefont {Parkin}}]{graf2011simple}%
  \BibitemOpen
  \bibfield  {author} {\bibinfo {author} {\bibfnamefont {T.}~\bibnamefont
  {Graf}}, \bibinfo {author} {\bibfnamefont {C.}~\bibnamefont {Felser}}, \ and\
  \bibinfo {author} {\bibfnamefont {S.~S.}\ \bibnamefont {Parkin}},\
  }\href@noop {} {\bibfield  {journal} {\bibinfo  {journal} {Progress in solid
  state chemistry}\ }\textbf {\bibinfo {volume} {39}},\ \bibinfo {pages} {1}
  (\bibinfo {year} {2011})}\BibitemShut {NoStop}%
\bibitem [{\citenamefont {Krenke}\ \emph {et~al.}(2005)\citenamefont {Krenke},
  \citenamefont {Duman}, \citenamefont {Acet}, \citenamefont {Wassermann},
  \citenamefont {Moya}, \citenamefont {Ma{\~n}osa},\ and\ \citenamefont
  {Planes}}]{krenke2005inverse}%
  \BibitemOpen
  \bibfield  {author} {\bibinfo {author} {\bibfnamefont {T.}~\bibnamefont
  {Krenke}}, \bibinfo {author} {\bibfnamefont {E.}~\bibnamefont {Duman}},
  \bibinfo {author} {\bibfnamefont {M.}~\bibnamefont {Acet}}, \bibinfo {author}
  {\bibfnamefont {E.~F.}\ \bibnamefont {Wassermann}}, \bibinfo {author}
  {\bibfnamefont {X.}~\bibnamefont {Moya}}, \bibinfo {author} {\bibfnamefont
  {L.}~\bibnamefont {Ma{\~n}osa}}, \ and\ \bibinfo {author} {\bibfnamefont
  {A.}~\bibnamefont {Planes}},\ }\href@noop {} {\bibfield  {journal} {\bibinfo
  {journal} {Nature materials}\ }\textbf {\bibinfo {volume} {4}},\ \bibinfo
  {pages} {450} (\bibinfo {year} {2005})}\BibitemShut {NoStop}%
\bibitem [{\citenamefont {Levin}\ \emph {et~al.}(2017)\citenamefont {Levin},
  \citenamefont {Bocarsly}, \citenamefont {Wyckoff}, \citenamefont {Pollock},\
  and\ \citenamefont {Seshadri}}]{levin2017tuning}%
  \BibitemOpen
  \bibfield  {author} {\bibinfo {author} {\bibfnamefont {E.~E.}\ \bibnamefont
  {Levin}}, \bibinfo {author} {\bibfnamefont {J.~D.}\ \bibnamefont {Bocarsly}},
  \bibinfo {author} {\bibfnamefont {K.~E.}\ \bibnamefont {Wyckoff}}, \bibinfo
  {author} {\bibfnamefont {T.~M.}\ \bibnamefont {Pollock}}, \ and\ \bibinfo
  {author} {\bibfnamefont {R.}~\bibnamefont {Seshadri}},\ }\href@noop {}
  {\bibfield  {journal} {\bibinfo  {journal} {Physical Review Materials}\
  }\textbf {\bibinfo {volume} {1}},\ \bibinfo {pages} {075003} (\bibinfo {year}
  {2017})}\BibitemShut {NoStop}%
\bibitem [{\citenamefont {Liu}\ \emph {et~al.}(2012{\natexlab{b}})\citenamefont
  {Liu}, \citenamefont {Gottschall}, \citenamefont {Skokov}, \citenamefont
  {Moore},\ and\ \citenamefont {Gutfleisch}}]{liu2012giant}%
  \BibitemOpen
  \bibfield  {author} {\bibinfo {author} {\bibfnamefont {J.}~\bibnamefont
  {Liu}}, \bibinfo {author} {\bibfnamefont {T.}~\bibnamefont {Gottschall}},
  \bibinfo {author} {\bibfnamefont {K.~P.}\ \bibnamefont {Skokov}}, \bibinfo
  {author} {\bibfnamefont {J.~D.}\ \bibnamefont {Moore}}, \ and\ \bibinfo
  {author} {\bibfnamefont {O.}~\bibnamefont {Gutfleisch}},\ }\href@noop {}
  {\bibfield  {journal} {\bibinfo  {journal} {Nature materials}\ }\textbf
  {\bibinfo {volume} {11}},\ \bibinfo {pages} {620} (\bibinfo {year}
  {2012}{\natexlab{b}})}\BibitemShut {NoStop}%
\bibitem [{\citenamefont {Buschow}\ and\ \citenamefont
  {Van~Engen}(1981)}]{buschow1981magnetic}%
  \BibitemOpen
  \bibfield  {author} {\bibinfo {author} {\bibfnamefont {K.}~\bibnamefont
  {Buschow}}\ and\ \bibinfo {author} {\bibfnamefont {P.}~\bibnamefont
  {Van~Engen}},\ }\href@noop {} {\bibfield  {journal} {\bibinfo  {journal}
  {Journal of Magnetism and Magnetic Materials}\ }\textbf {\bibinfo {volume}
  {25}},\ \bibinfo {pages} {90} (\bibinfo {year} {1981})}\BibitemShut {NoStop}%
\bibitem [{\citenamefont {Picozzi}\ \emph {et~al.}(2006)\citenamefont
  {Picozzi}, \citenamefont {Continenza},\ and\ \citenamefont
  {Freeman}}]{picozzi2006magneto}%
  \BibitemOpen
  \bibfield  {author} {\bibinfo {author} {\bibfnamefont {S.}~\bibnamefont
  {Picozzi}}, \bibinfo {author} {\bibfnamefont {A.}~\bibnamefont {Continenza}},
  \ and\ \bibinfo {author} {\bibfnamefont {A.~J.}\ \bibnamefont {Freeman}},\
  }\href@noop {} {\bibfield  {journal} {\bibinfo  {journal} {Journal of Physics
  D: Applied Physics}\ }\textbf {\bibinfo {volume} {39}},\ \bibinfo {pages}
  {851} (\bibinfo {year} {2006})}\BibitemShut {NoStop}%
\bibitem [{\citenamefont {Sanvito}\ \emph {et~al.}(2017)\citenamefont
  {Sanvito}, \citenamefont {Oses}, \citenamefont {Xue}, \citenamefont {Tiwari},
  \citenamefont {Zic}, \citenamefont {Archer}, \citenamefont {Tozman},
  \citenamefont {Venkatesan}, \citenamefont {Coey},\ and\ \citenamefont
  {Curtarolo}}]{sanvito2017accelerated}%
  \BibitemOpen
  \bibfield  {author} {\bibinfo {author} {\bibfnamefont {S.}~\bibnamefont
  {Sanvito}}, \bibinfo {author} {\bibfnamefont {C.}~\bibnamefont {Oses}},
  \bibinfo {author} {\bibfnamefont {J.}~\bibnamefont {Xue}}, \bibinfo {author}
  {\bibfnamefont {A.}~\bibnamefont {Tiwari}}, \bibinfo {author} {\bibfnamefont
  {M.}~\bibnamefont {Zic}}, \bibinfo {author} {\bibfnamefont {T.}~\bibnamefont
  {Archer}}, \bibinfo {author} {\bibfnamefont {P.}~\bibnamefont {Tozman}},
  \bibinfo {author} {\bibfnamefont {M.}~\bibnamefont {Venkatesan}}, \bibinfo
  {author} {\bibfnamefont {M.}~\bibnamefont {Coey}}, \ and\ \bibinfo {author}
  {\bibfnamefont {S.}~\bibnamefont {Curtarolo}},\ }\href@noop {} {\bibfield
  {journal} {\bibinfo  {journal} {Science advances}\ }\textbf {\bibinfo
  {volume} {3}},\ \bibinfo {pages} {e1602241} (\bibinfo {year}
  {2017})}\BibitemShut {NoStop}%
\bibitem [{\citenamefont {Kundu}\ \emph {et~al.}(2017)\citenamefont {Kundu},
  \citenamefont {Ghosh}, \citenamefont {Banerjee}, \citenamefont {Ghosh},\ and\
  \citenamefont {Sanyal}}]{kundu2017new}%
  \BibitemOpen
  \bibfield  {author} {\bibinfo {author} {\bibfnamefont {A.}~\bibnamefont
  {Kundu}}, \bibinfo {author} {\bibfnamefont {S.}~\bibnamefont {Ghosh}},
  \bibinfo {author} {\bibfnamefont {R.}~\bibnamefont {Banerjee}}, \bibinfo
  {author} {\bibfnamefont {S.}~\bibnamefont {Ghosh}}, \ and\ \bibinfo {author}
  {\bibfnamefont {B.}~\bibnamefont {Sanyal}},\ }\href@noop {} {\bibfield
  {journal} {\bibinfo  {journal} {Scientific reports}\ }\textbf {\bibinfo
  {volume} {7}},\ \bibinfo {pages} {1} (\bibinfo {year} {2017})}\BibitemShut
  {NoStop}%
\bibitem [{\citenamefont {Blum}\ \emph {et~al.}(2009)\citenamefont {Blum},
  \citenamefont {Jenkins}, \citenamefont {Barth}, \citenamefont {Felser},
  \citenamefont {Wurmehl}, \citenamefont {Friemel}, \citenamefont {Hess},
  \citenamefont {Behr}, \citenamefont {B{\"u}chner}, \citenamefont {Reller}
  \emph {et~al.}}]{blum2009highly}%
  \BibitemOpen
  \bibfield  {author} {\bibinfo {author} {\bibfnamefont {C.}~\bibnamefont
  {Blum}}, \bibinfo {author} {\bibfnamefont {C.}~\bibnamefont {Jenkins}},
  \bibinfo {author} {\bibfnamefont {J.}~\bibnamefont {Barth}}, \bibinfo
  {author} {\bibfnamefont {C.}~\bibnamefont {Felser}}, \bibinfo {author}
  {\bibfnamefont {S.}~\bibnamefont {Wurmehl}}, \bibinfo {author} {\bibfnamefont
  {G.}~\bibnamefont {Friemel}}, \bibinfo {author} {\bibfnamefont
  {C.}~\bibnamefont {Hess}}, \bibinfo {author} {\bibfnamefont {G.}~\bibnamefont
  {Behr}}, \bibinfo {author} {\bibfnamefont {B.}~\bibnamefont {B{\"u}chner}},
  \bibinfo {author} {\bibfnamefont {A.}~\bibnamefont {Reller}},  \emph
  {et~al.},\ }\href@noop {} {\bibfield  {journal} {\bibinfo  {journal} {Applied
  Physics Letters}\ }\textbf {\bibinfo {volume} {95}},\ \bibinfo {pages}
  {161903} (\bibinfo {year} {2009})}\BibitemShut {NoStop}%
\bibitem [{\citenamefont {Wurmehl}\ \emph {et~al.}(2006)\citenamefont
  {Wurmehl}, \citenamefont {Fecher}, \citenamefont {Kandpal}, \citenamefont
  {Ksenofontov}, \citenamefont {Felser},\ and\ \citenamefont
  {Lin}}]{wurmehl2006investigation}%
  \BibitemOpen
  \bibfield  {author} {\bibinfo {author} {\bibfnamefont {S.}~\bibnamefont
  {Wurmehl}}, \bibinfo {author} {\bibfnamefont {G.~H.}\ \bibnamefont {Fecher}},
  \bibinfo {author} {\bibfnamefont {H.~C.}\ \bibnamefont {Kandpal}}, \bibinfo
  {author} {\bibfnamefont {V.}~\bibnamefont {Ksenofontov}}, \bibinfo {author}
  {\bibfnamefont {C.}~\bibnamefont {Felser}}, \ and\ \bibinfo {author}
  {\bibfnamefont {H.-J.}\ \bibnamefont {Lin}},\ }\href@noop {} {\bibfield
  {journal} {\bibinfo  {journal} {Applied physics letters}\ }\textbf {\bibinfo
  {volume} {88}},\ \bibinfo {pages} {032503} (\bibinfo {year}
  {2006})}\BibitemShut {NoStop}%
\bibitem [{\citenamefont {Huang}\ \emph {et~al.}(2018)\citenamefont {Huang},
  \citenamefont {Liu}, \citenamefont {Zheng}, \citenamefont {Guo},
  \citenamefont {Xiong}, \citenamefont {Wang},\ and\ \citenamefont
  {Shi}}]{huang2018dramatically}%
  \BibitemOpen
  \bibfield  {author} {\bibinfo {author} {\bibfnamefont {S.}~\bibnamefont
  {Huang}}, \bibinfo {author} {\bibfnamefont {X.}~\bibnamefont {Liu}}, \bibinfo
  {author} {\bibfnamefont {W.}~\bibnamefont {Zheng}}, \bibinfo {author}
  {\bibfnamefont {J.}~\bibnamefont {Guo}}, \bibinfo {author} {\bibfnamefont
  {R.}~\bibnamefont {Xiong}}, \bibinfo {author} {\bibfnamefont
  {Z.}~\bibnamefont {Wang}}, \ and\ \bibinfo {author} {\bibfnamefont
  {J.}~\bibnamefont {Shi}},\ }\href@noop {} {\bibfield  {journal} {\bibinfo
  {journal} {Journal of Materials Chemistry A}\ }\textbf {\bibinfo {volume}
  {6}},\ \bibinfo {pages} {20069} (\bibinfo {year} {2018})}\BibitemShut
  {NoStop}%
\bibitem [{\citenamefont {Krishnaveni}\ \emph {et~al.}(2016)\citenamefont
  {Krishnaveni}, \citenamefont {Sundareswari}, \citenamefont {Deshmukh},
  \citenamefont {Valluri},\ and\ \citenamefont
  {Roberts}}]{krishnaveni2016band}%
  \BibitemOpen
  \bibfield  {author} {\bibinfo {author} {\bibfnamefont {S.}~\bibnamefont
  {Krishnaveni}}, \bibinfo {author} {\bibfnamefont {M.}~\bibnamefont
  {Sundareswari}}, \bibinfo {author} {\bibfnamefont {P.}~\bibnamefont
  {Deshmukh}}, \bibinfo {author} {\bibfnamefont {S.}~\bibnamefont {Valluri}}, \
  and\ \bibinfo {author} {\bibfnamefont {K.}~\bibnamefont {Roberts}},\
  }\href@noop {} {\bibfield  {journal} {\bibinfo  {journal} {Journal of
  Materials Research}\ }\textbf {\bibinfo {volume} {31}},\ \bibinfo {pages}
  {1306} (\bibinfo {year} {2016})}\BibitemShut {NoStop}%
\bibitem [{\citenamefont {Yang}\ \emph {et~al.}(2008)\citenamefont {Yang},
  \citenamefont {Li}, \citenamefont {Wu}, \citenamefont {Zhang}, \citenamefont
  {Chen},\ and\ \citenamefont {Yang}}]{yang2008evaluation}%
  \BibitemOpen
  \bibfield  {author} {\bibinfo {author} {\bibfnamefont {J.}~\bibnamefont
  {Yang}}, \bibinfo {author} {\bibfnamefont {H.}~\bibnamefont {Li}}, \bibinfo
  {author} {\bibfnamefont {T.}~\bibnamefont {Wu}}, \bibinfo {author}
  {\bibfnamefont {W.}~\bibnamefont {Zhang}}, \bibinfo {author} {\bibfnamefont
  {L.}~\bibnamefont {Chen}}, \ and\ \bibinfo {author} {\bibfnamefont
  {J.}~\bibnamefont {Yang}},\ }\href@noop {} {\bibfield  {journal} {\bibinfo
  {journal} {Advanced Functional Materials}\ }\textbf {\bibinfo {volume}
  {18}},\ \bibinfo {pages} {2880} (\bibinfo {year} {2008})}\BibitemShut
  {NoStop}%
\bibitem [{\citenamefont {Gofryk}\ \emph {et~al.}(2011)\citenamefont {Gofryk},
  \citenamefont {Kaczorowski}, \citenamefont {Plackowski}, \citenamefont
  {Leithe-Jasper},\ and\ \citenamefont {Grin}}]{gofryk2011magnetic}%
  \BibitemOpen
  \bibfield  {author} {\bibinfo {author} {\bibfnamefont {K.}~\bibnamefont
  {Gofryk}}, \bibinfo {author} {\bibfnamefont {D.}~\bibnamefont {Kaczorowski}},
  \bibinfo {author} {\bibfnamefont {T.}~\bibnamefont {Plackowski}}, \bibinfo
  {author} {\bibfnamefont {A.}~\bibnamefont {Leithe-Jasper}}, \ and\ \bibinfo
  {author} {\bibfnamefont {Y.}~\bibnamefont {Grin}},\ }\href@noop {} {\bibfield
   {journal} {\bibinfo  {journal} {Physical Review B}\ }\textbf {\bibinfo
  {volume} {84}},\ \bibinfo {pages} {035208} (\bibinfo {year}
  {2011})}\BibitemShut {NoStop}%
\bibitem [{\citenamefont {Radmanesh}\ \emph {et~al.}(2018)\citenamefont
  {Radmanesh}, \citenamefont {Martin}, \citenamefont {Zhu}, \citenamefont
  {Yin}, \citenamefont {Xiao}, \citenamefont {Mao},\ and\ \citenamefont
  {Spinu}}]{radmanesh2018evidence}%
  \BibitemOpen
  \bibfield  {author} {\bibinfo {author} {\bibfnamefont {S.}~\bibnamefont
  {Radmanesh}}, \bibinfo {author} {\bibfnamefont {C.}~\bibnamefont {Martin}},
  \bibinfo {author} {\bibfnamefont {Y.}~\bibnamefont {Zhu}}, \bibinfo {author}
  {\bibfnamefont {X.}~\bibnamefont {Yin}}, \bibinfo {author} {\bibfnamefont
  {H.}~\bibnamefont {Xiao}}, \bibinfo {author} {\bibfnamefont {Z.}~\bibnamefont
  {Mao}}, \ and\ \bibinfo {author} {\bibfnamefont {L.}~\bibnamefont {Spinu}},\
  }\href@noop {} {\bibfield  {journal} {\bibinfo  {journal} {Physical Review
  B}\ }\textbf {\bibinfo {volume} {98}},\ \bibinfo {pages} {241111} (\bibinfo
  {year} {2018})}\BibitemShut {NoStop}%
\bibitem [{\citenamefont {Pavlosiuk}\ \emph {et~al.}(2016)\citenamefont
  {Pavlosiuk}, \citenamefont {Kaczorowski}, \citenamefont {Fabreges},
  \citenamefont {Gukasov},\ and\ \citenamefont
  {Wi{\'s}niewski}}]{pavlosiuk2016antiferromagnetism}%
  \BibitemOpen
  \bibfield  {author} {\bibinfo {author} {\bibfnamefont {O.}~\bibnamefont
  {Pavlosiuk}}, \bibinfo {author} {\bibfnamefont {D.}~\bibnamefont
  {Kaczorowski}}, \bibinfo {author} {\bibfnamefont {X.}~\bibnamefont
  {Fabreges}}, \bibinfo {author} {\bibfnamefont {A.}~\bibnamefont {Gukasov}}, \
  and\ \bibinfo {author} {\bibfnamefont {P.}~\bibnamefont {Wi{\'s}niewski}},\
  }\href@noop {} {\bibfield  {journal} {\bibinfo  {journal} {Scientific
  reports}\ }\textbf {\bibinfo {volume} {6}},\ \bibinfo {pages} {18797}
  (\bibinfo {year} {2016})}\BibitemShut {NoStop}%
\bibitem [{\citenamefont {Xu}\ \emph {et~al.}(2014)\citenamefont {Xu},
  \citenamefont {Wang}, \citenamefont {Zhang}, \citenamefont {Du},
  \citenamefont {Liu}, \citenamefont {Wang}, \citenamefont {Wu}, \citenamefont
  {Liu},\ and\ \citenamefont {Zhang}}]{xu2014weak}%
  \BibitemOpen
  \bibfield  {author} {\bibinfo {author} {\bibfnamefont {G.}~\bibnamefont
  {Xu}}, \bibinfo {author} {\bibfnamefont {W.}~\bibnamefont {Wang}}, \bibinfo
  {author} {\bibfnamefont {X.}~\bibnamefont {Zhang}}, \bibinfo {author}
  {\bibfnamefont {Y.}~\bibnamefont {Du}}, \bibinfo {author} {\bibfnamefont
  {E.}~\bibnamefont {Liu}}, \bibinfo {author} {\bibfnamefont {S.}~\bibnamefont
  {Wang}}, \bibinfo {author} {\bibfnamefont {G.}~\bibnamefont {Wu}}, \bibinfo
  {author} {\bibfnamefont {Z.}~\bibnamefont {Liu}}, \ and\ \bibinfo {author}
  {\bibfnamefont {X.~X.}\ \bibnamefont {Zhang}},\ }\href@noop {} {\bibfield
  {journal} {\bibinfo  {journal} {Scientific reports}\ }\textbf {\bibinfo
  {volume} {4}},\ \bibinfo {pages} {5709} (\bibinfo {year} {2014})}\BibitemShut
  {NoStop}%
\bibitem [{\citenamefont {Pan}\ \emph {et~al.}(2013)\citenamefont {Pan},
  \citenamefont {Nikitin}, \citenamefont {Bay}, \citenamefont {Huang},
  \citenamefont {Paulsen}, \citenamefont {Yan},\ and\ \citenamefont
  {De~Visser}}]{pan2013superconductivity}%
  \BibitemOpen
  \bibfield  {author} {\bibinfo {author} {\bibfnamefont {Y.}~\bibnamefont
  {Pan}}, \bibinfo {author} {\bibfnamefont {A.}~\bibnamefont {Nikitin}},
  \bibinfo {author} {\bibfnamefont {T.}~\bibnamefont {Bay}}, \bibinfo {author}
  {\bibfnamefont {Y.}~\bibnamefont {Huang}}, \bibinfo {author} {\bibfnamefont
  {C.}~\bibnamefont {Paulsen}}, \bibinfo {author} {\bibfnamefont
  {B.}~\bibnamefont {Yan}}, \ and\ \bibinfo {author} {\bibfnamefont
  {A.}~\bibnamefont {De~Visser}},\ }\href@noop {} {\bibfield  {journal}
  {\bibinfo  {journal} {EPL (Europhysics Letters)}\ }\textbf {\bibinfo {volume}
  {104}},\ \bibinfo {pages} {27001} (\bibinfo {year} {2013})}\BibitemShut
  {NoStop}%
\bibitem [{\citenamefont {Nakajima}\ \emph {et~al.}(2015)\citenamefont
  {Nakajima}, \citenamefont {Hu}, \citenamefont {Kirshenbaum}, \citenamefont
  {Hughes}, \citenamefont {Syers}, \citenamefont {Wang}, \citenamefont {Wang},
  \citenamefont {Wang}, \citenamefont {Saha}, \citenamefont {Pratt} \emph
  {et~al.}}]{nakajima2015topological}%
  \BibitemOpen
  \bibfield  {author} {\bibinfo {author} {\bibfnamefont {Y.}~\bibnamefont
  {Nakajima}}, \bibinfo {author} {\bibfnamefont {R.}~\bibnamefont {Hu}},
  \bibinfo {author} {\bibfnamefont {K.}~\bibnamefont {Kirshenbaum}}, \bibinfo
  {author} {\bibfnamefont {A.}~\bibnamefont {Hughes}}, \bibinfo {author}
  {\bibfnamefont {P.}~\bibnamefont {Syers}}, \bibinfo {author} {\bibfnamefont
  {X.}~\bibnamefont {Wang}}, \bibinfo {author} {\bibfnamefont {K.}~\bibnamefont
  {Wang}}, \bibinfo {author} {\bibfnamefont {R.}~\bibnamefont {Wang}}, \bibinfo
  {author} {\bibfnamefont {S.~R.}\ \bibnamefont {Saha}}, \bibinfo {author}
  {\bibfnamefont {D.}~\bibnamefont {Pratt}},  \emph {et~al.},\ }\href@noop {}
  {\bibfield  {journal} {\bibinfo  {journal} {Science advances}\ }\textbf
  {\bibinfo {volume} {1}},\ \bibinfo {pages} {e1500242} (\bibinfo {year}
  {2015})}\BibitemShut {NoStop}%
\bibitem [{\citenamefont {Pavlosiuk}\ \emph {et~al.}(2018)\citenamefont
  {Pavlosiuk}, \citenamefont {Fabreges}, \citenamefont {Gukasov}, \citenamefont
  {Meven}, \citenamefont {Kaczorowski},\ and\ \citenamefont {Wi~{\ 's}
  niewski}}]{pavlosiuk2018magnetic}%
  \BibitemOpen
  \bibfield  {author} {\bibinfo {author} {\bibfnamefont {O.}~\bibnamefont
  {Pavlosiuk}}, \bibinfo {author} {\bibfnamefont {X.}~\bibnamefont {Fabreges}},
  \bibinfo {author} {\bibfnamefont {A.}~\bibnamefont {Gukasov}}, \bibinfo
  {author} {\bibfnamefont {M.}~\bibnamefont {Meven}}, \bibinfo {author}
  {\bibfnamefont {D.}~\bibnamefont {Kaczorowski}}, \ and\ \bibinfo {author}
  {\bibfnamefont {P.}~\bibnamefont {Wi~{\ 's} niewski}},\ }\href@noop {}
  {\bibfield  {journal} {\bibinfo  {journal} {Physica B: Condensed Matter}\
  }\textbf {\bibinfo {volume} {536}},\ \bibinfo {pages} {56} (\bibinfo {year}
  {2018})}\BibitemShut {NoStop}%
\bibitem [{\citenamefont {Suzuki}\ \emph {et~al.}(2016)\citenamefont {Suzuki},
  \citenamefont {Chisnell}, \citenamefont {Devarakonda}, \citenamefont {Liu},
  \citenamefont {Feng}, \citenamefont {Xiao}, \citenamefont {Lynn},\ and\
  \citenamefont {Checkelsky}}]{suzuki2016large}%
  \BibitemOpen
  \bibfield  {author} {\bibinfo {author} {\bibfnamefont {T.}~\bibnamefont
  {Suzuki}}, \bibinfo {author} {\bibfnamefont {R.}~\bibnamefont {Chisnell}},
  \bibinfo {author} {\bibfnamefont {A.}~\bibnamefont {Devarakonda}}, \bibinfo
  {author} {\bibfnamefont {Y.-T.}\ \bibnamefont {Liu}}, \bibinfo {author}
  {\bibfnamefont {W.}~\bibnamefont {Feng}}, \bibinfo {author} {\bibfnamefont
  {D.}~\bibnamefont {Xiao}}, \bibinfo {author} {\bibfnamefont {J.~W.}\
  \bibnamefont {Lynn}}, \ and\ \bibinfo {author} {\bibfnamefont
  {J.}~\bibnamefont {Checkelsky}},\ }\href@noop {} {\bibfield  {journal}
  {\bibinfo  {journal} {Nature Physics}\ }\textbf {\bibinfo {volume} {12}},\
  \bibinfo {pages} {1119} (\bibinfo {year} {2016})}\BibitemShut {NoStop}%
\bibitem [{\citenamefont {Manna}\ \emph {et~al.}(2018)\citenamefont {Manna},
  \citenamefont {Sun}, \citenamefont {Muechler}, \citenamefont {K{\"u}bler},\
  and\ \citenamefont {Felser}}]{manna2018heusler}%
  \BibitemOpen
  \bibfield  {author} {\bibinfo {author} {\bibfnamefont {K.}~\bibnamefont
  {Manna}}, \bibinfo {author} {\bibfnamefont {Y.}~\bibnamefont {Sun}}, \bibinfo
  {author} {\bibfnamefont {L.}~\bibnamefont {Muechler}}, \bibinfo {author}
  {\bibfnamefont {J.}~\bibnamefont {K{\"u}bler}}, \ and\ \bibinfo {author}
  {\bibfnamefont {C.}~\bibnamefont {Felser}},\ }\href@noop {} {\bibfield
  {journal} {\bibinfo  {journal} {Nature Reviews Materials}\ }\textbf {\bibinfo
  {volume} {3}},\ \bibinfo {pages} {244} (\bibinfo {year} {2018})}\BibitemShut
  {NoStop}%
\bibitem [{\citenamefont {Shi}\ \emph {et~al.}(2018)\citenamefont {Shi},
  \citenamefont {Jia}, \citenamefont {Si}, \citenamefont {Zhang}, \citenamefont
  {Xie}, \citenamefont {Xiao}, \citenamefont {Yang}, \citenamefont {Shi},\ and\
  \citenamefont {Luo}}]{shi2018tunable}%
  \BibitemOpen
  \bibfield  {author} {\bibinfo {author} {\bibfnamefont {F.}~\bibnamefont
  {Shi}}, \bibinfo {author} {\bibfnamefont {L.}~\bibnamefont {Jia}}, \bibinfo
  {author} {\bibfnamefont {M.}~\bibnamefont {Si}}, \bibinfo {author}
  {\bibfnamefont {Z.}~\bibnamefont {Zhang}}, \bibinfo {author} {\bibfnamefont
  {J.}~\bibnamefont {Xie}}, \bibinfo {author} {\bibfnamefont {C.}~\bibnamefont
  {Xiao}}, \bibinfo {author} {\bibfnamefont {D.}~\bibnamefont {Yang}}, \bibinfo
  {author} {\bibfnamefont {H.}~\bibnamefont {Shi}}, \ and\ \bibinfo {author}
  {\bibfnamefont {Q.}~\bibnamefont {Luo}},\ }\href@noop {} {\bibfield
  {journal} {\bibinfo  {journal} {Applied Physics Express}\ }\textbf {\bibinfo
  {volume} {11}},\ \bibinfo {pages} {095701} (\bibinfo {year}
  {2018})}\BibitemShut {NoStop}%
\bibitem [{\citenamefont {Liu}\ \emph {et~al.}(2016)\citenamefont {Liu},
  \citenamefont {Yang}, \citenamefont {Wu}, \citenamefont {Shekhar},
  \citenamefont {Jiang}, \citenamefont {Yang}, \citenamefont {Zhang},
  \citenamefont {Mo}, \citenamefont {Hussain}, \citenamefont {Yan} \emph
  {et~al.}}]{liu2016observation}%
  \BibitemOpen
  \bibfield  {author} {\bibinfo {author} {\bibfnamefont {Z.}~\bibnamefont
  {Liu}}, \bibinfo {author} {\bibfnamefont {L.}~\bibnamefont {Yang}}, \bibinfo
  {author} {\bibfnamefont {S.-C.}\ \bibnamefont {Wu}}, \bibinfo {author}
  {\bibfnamefont {C.}~\bibnamefont {Shekhar}}, \bibinfo {author} {\bibfnamefont
  {J.}~\bibnamefont {Jiang}}, \bibinfo {author} {\bibfnamefont
  {H.}~\bibnamefont {Yang}}, \bibinfo {author} {\bibfnamefont {Y.}~\bibnamefont
  {Zhang}}, \bibinfo {author} {\bibfnamefont {S.-K.}\ \bibnamefont {Mo}},
  \bibinfo {author} {\bibfnamefont {Z.}~\bibnamefont {Hussain}}, \bibinfo
  {author} {\bibfnamefont {B.}~\bibnamefont {Yan}},  \emph {et~al.},\
  }\href@noop {} {\bibfield  {journal} {\bibinfo  {journal} {Nature
  communications}\ }\textbf {\bibinfo {volume} {7}},\ \bibinfo {pages} {1}
  (\bibinfo {year} {2016})}\BibitemShut {NoStop}%
\bibitem [{\citenamefont {Xiao}\ \emph {et~al.}(2010)\citenamefont {Xiao},
  \citenamefont {Yao}, \citenamefont {Feng}, \citenamefont {Wen}, \citenamefont
  {Zhu}, \citenamefont {Chen}, \citenamefont {Stocks},\ and\ \citenamefont
  {Zhang}}]{xiao2010half}%
  \BibitemOpen
  \bibfield  {author} {\bibinfo {author} {\bibfnamefont {D.}~\bibnamefont
  {Xiao}}, \bibinfo {author} {\bibfnamefont {Y.}~\bibnamefont {Yao}}, \bibinfo
  {author} {\bibfnamefont {W.}~\bibnamefont {Feng}}, \bibinfo {author}
  {\bibfnamefont {J.}~\bibnamefont {Wen}}, \bibinfo {author} {\bibfnamefont
  {W.}~\bibnamefont {Zhu}}, \bibinfo {author} {\bibfnamefont {X.-Q.}\
  \bibnamefont {Chen}}, \bibinfo {author} {\bibfnamefont {G.~M.}\ \bibnamefont
  {Stocks}}, \ and\ \bibinfo {author} {\bibfnamefont {Z.}~\bibnamefont
  {Zhang}},\ }\href@noop {} {\bibfield  {journal} {\bibinfo  {journal}
  {Physical review letters}\ }\textbf {\bibinfo {volume} {105}},\ \bibinfo
  {pages} {096404} (\bibinfo {year} {2010})}\BibitemShut {NoStop}%
\bibitem [{\citenamefont {Jung}\ \emph {et~al.}(2000)\citenamefont {Jung},
  \citenamefont {Koo},\ and\ \citenamefont {Whangbo}}]{jung2000study}%
  \BibitemOpen
  \bibfield  {author} {\bibinfo {author} {\bibfnamefont {D.}~\bibnamefont
  {Jung}}, \bibinfo {author} {\bibfnamefont {H.-J.}\ \bibnamefont {Koo}}, \
  and\ \bibinfo {author} {\bibfnamefont {M.-H.}\ \bibnamefont {Whangbo}},\
  }\href@noop {} {\bibfield  {journal} {\bibinfo  {journal} {Journal of
  Molecular Structure: THEOCHEM}\ }\textbf {\bibinfo {volume} {527}},\ \bibinfo
  {pages} {113} (\bibinfo {year} {2000})}\BibitemShut {NoStop}%
\bibitem [{\citenamefont {Hao}\ \emph {et~al.}(2019)\citenamefont {Hao},
  \citenamefont {Dravid}, \citenamefont {Kanatzidis},\ and\ \citenamefont
  {Wolverton}}]{hao2019computational}%
  \BibitemOpen
  \bibfield  {author} {\bibinfo {author} {\bibfnamefont {S.}~\bibnamefont
  {Hao}}, \bibinfo {author} {\bibfnamefont {V.~P.}\ \bibnamefont {Dravid}},
  \bibinfo {author} {\bibfnamefont {M.~G.}\ \bibnamefont {Kanatzidis}}, \ and\
  \bibinfo {author} {\bibfnamefont {C.}~\bibnamefont {Wolverton}},\ }\href@noop
  {} {\bibfield  {journal} {\bibinfo  {journal} {npj Computational Materials}\
  }\textbf {\bibinfo {volume} {5}},\ \bibinfo {pages} {58} (\bibinfo {year}
  {2019})}\BibitemShut {NoStop}%
\bibitem [{\citenamefont {Lee}\ \emph {et~al.}(2020)\citenamefont {Lee},
  \citenamefont {Kim}, \citenamefont {Kim},\ and\ \citenamefont
  {Kim}}]{lee2020band}%
  \BibitemOpen
  \bibfield  {author} {\bibinfo {author} {\bibfnamefont {K.~H.}\ \bibnamefont
  {Lee}}, \bibinfo {author} {\bibfnamefont {S.-i.}\ \bibnamefont {Kim}},
  \bibinfo {author} {\bibfnamefont {H.-S.}\ \bibnamefont {Kim}}, \ and\
  \bibinfo {author} {\bibfnamefont {S.~W.}\ \bibnamefont {Kim}},\ }\href@noop
  {} {\bibfield  {journal} {\bibinfo  {journal} {ACS Applied Energy Materials}\
  } (\bibinfo {year} {2020})}\BibitemShut {NoStop}%
\bibitem [{\citenamefont {Xiao}\ and\ \citenamefont
  {Zhao}(2018)}]{xiao2018charge}%
  \BibitemOpen
  \bibfield  {author} {\bibinfo {author} {\bibfnamefont {Y.}~\bibnamefont
  {Xiao}}\ and\ \bibinfo {author} {\bibfnamefont {L.-D.}\ \bibnamefont
  {Zhao}},\ }\href@noop {} {\bibfield  {journal} {\bibinfo  {journal} {npj
  Quantum Materials}\ }\textbf {\bibinfo {volume} {3}},\ \bibinfo {pages} {1}
  (\bibinfo {year} {2018})}\BibitemShut {NoStop}%
\bibitem [{\citenamefont {Zhu}\ \emph {et~al.}(2018)\citenamefont {Zhu},
  \citenamefont {He}, \citenamefont {Mao}, \citenamefont {Zhu}, \citenamefont
  {Li}, \citenamefont {Sun}, \citenamefont {Ren}, \citenamefont {Wang},
  \citenamefont {Liu}, \citenamefont {Tang} \emph {et~al.}}]{zhu2018discovery}%
  \BibitemOpen
  \bibfield  {author} {\bibinfo {author} {\bibfnamefont {H.}~\bibnamefont
  {Zhu}}, \bibinfo {author} {\bibfnamefont {R.}~\bibnamefont {He}}, \bibinfo
  {author} {\bibfnamefont {J.}~\bibnamefont {Mao}}, \bibinfo {author}
  {\bibfnamefont {Q.}~\bibnamefont {Zhu}}, \bibinfo {author} {\bibfnamefont
  {C.}~\bibnamefont {Li}}, \bibinfo {author} {\bibfnamefont {J.}~\bibnamefont
  {Sun}}, \bibinfo {author} {\bibfnamefont {W.}~\bibnamefont {Ren}}, \bibinfo
  {author} {\bibfnamefont {Y.}~\bibnamefont {Wang}}, \bibinfo {author}
  {\bibfnamefont {Z.}~\bibnamefont {Liu}}, \bibinfo {author} {\bibfnamefont
  {Z.}~\bibnamefont {Tang}},  \emph {et~al.},\ }\href@noop {} {\bibfield
  {journal} {\bibinfo  {journal} {Nature communications}\ }\textbf {\bibinfo
  {volume} {9}},\ \bibinfo {pages} {1} (\bibinfo {year} {2018})}\BibitemShut
  {NoStop}%
\bibitem [{\citenamefont {Simon}(1964)}]{simon1964maximum}%
  \BibitemOpen
  \bibfield  {author} {\bibinfo {author} {\bibfnamefont {R.}~\bibnamefont
  {Simon}},\ }\href@noop {} {\bibfield  {journal} {\bibinfo  {journal}
  {Solid-State Electronics}\ }\textbf {\bibinfo {volume} {7}},\ \bibinfo
  {pages} {397} (\bibinfo {year} {1964})}\BibitemShut {NoStop}%
\bibitem [{\citenamefont {Slack}\ and\ \citenamefont
  {Hussain}(1991)}]{slack1991maximum}%
  \BibitemOpen
  \bibfield  {author} {\bibinfo {author} {\bibfnamefont {G.~A.}\ \bibnamefont
  {Slack}}\ and\ \bibinfo {author} {\bibfnamefont {M.~A.}\ \bibnamefont
  {Hussain}},\ }\href@noop {} {\bibfield  {journal} {\bibinfo  {journal}
  {Journal of applied physics}\ }\textbf {\bibinfo {volume} {70}},\ \bibinfo
  {pages} {2694} (\bibinfo {year} {1991})}\BibitemShut {NoStop}%
\bibitem [{\citenamefont {Liu}\ \emph {et~al.}(2008)\citenamefont {Liu},
  \citenamefont {Zhao}, \citenamefont {Zhang}, \citenamefont {Zhang},\ and\
  \citenamefont {Li}}]{liu2008enhanced}%
  \BibitemOpen
  \bibfield  {author} {\bibinfo {author} {\bibfnamefont {W.-S.}\ \bibnamefont
  {Liu}}, \bibinfo {author} {\bibfnamefont {L.-D.}\ \bibnamefont {Zhao}},
  \bibinfo {author} {\bibfnamefont {B.-P.}\ \bibnamefont {Zhang}}, \bibinfo
  {author} {\bibfnamefont {H.-L.}\ \bibnamefont {Zhang}}, \ and\ \bibinfo
  {author} {\bibfnamefont {J.-F.}\ \bibnamefont {Li}},\ }\href@noop {}
  {\bibfield  {journal} {\bibinfo  {journal} {Applied Physics Letters}\
  }\textbf {\bibinfo {volume} {93}},\ \bibinfo {pages} {042109} (\bibinfo
  {year} {2008})}\BibitemShut {NoStop}%
\bibitem [{\citenamefont {Pei}\ \emph {et~al.}(2011{\natexlab{b}})\citenamefont
  {Pei}, \citenamefont {Shi}, \citenamefont {LaLonde}, \citenamefont {Wang},
  \citenamefont {Chen},\ and\ \citenamefont {Snyder}}]{pei2011convergence}%
  \BibitemOpen
  \bibfield  {author} {\bibinfo {author} {\bibfnamefont {Y.}~\bibnamefont
  {Pei}}, \bibinfo {author} {\bibfnamefont {X.}~\bibnamefont {Shi}}, \bibinfo
  {author} {\bibfnamefont {A.}~\bibnamefont {LaLonde}}, \bibinfo {author}
  {\bibfnamefont {H.}~\bibnamefont {Wang}}, \bibinfo {author} {\bibfnamefont
  {L.}~\bibnamefont {Chen}}, \ and\ \bibinfo {author} {\bibfnamefont {G.~J.}\
  \bibnamefont {Snyder}},\ }\href@noop {} {\bibfield  {journal} {\bibinfo
  {journal} {Nature}\ }\textbf {\bibinfo {volume} {473}},\ \bibinfo {pages}
  {66} (\bibinfo {year} {2011}{\natexlab{b}})}\BibitemShut {NoStop}%
\bibitem [{\citenamefont {Birkel}\ \emph {et~al.}(2013)\citenamefont {Birkel},
  \citenamefont {Douglas}, \citenamefont {Lettiere}, \citenamefont {Seward},
  \citenamefont {Verma}, \citenamefont {Zhang}, \citenamefont {Pollock},
  \citenamefont {Seshadri},\ and\ \citenamefont
  {Stucky}}]{birkel2013improving}%
  \BibitemOpen
  \bibfield  {author} {\bibinfo {author} {\bibfnamefont {C.~S.}\ \bibnamefont
  {Birkel}}, \bibinfo {author} {\bibfnamefont {J.~E.}\ \bibnamefont {Douglas}},
  \bibinfo {author} {\bibfnamefont {B.~R.}\ \bibnamefont {Lettiere}}, \bibinfo
  {author} {\bibfnamefont {G.}~\bibnamefont {Seward}}, \bibinfo {author}
  {\bibfnamefont {N.}~\bibnamefont {Verma}}, \bibinfo {author} {\bibfnamefont
  {Y.}~\bibnamefont {Zhang}}, \bibinfo {author} {\bibfnamefont {T.~M.}\
  \bibnamefont {Pollock}}, \bibinfo {author} {\bibfnamefont {R.}~\bibnamefont
  {Seshadri}}, \ and\ \bibinfo {author} {\bibfnamefont {G.~D.}\ \bibnamefont
  {Stucky}},\ }\href@noop {} {\bibfield  {journal} {\bibinfo  {journal}
  {Physical Chemistry Chemical Physics}\ }\textbf {\bibinfo {volume} {15}},\
  \bibinfo {pages} {6990} (\bibinfo {year} {2013})}\BibitemShut {NoStop}%
\bibitem [{\citenamefont {Kim}\ \emph {et~al.}(2007)\citenamefont {Kim},
  \citenamefont {Kimura},\ and\ \citenamefont {Mishima}}]{kim2007high}%
  \BibitemOpen
  \bibfield  {author} {\bibinfo {author} {\bibfnamefont {S.-W.}\ \bibnamefont
  {Kim}}, \bibinfo {author} {\bibfnamefont {Y.}~\bibnamefont {Kimura}}, \ and\
  \bibinfo {author} {\bibfnamefont {Y.}~\bibnamefont {Mishima}},\ }\href@noop
  {} {\bibfield  {journal} {\bibinfo  {journal} {Intermetallics}\ }\textbf
  {\bibinfo {volume} {15}},\ \bibinfo {pages} {349} (\bibinfo {year}
  {2007})}\BibitemShut {NoStop}%
\bibitem [{\citenamefont {Debye}(1912)}]{debye1912theorie}%
  \BibitemOpen
  \bibfield  {author} {\bibinfo {author} {\bibfnamefont {P.}~\bibnamefont
  {Debye}},\ }\href@noop {} {\bibfield  {journal} {\bibinfo  {journal} {Annalen
  der Physik}\ }\textbf {\bibinfo {volume} {344}},\ \bibinfo {pages} {789}
  (\bibinfo {year} {1912})}\BibitemShut {NoStop}%
\bibitem [{\citenamefont {Peierls}(1955)}]{peierls1955quantum}%
  \BibitemOpen
  \bibfield  {author} {\bibinfo {author} {\bibfnamefont {R.}~\bibnamefont
  {Peierls}},\ }\href@noop {} {\bibfield  {journal} {\bibinfo  {journal}
  {London, England}\ ,\ \bibinfo {pages} {108}} (\bibinfo {year}
  {1955})}\BibitemShut {NoStop}%
\bibitem [{\citenamefont {Ziman}(1960)}]{ziman1960electrons}%
  \BibitemOpen
  \bibfield  {author} {\bibinfo {author} {\bibfnamefont {J.}~\bibnamefont
  {Ziman}},\ }\href@noop {} {\enquote {\bibinfo {title} {Electrons and phonons,
  oxford university press},}\ } (\bibinfo {year} {1960})\BibitemShut {NoStop}%
\bibitem [{\citenamefont {Broido}\ \emph {et~al.}(2007)\citenamefont {Broido},
  \citenamefont {Malorny}, \citenamefont {Birner}, \citenamefont {Mingo},\ and\
  \citenamefont {Stewart}}]{broido2007intrinsic}%
  \BibitemOpen
  \bibfield  {author} {\bibinfo {author} {\bibfnamefont {D.~A.}\ \bibnamefont
  {Broido}}, \bibinfo {author} {\bibfnamefont {M.}~\bibnamefont {Malorny}},
  \bibinfo {author} {\bibfnamefont {G.}~\bibnamefont {Birner}}, \bibinfo
  {author} {\bibfnamefont {N.}~\bibnamefont {Mingo}}, \ and\ \bibinfo {author}
  {\bibfnamefont {D.}~\bibnamefont {Stewart}},\ }\href@noop {} {\bibfield
  {journal} {\bibinfo  {journal} {Applied Physics Letters}\ }\textbf {\bibinfo
  {volume} {91}},\ \bibinfo {pages} {231922} (\bibinfo {year}
  {2007})}\BibitemShut {NoStop}%
\bibitem [{\citenamefont {Ward}\ and\ \citenamefont
  {Broido}(2010)}]{ward2010intrinsic}%
  \BibitemOpen
  \bibfield  {author} {\bibinfo {author} {\bibfnamefont {A.}~\bibnamefont
  {Ward}}\ and\ \bibinfo {author} {\bibfnamefont {D.}~\bibnamefont {Broido}},\
  }\href@noop {} {\bibfield  {journal} {\bibinfo  {journal} {Physical Review
  B}\ }\textbf {\bibinfo {volume} {81}},\ \bibinfo {pages} {085205} (\bibinfo
  {year} {2010})}\BibitemShut {NoStop}%
\bibitem [{\citenamefont {Tang}\ and\ \citenamefont
  {Dong}(2010)}]{tang2010lattice}%
  \BibitemOpen
  \bibfield  {author} {\bibinfo {author} {\bibfnamefont {X.}~\bibnamefont
  {Tang}}\ and\ \bibinfo {author} {\bibfnamefont {J.}~\bibnamefont {Dong}},\
  }\href@noop {} {\bibfield  {journal} {\bibinfo  {journal} {Proceedings of the
  National Academy of Sciences}\ }\textbf {\bibinfo {volume} {107}},\ \bibinfo
  {pages} {4539} (\bibinfo {year} {2010})}\BibitemShut {NoStop}%
\bibitem [{\citenamefont {Callaway}(1959)}]{callaway1959model}%
  \BibitemOpen
  \bibfield  {author} {\bibinfo {author} {\bibfnamefont {J.}~\bibnamefont
  {Callaway}},\ }\href@noop {} {\bibfield  {journal} {\bibinfo  {journal}
  {Physical Review}\ }\textbf {\bibinfo {volume} {113}},\ \bibinfo {pages}
  {1046} (\bibinfo {year} {1959})}\BibitemShut {NoStop}%
\bibitem [{\citenamefont {Allen}(2013)}]{allen2013improved}%
  \BibitemOpen
  \bibfield  {author} {\bibinfo {author} {\bibfnamefont {P.~B.}\ \bibnamefont
  {Allen}},\ }\href@noop {} {\bibfield  {journal} {\bibinfo  {journal}
  {Physical Review B}\ }\textbf {\bibinfo {volume} {88}},\ \bibinfo {pages}
  {144302} (\bibinfo {year} {2013})}\BibitemShut {NoStop}%
\bibitem [{\citenamefont {Deinzer}\ \emph {et~al.}(2003)\citenamefont
  {Deinzer}, \citenamefont {Birner},\ and\ \citenamefont
  {Strauch}}]{deinzer2003ab}%
  \BibitemOpen
  \bibfield  {author} {\bibinfo {author} {\bibfnamefont {G.}~\bibnamefont
  {Deinzer}}, \bibinfo {author} {\bibfnamefont {G.}~\bibnamefont {Birner}}, \
  and\ \bibinfo {author} {\bibfnamefont {D.}~\bibnamefont {Strauch}},\
  }\href@noop {} {\bibfield  {journal} {\bibinfo  {journal} {Physical Review
  B}\ }\textbf {\bibinfo {volume} {67}},\ \bibinfo {pages} {144304} (\bibinfo
  {year} {2003})}\BibitemShut {NoStop}%
\bibitem [{\citenamefont {Chernatynskiy}\ and\ \citenamefont
  {Phillpot}(2015)}]{chernatynskiy2015phonon}%
  \BibitemOpen
  \bibfield  {author} {\bibinfo {author} {\bibfnamefont {A.}~\bibnamefont
  {Chernatynskiy}}\ and\ \bibinfo {author} {\bibfnamefont {S.~R.}\ \bibnamefont
  {Phillpot}},\ }\href@noop {} {\bibfield  {journal} {\bibinfo  {journal}
  {Computer Physics Communications}\ }\textbf {\bibinfo {volume} {192}},\
  \bibinfo {pages} {196} (\bibinfo {year} {2015})}\BibitemShut {NoStop}%
\bibitem [{\citenamefont {Tadano}\ \emph {et~al.}(2014)\citenamefont {Tadano},
  \citenamefont {Gohda},\ and\ \citenamefont
  {Tsuneyuki}}]{tadano2014anharmonic}%
  \BibitemOpen
  \bibfield  {author} {\bibinfo {author} {\bibfnamefont {T.}~\bibnamefont
  {Tadano}}, \bibinfo {author} {\bibfnamefont {Y.}~\bibnamefont {Gohda}}, \
  and\ \bibinfo {author} {\bibfnamefont {S.}~\bibnamefont {Tsuneyuki}},\
  }\href@noop {} {\bibfield  {journal} {\bibinfo  {journal} {Journal of
  Physics: Condensed Matter}\ }\textbf {\bibinfo {volume} {26}},\ \bibinfo
  {pages} {225402} (\bibinfo {year} {2014})}\BibitemShut {NoStop}%
\bibitem [{\citenamefont {Li}\ \emph {et~al.}(2014)\citenamefont {Li},
  \citenamefont {Carrete}, \citenamefont {Katcho},\ and\ \citenamefont
  {Mingo}}]{li2014shengbte}%
  \BibitemOpen
  \bibfield  {author} {\bibinfo {author} {\bibfnamefont {W.}~\bibnamefont
  {Li}}, \bibinfo {author} {\bibfnamefont {J.}~\bibnamefont {Carrete}},
  \bibinfo {author} {\bibfnamefont {N.~A.}\ \bibnamefont {Katcho}}, \ and\
  \bibinfo {author} {\bibfnamefont {N.}~\bibnamefont {Mingo}},\ }\href@noop {}
  {\bibfield  {journal} {\bibinfo  {journal} {Computer Physics Communications}\
  }\textbf {\bibinfo {volume} {185}},\ \bibinfo {pages} {1747} (\bibinfo {year}
  {2014})}\BibitemShut {NoStop}%
\bibitem [{\citenamefont {Kachanov}\ \emph {et~al.}(2003)\citenamefont
  {Kachanov}, \citenamefont {Shafiro},\ and\ \citenamefont
  {Tsukrov}}]{kachanov2003handbook}%
  \BibitemOpen
  \bibfield  {author} {\bibinfo {author} {\bibfnamefont {M.~L.}\ \bibnamefont
  {Kachanov}}, \bibinfo {author} {\bibfnamefont {B.}~\bibnamefont {Shafiro}}, \
  and\ \bibinfo {author} {\bibfnamefont {I.}~\bibnamefont {Tsukrov}},\
  }\href@noop {} {\emph {\bibinfo {title} {Handbook of elasticity solutions}}}\
  (\bibinfo  {publisher} {Springer Science \& Business Media},\ \bibinfo {year}
  {2003})\BibitemShut {NoStop}%
\bibitem [{\citenamefont {Den~Toonder}\ \emph {et~al.}(1999)\citenamefont
  {Den~Toonder}, \citenamefont {Van~Dommelen},\ and\ \citenamefont
  {Baaijens}}]{den1999relation}%
  \BibitemOpen
  \bibfield  {author} {\bibinfo {author} {\bibfnamefont {J.}~\bibnamefont
  {Den~Toonder}}, \bibinfo {author} {\bibfnamefont {J.}~\bibnamefont
  {Van~Dommelen}}, \ and\ \bibinfo {author} {\bibfnamefont {F.}~\bibnamefont
  {Baaijens}},\ }\href@noop {} {\bibfield  {journal} {\bibinfo  {journal}
  {Modelling and Simulation in Materials Science and Engineering}\ }\textbf
  {\bibinfo {volume} {7}},\ \bibinfo {pages} {909} (\bibinfo {year}
  {1999})}\BibitemShut {NoStop}%
\bibitem [{\citenamefont {Ericksen}(1984)}]{ericksen1984phase}%
  \BibitemOpen
  \bibfield  {author} {\bibinfo {author} {\bibfnamefont {J.}~\bibnamefont
  {Ericksen}},\ }\href@noop {} {\bibfield  {journal} {\bibinfo  {journal} {The
  Cauchy and Born Hypotheses for Crystals. Academic Press, London}\ ,\ \bibinfo
  {pages} {61}} (\bibinfo {year} {1984})}\BibitemShut {NoStop}%
\bibitem [{\citenamefont {Born}(1940)}]{born1940stability}%
  \BibitemOpen
  \bibfield  {author} {\bibinfo {author} {\bibfnamefont {M.}~\bibnamefont
  {Born}},\ }in\ \href@noop {} {\emph {\bibinfo {booktitle} {Mathematical
  Proceedings of the Cambridge Philosophical Society}}},\ Vol.~\bibinfo
  {volume} {36}\ (\bibinfo {organization} {Cambridge University Press},\
  \bibinfo {year} {1940})\ pp.\ \bibinfo {pages} {160--172}\BibitemShut
  {NoStop}%
\bibitem [{\citenamefont {Bhattacharya}\ \emph {et~al.}(2002)\citenamefont
  {Bhattacharya}, \citenamefont {Tritt}, \citenamefont {Xia}, \citenamefont
  {Ponnambalam}, \citenamefont {Poon},\ and\ \citenamefont
  {Thadhani}}]{bhattacharya2002grain}%
  \BibitemOpen
  \bibfield  {author} {\bibinfo {author} {\bibfnamefont {S.}~\bibnamefont
  {Bhattacharya}}, \bibinfo {author} {\bibfnamefont {T.~M.}\ \bibnamefont
  {Tritt}}, \bibinfo {author} {\bibfnamefont {Y.}~\bibnamefont {Xia}}, \bibinfo
  {author} {\bibfnamefont {V.}~\bibnamefont {Ponnambalam}}, \bibinfo {author}
  {\bibfnamefont {S.}~\bibnamefont {Poon}}, \ and\ \bibinfo {author}
  {\bibfnamefont {N.}~\bibnamefont {Thadhani}},\ }\href@noop {} {\bibfield
  {journal} {\bibinfo  {journal} {Applied physics letters}\ }\textbf {\bibinfo
  {volume} {81}},\ \bibinfo {pages} {43} (\bibinfo {year} {2002})}\BibitemShut
  {NoStop}%
\bibitem [{\citenamefont {Downie}\ \emph {et~al.}(2013)\citenamefont {Downie},
  \citenamefont {MacLaren}, \citenamefont {Smith},\ and\ \citenamefont
  {Bos}}]{downie2013enhanced}%
  \BibitemOpen
  \bibfield  {author} {\bibinfo {author} {\bibfnamefont {R.~A.}\ \bibnamefont
  {Downie}}, \bibinfo {author} {\bibfnamefont {D.}~\bibnamefont {MacLaren}},
  \bibinfo {author} {\bibfnamefont {R.}~\bibnamefont {Smith}}, \ and\ \bibinfo
  {author} {\bibfnamefont {J.}~\bibnamefont {Bos}},\ }\href@noop {} {\bibfield
  {journal} {\bibinfo  {journal} {Chemical communications}\ }\textbf {\bibinfo
  {volume} {49}},\ \bibinfo {pages} {4184} (\bibinfo {year}
  {2013})}\BibitemShut {NoStop}%
\bibitem [{\citenamefont {Kirievsky}\ \emph {et~al.}(2013)\citenamefont
  {Kirievsky}, \citenamefont {Gelbstein},\ and\ \citenamefont
  {Fuks}}]{kirievsky2013phase}%
  \BibitemOpen
  \bibfield  {author} {\bibinfo {author} {\bibfnamefont {K.}~\bibnamefont
  {Kirievsky}}, \bibinfo {author} {\bibfnamefont {Y.}~\bibnamefont
  {Gelbstein}}, \ and\ \bibinfo {author} {\bibfnamefont {D.}~\bibnamefont
  {Fuks}},\ }\href@noop {} {\bibfield  {journal} {\bibinfo  {journal} {Journal
  of Solid State Chemistry}\ }\textbf {\bibinfo {volume} {203}},\ \bibinfo
  {pages} {247} (\bibinfo {year} {2013})}\BibitemShut {NoStop}%
\bibitem [{\citenamefont {Berche}\ and\ \citenamefont
  {Jund}(2018)}]{berche2018oxidation}%
  \BibitemOpen
  \bibfield  {author} {\bibinfo {author} {\bibfnamefont {A.}~\bibnamefont
  {Berche}}\ and\ \bibinfo {author} {\bibfnamefont {P.}~\bibnamefont {Jund}},\
  }\href@noop {} {\bibfield  {journal} {\bibinfo  {journal} {Intermetallics}\
  }\textbf {\bibinfo {volume} {92}},\ \bibinfo {pages} {62} (\bibinfo {year}
  {2018})}\BibitemShut {NoStop}%
\bibitem [{\citenamefont {Young}\ and\ \citenamefont
  {Reddy}(2019)}]{young2019processing}%
  \BibitemOpen
  \bibfield  {author} {\bibinfo {author} {\bibfnamefont {J.}~\bibnamefont
  {Young}}\ and\ \bibinfo {author} {\bibfnamefont {R.}~\bibnamefont {Reddy}},\
  }\href@noop {} {\bibfield  {journal} {\bibinfo  {journal} {Journal of
  Materials Engineering and Performance}\ }\textbf {\bibinfo {volume} {28}},\
  \bibinfo {pages} {5917} (\bibinfo {year} {2019})}\BibitemShut {NoStop}%
\end{thebibliography}%
\end{document}